\documentclass[12pt]{article}
\usepackage{amsmath}
\usepackage{amssymb}
\begin{document}
\begin{center}
{\bf {\large{Constraints and Conserved Charges for Modified Massive  and Massless  Abelian 1-Form and 2-Form Theories: A Brief Review}}} 

\vskip 3 cm

{\sf  A. K. Rao$^{(a)}$, B. Chauhan$^{(a)}$, R. P. Malik$^{(a,b)}$}\\
$^{(a)}$ {\it Physics Department, Institute of Science,}\\
{\it Banaras Hindu University, Varanasi - 221 005, (U.P.), India}\\

\vskip 0.1cm

$^{(b)}$ {\it DST Centre for Interdisciplinary Mathematical Sciences,}\\
{\it Institute of Science, Banaras Hindu University, Varanasi - 221 005, India}\\
{\small {\sf {e-mails:  amit.akrao@gmail.com; bchauhan501gmail.com; rpmalik1995@gmail.com}}}
\end{center}

\vskip 2.0 cm

\noindent
{\bf Abstract:}
We demonstrate that the generators for the local, continuous and infinitesimal
{\it classical} gauge symmetry transformations in the cases of (i) the St$\ddot u$ckelberg-modified {\it massive} Abelian 1-form and 2-form
theories, and (ii)  the {\it massless} Abelian 1-form and
2-form free theories owe their origin to the first-class constraints (on {\it these}  theories).  
We discuss the appearance of {\it these}  constraints at the quantum level, within the framework of
Becchi-Rouet-Stora-Tyutin (BRST) formalism, through the physicality criteria w.r.t. the
conserved and nilpotent (anti-)BRST charges. 
One of the highlights of our present investigation is the derivation of the nilpotent versions of the (anti-)BRST charges
(from the standard non-nilpotent Noether (anti-)BRST charges) which lead to the appearance of the operator forms of the first-class constraints 
through the physicality criteria at the {\it quantum} level
in the context of the   {\it modified} massive and {\it massless} Abelian 2-form theories. We also comment on (i)
the existence of the Curci-Ferrari (CF) type 
restrictions on the Abelian 2-form theories ({\it with} and {\it without} mass), (ii) the modifications in the   
St$\ddot u$ckelberg-technique for the massive 2D Abelian 1-form and 4D Abelian 2-form theories and their consequences, and (iii) the off-shell nilpotent version of the conserved co-BRST charge and its role in the physicality criteria  for the St$\ddot u$ckelberg-modified 
4D massive Abelian 2-form theory.



\newpage
\section {Introduction}
The ideas of symmetries in physics have brought together the sophisticated concepts of mathematics
and the physical descriptions of the natural realities on a platform where the deep 
insights into the working principle behind  a natural phenomenon (learnt and understood by 
the theoretical physicists and mathematicians) have been enriched in  a meaningful manner 
(see, e.g. [1-4] and references therein). In fact, the principles of the local symmetries dictate interactions [1]. One such symmetry, 
in the context of the above, 
 is the {\it gauge} symmetry which has been able to play a decisive role in the precise theoretical description 
of the {\it three} out of  four fundamental interactions  of nature.
 The most modern definition of a gauge symmetry is elucidated in terms of
the first-class constraints in the terminology of Dirac's prescription for the classification scheme of 
constraints (see, e.g. [5-8] for details). In other words, {\it those} continuous and local symmetry transformations 
 are called as  the {\it gauge} symmetry transformations that are generated by the 
first-class constraints. In this context, mention can be made of the $U(1)$ 
Abelian local gauge symmetry transformations  that describe and dictate  
the electromagnetic interaction and the $SU(N)$ non-Abelian gauge symmetry transformations of the 
Yang-Mills theories that play very important roles in the precise theoretical description of the weak and strong interactions of nature. 
The culmination  of the success story of the gauge symmetry transformations is the systematic construction of the
standard model of particle physics (SMPP) where there is stunning degree of agreements between theory and experiment
(see, e.g. [9-13] for details).

Despite many successes, the standard model of particle physics (based on the {\it interacting} non-Abelian 1-form gauge theory)
is {\it not} a complete theory due to the experimental evidence of the masses of all the  three varieties of neutrinos. Hence, one has to 
go beyond the realm of the validity of the SMPP. One  of the most promising 
candidates, for such an endeavor in theoretical high energy physics, is the idea of superstring theories 
(see, e.g. [14-18] and references therein) whose quantum excitations lead to the existence of the higher $p$-form 
$(p = 2, 3, ...)$ gauge fields. Thus, the study of the higher $p$-form gauge theories is important and essential. 
In our present endeavor, we have concentrated on the study of the {\it modified} 
massive and massless Abelian 2-form gauge  theories  which are the  modest examples as well as  the 
lowest rungs\footnote{The
idea of an antisymmetric tensor  Abelian 2-form (i.e. $B^{(2)} = [(d\,x^\mu \wedge d\,x^\nu)/{2!}]\,B_{\mu\nu}$) 
gauge  field was introduced by V. I. Ogievetsky  and 
 I. V. Palubarinov: {\it Sov. J. Nucl. Phys.} (Yad. Fiz.) {\bf 4}, 156  (1967) 
who coined the nomenclature ``notoph" (i.e. the opposite of ``photon") for this field whose Dirac's quantization scheme and  
 constraint structures were studied 
by R. K. Kaul: {\it Phys. Rev.} D {\bf 18}, 1127 (1978). This Abelian 2-form theory, in modern times, is popularly known as the 
Kalb-Ramond theory.}
of the tower of  the
 higher $p$-form  $(p = 2, 3, ...)$ gauge fields that appear in the quantum excitations of the (super)string theories. 
We have purposely  discussed the {\it modified} massive and massless Abelian $p = 1, 2$ form gauge theories 
{\it together} so that the decisive differences and striking  similarities between  the two types of  
theories, at the classical and  quantum levels, could become clear. As far as 
the discussions on the above two types of theories  at the {\it quantum} level  is concerned, 
we have chosen the mathematically rich and physically very  intuitive theoretical 
method of BRST formalism [19-22] where the sacrosanct requirements of unitarity and {\it quantum} gauge  
[i.e.  (anti-)BRST] invariance are respected {\it together}  at any arbitrary order of  perturbative computations for a given 
physical process  that is allowed by the (anti-)BRST invariant  theory (see, e.g. [23-26, 8]).

In our present endeavor, we have laid a great deal of emphasis on the existence of the 
first-class constraints and the associated gauge as well as the (anti-)BRST symmetry transformations 
and established a deep connection between the first-class constraints and the Noether conserved charges. 
The {\it latter} are derived from (i) the infinitesimal, local and continuous {\it gauge} 
symmetry transformations that are generated by the first-class constraints at the {\it classical} level, and 
(ii) the (anti-)BRST symmetry transformations which are the generalizations of the 
{\it classical} local gauge symmetry transformations to the {\it quantum} level. 
The appearance of the operator forms of the constraints of the {\it classical} gauge theories 
({\it with} mass and {\it without} mass), at the quantum level, has been established by the requirement 
of the physicality criteria where we demand that the physical states (in the total Hilbert space of quantum states)
are {\it those} that are annihilated by the conserved and off-shell nilpotent (anti-)BRST charges. 
Furthermore, it has been pointed out that the 2D {\it modified} Proca  and 
4D {\it modified} massive Abelian 2-form theories are field-theoretic models of
Hodge theory [27-30] because the physical states are annihilated {\it together}  by the 
(i) the first-class constraints, and (ii) the {\it dual} versions of the first-class constraints
of the {\it classical} gauge theories (of the {\it modified} massive and massless varieties) within the 
framework of BRST formalism. The above statements  are correct  because the physical states are annihilated 
by the conserved and off-shell nilpotent versions of the (anti-)BRST and (anti-)co-BRST charges {\it together}.   
Hence, these specific theories in 2D and 4D are very {\it special} in some sense 
because  the ideas from the topological field theories [31, 32] {\it also} play some crucial roles  
in  the {\it massless} versions of these Abelian 1-form and 2-form gauge theories [33, 34].

Our present brief review article has been spurred by the following key motivating 
factors. First, we have discussed various aspects of the {\it massless} and modified {\it massive} versions of the
Abelian 3-form gauge theories in our earlier works within the purview of BRST [35, 36]  and related superfield 
formalism [37]. Hence, it has been a challenge for us to assimilate our insights into the
various aspects of the {\it massless} and modified {\it massive} Abelian 1-form and 2-form theories [38, 39, 27-30, 33, 34] in a coherent 
fashion at one place and discuss the deep connections that exist between the conserved charges and the first-class constraints of {\it these}
theories at the classical and quantum levels. 
Second, the 2D {\it modified} massive Abelian 1-form and 4D {\it modified} massive Abelian 2-form 
theories have been very {\it special} in our studies over the years [27-30, 33, 34, 38, 39]. It has been a challenge for us to state 
something worthwhile about these theories coherently in a single article at one place.  
We have accomplished this goal in our present endeavor. 
Third, in  our earlier work [40], we have systematically developed a theoretical method 
to obtain the nilpotent versions of the (anti-)BRST charges from the standard conserved {\it but}
non-nilpotent versions of the Noether conserved (anti-)BRST charges. However, we have not 
discussed anything  about the (anti-)co-BRST charges that exist for the field-theoretic models 
of Hodge theory (see, e.g. [27-30]). We have accomplished this goal, too, in our present endeavor.   
Finally, in all our previous works on the Abelian $p = 1, 2, 3$ form theories of massless and {\it modified} massive
varieties, we have {\it not} concentrated on the constraints. In our present endeavor, we have given a great deal of 
importance to the first-class constraints.

The theoretical contents of our present endeavor  are organized as follows. First of all, as
a warm-up exercise, we consider the case of the {\it modified} massive and  {\it massless}  Abelian 1-form
theories (i) to discuss {\it their} first-class constraints and associated generators, 
(ii) to establish  connections of the {\it latter}  with the Noether conserved charges, and 
(iii) to show the existence of {\it these} constraints at the {\it quantum}  level within the framework of
BRST formalism in our Sec. 2. Our Sec. 3 is devoted to the discussion on the existence
of the first-class constraints for the St$\ddot u$ckelberg-modified massive Abelian 2-form theory  
whose {\it massless}  limit is valid for the {\it free} Abelian 2-form {\it gauge} theory. In Sec. 4, we concentrate on the construction of the generator, in terms of the first-class constraints, for the {\it modified} 
massive Abelian 2-form theory whose {\it massless}  limit is {\it true} for the Abelian 2-form {\it gauge} theory.
Our Sec. 5 deals with the derivations of the Noether conserved currents and charges for the
{\it above} Abelian 2-form theories ({\it without} mass and {\it with} mass) within the framework of BRST formalism. 
In Sec. 6, we discuss the modifications in the St$\ddot u$ckelberg-technique for the 2D  Proca and 4D
 massive Abelian 2-form theories.
Finally, we make some concluding remarks and point out a few future directions for further investigations in our Sec. 7.

Our Appendix A is devoted to the derivation of the proper 2D Lagrangian density for the {\it modified } 2D Proca theory where only 
the {\it covariant} notations have been used.
In our Appendix  B, we comment briefly on the existence of the  CF-type restrictions for the massless and 
{\it modified} massive Abelian 2-form theories from the symmetry considerations.  \\

{\it Convention and Notations:} 
 We adopt the convention of the left-derivative w.r.t. all the
{\it fermionic} fields of our theories. The background D-dimensional  flat Minkowskian spacetime manifold is
endowed with the metric tensor  $\eta_{\mu\nu} =$ diag $(+1, -1, -1,...)$  so that the dot product between
two non-null Lorentz vectors $S_\mu$ and $T_\mu$ is defined as: $S \cdot T = \eta_{\mu\nu}\, S^\mu \, T^\nu  = S_0 \, T_0 - S_i \,  T_i$
where the Greek indices $\mu,\,\nu,\,\lambda,... = 0, 1, 2,..., D-1 $  denote the time and space directions
and the Latin indices $i,\,j,\,k... = 1, 2,... D-1$ stand for the space directions {\it only}. Einstein's
summation convention has been adopted in the whole body of our text. We always denote
the (anti-)BRST symmetry transformations by the symbols $s_{(a)b}$. We adopt the notation $\delta_g$ for 
the infinitesimal, continuous and local  {\it classical} gauge symmetry transformations in the entire text. The over-dot on a generic field $\Psi$  
denotes the partial time derivative as: $\dot\Psi = (\partial\, \Psi / \partial\, t)$.\\

\section {St$\ddot u$ckelberg Modified Massive and Massless Abelian 1-Form Theories: A Bird's Eye View}

Our present section is divided into {\it three} parts. In subsection 2.1, we discuss the constraint structures
of the St$\ddot u$ckelberg-modified Proca (i.e. massive Abelian 1-form) and
{\it massless} Abelian 1-form {\it gauge}  theories  and establish that
they belong to the first-class variety (see, e.g. [5-8]).  These constraints are present in the standard
expressions for the generators of the gauge transformations.  In subsection 2.2, we derive the Noether conserved charges
from  the local, infinitesimal and continuous {\it classical}  gauge symmetry transformations and 
establish their connections with the {\it classical} symmetry generators (cf. Subsec. 2.1). Finally, in subsection
2.3, we demand  the physicality criteria  w.r.t the conserved and off-shell nilpotent (anti-)BRST charges 
and demonstrate  the existence of the first-class constraints 
of the {\it original} St$\ddot u$ckelberg-modified {\it massive} and {\it massless} gauge 
theories (in their operator forms) at the {\it quantum}  level
within the framework of BRST formalism which are found to be consistent with Dirac's
quantization conditions  for the systems endowed with any kinds of constraints (see, e.g. [6, 8] for details).

\subsection{First-Class Constraints and Generators}

In this subsection, we begin with the Lagrangian density for the D-dimensional Proca (i.e. {\it massive} Abelian
1-form) theory and discuss its {\it massless} limit in a concise manner so that the essence of
the Maxwell's theory can {\it also} be captured (as far as the constraint structures are  concerned).
We begin with the following Lagrangian density (with rest mass $m$ of the gauge boson) 
for the Proca theory (see, e.g. [41, 38, 39] for details)
\begin{eqnarray}
{\cal L}_{(P)} = -\frac {1}{4} F_{\mu\nu} F^{\mu\nu} +  \frac {m^2}{2}  A_\mu \, A^\mu,
\end{eqnarray}
where the 2-form $F^{(2)} = d A^{(1)} =  \frac{1} {2!}\,F_{\mu\nu}\; (d x^\mu \wedge  d x^\nu)$ 
defines the field-strength tensor  $F_{\mu\nu}=\partial_\mu A_\nu - \partial_\nu A_\mu$   
in terms of the Abelian 1-form ($A^{(1)} =  A_\mu \;  d x^\mu$)  {\it massive} vector field $A_\mu$.
Here the mathematical symbol $d$ stands for the exterior derivative $d = \partial_\mu \; dx^\mu$
[with $d^2 \equiv \frac{1} {2!}\,(\partial_\mu \, \partial_\nu - \partial_\nu \, \partial_\mu)\; 
(d x^\mu\; \wedge  d x^\nu) = 0 \,\mbox{where} \,\mu, \nu = 0, 1, 2,...D-1$].
The above Lagrangian density respects {\it no} gauge symmetry transformations because it is
endowed with the second-class constraints in the terminology of Dirac's prescription for
the classification scheme of constraints (see, e.g. [5-8] for details). To restore the gauge symmetry
transformations, we exploit the beauty and theoretical strength of the St$\ddot u$ckelberg-technique of the compensating
field ($\phi$) and take into account  the replacement: $A_\mu \longrightarrow A_\mu \mp \frac {1}{m}\, \partial_\mu \phi$
where the field $\phi$ is a {\it pure} scalar field. The resulting St$\ddot u$ckelberg-modified Lagrangian 
density  [${\cal L}_{(S)}$]  for the {\it modified} Proca theory is as follows (see, e.g. [41, 38, 39] for details): 
\begin{eqnarray} 
{\cal L}_{(P)} \longrightarrow {\cal L}_{(S)} = -\frac {1}{4} F_{\mu\nu} F^{\mu\nu} +  \frac {m^2}{2}  A_\mu \, A^\mu  
 \mp m \, A_\mu \partial^\mu \phi + \frac {1}{2}\partial_\mu\phi \, \partial^\mu\phi,
\end{eqnarray}
The above {\it modified} Lagrangian density [${\cal L}_{(S)}$] is endowed with the first-class constraints. 
To corroborate  this claim, first of all, we compute the conjugate momenta w.r.t. the basic 
fields $A_\mu$ and $\phi$ (of our St$\ddot u$ckelberg-modified massive Abelian 1-form theory) as:  
\begin{eqnarray} 
&&\Pi^\mu_{(A)}   = \frac{\partial\, {\cal L}_{(S)}}{\partial\, (\partial_0\, A_\mu)} = -\, F^{0\mu}
\; \Longrightarrow \;  \Pi^0_{(A)}  =  -\, F^{00} \approx 0   \quad  \Longrightarrow \quad (P. C.),   \nonumber\\ 
&& \Pi_{(\phi)} =  \frac{\partial\, {\cal L}_{(S)}}{\partial\, (\partial_0\, \phi)} 
= \mp  m\, A^0 + \partial^0\, \phi. 
\end{eqnarray}
In the above, we have adopted Dirac's notation for the concept of {\it weakly}  zero (in $\Pi^0_{(A)} = 
 -\, F^{00} \approx 0$) on our theory to
define the primary constraint (P. C.). The time-evolution invariance [24]
[i.e. $({\partial\, \Pi^0_{(A)} }/{\partial\, t}) \approx 0$] of the naturally present P. C. 
(i.e. $\Pi^0_{(A)} \approx 0$) on our theory can be computed  (see, e.g. [24]) from the following Euler-Lagrange equation of motion
(EL-EoM) that is derived (w.r.t. the basic field $A_\mu$) from  ${\cal L}_{(S)} $, namely;
\begin{eqnarray} 
\partial_\mu\, F^{\mu\nu} + m^2 \, A^\nu \mp  m\, \partial^\nu\, \phi = 0, 
\end{eqnarray}
provided we take into account $\nu = 0$,  in the above equation,  to obtain: 
\begin{eqnarray} 
\partial_0 \, F^{00} &+& \partial_i \, F^{i0} + m^2 \, A^0 \mp  m\, \partial^0 \, \phi = 0, \nonumber\\
\frac{\partial\, \Pi^0_{(A)} }{\partial\, t} &=& \partial_i\, E_i \mp   m\, \Pi_{(\phi)} \approx 0 \nonumber\\ 
&  \equiv& \partial_i\, \Pi^i_{(A)} \mp m\, \Pi_{\phi} \approx 0  \quad \Longrightarrow  \quad (S. C.). 
\end{eqnarray}
The above relationship is nothing but the derivation of secondary constraint (S.C.) on our {\it modified} Proca theory
(see, e.g. [24]) where $\Pi^i_{(A)} = -\, F^{0i} = E_i$ are  the components of the  electric fields  (i.e. $E_i \equiv \vec E$) 
which are the canonical conjugate momenta 
w.r.t. the vector field $A_i \equiv \vec A$. There are {\it no} further constraints on the theory. Since
the P. C. and S. C. are expressed in terms of the components of the conjugate momenta, it is evident 
that {\it both}  will commute  with each-other thereby leading to their classification as the first-class 
constraints in the terminology of Dirac's prescription for the classification scheme of constraints [5-8].

Following the standard formula [42, 43, 26, 8], we can
write down the generator for the local, infinitesimal and continuous {\it classical}  gauge symmetry
transformations for the  St$\ddot u$ckelberg-modified Lagrangian density [${\cal L}_{(S)}$], namely; 
\begin{eqnarray}
\delta_g\, A_{\mu} = \partial_\mu\,\chi,\qquad \delta_g\,\phi = \pm\,m\,\chi,\qquad \delta_g\, F_{\mu\nu} = 0,
\end{eqnarray}
in terms of the first-class constraints as (see, e.g. [42, 43] for details)
\begin{eqnarray} 
G = \int d^{D-1} x \, \Big[\big(\partial_0\, \chi \big)\, \Pi^0_{(A)} - \chi\, \big( \partial_i\, \Pi^i_{(A)} \mp m\, \Pi_{(\phi)} \big)  \Big],  
\end{eqnarray}
where $\chi\, (\vec x, t)$ is the physically well-defined infinitesimal, continuous  and local gauge symmetry transformation parameter. 
Using the Gauss divergence theorem, the above generator can be re-expressed (in a very transparent and useful form) as follows: 
\begin{eqnarray} 
G =  \int d^{D-1} x \, \Big[(\partial_0 \,  \chi )\, \Pi^0_{(A)} + \big(\partial_i\, \chi \big)\, \Pi^i_{(A)} \pm m\, \chi\, \Pi_{(\phi)} \Big ]. 
\end{eqnarray}
At this juncture, we define the following {\it non-trivial} equal-time commutators for the basic fields $A_\mu$ and $\phi$
of our {\it modified} Proca theory, in the natural units ($\hbar = c = 1$), as: 
\begin{eqnarray} 
&&[A_0\, (\vec x, t),\,  \Pi^0_{(A)}\, (\vec y, t)] = i\, \delta^{(D- 1)}\, (\vec x - \vec y), \nonumber\\
&&[A_i\, (\vec x, t),\,  E_j\, (\vec y, t)] = i\,\delta_{ij} \, \delta^{(D- 1)}\, (\vec x - \vec y), \nonumber\\
&&[\phi \, (\vec x, t), \, \Pi_{(\phi)} \, (\vec y, t)] = i\, \delta^{(D- 1)}\, (\vec x - \vec y). 
\end{eqnarray}
All the rest of the other equal-time brackets are {\it trivially} zero. It is crystal clear now that, we have the following 
infinitesimal, local and continuous {\it classical} gauge symmetry transformations for the basic fields in terms 
of the commutators between the basic fields and the 
generator $(G)$ of our  St$\ddot u$ckelberg-modified Proca theory, namely; 
\begin{eqnarray} 
&&\delta_g\,\phi \, (\vec x, t) =  -\, i \, [\, \phi \, (\vec x, t),\,  G] = \pm  m\ \chi\, (\vec x, t), \nonumber\\ 
&&\delta_g\, A_0\, (\vec x, t) = -\, i \, [A_0\, (\vec x, t), \, G] = \partial_0\, \chi\,(\vec x, t), \nonumber\\
&&\delta_g\, A_i\, (\vec x, t) = -\, i \,  [A_i\, (\vec x, t), \, G] = \partial_i\, \chi\, (\vec x, t). 
\end{eqnarray} 
Hence, we have derived the local, infinitesimal and continuous {\it classical} gauge 
symmetry transformations (6) for the {\it basic} fields which, automatically, imply that 
$\delta_g\, F_{\mu\nu} = 0$.

We conclude this subsection with the following remarks. First, it is clear that, in the {\it massless} limit 
(i.e. $m = 0, \,\phi = 0$) of the Proca theory, we obtain the usual $U (1)$ gauge invariant Maxwell's theory
whose gauge symmetry generator $(G^{(M)})$ can be obtained  from $G$ [cf. Eq. (8), with $m = 0$] as follows 
\begin{eqnarray} 
G^{(M)} &=& \int d^{D-1} x \, \Big[\big(\partial_0\, \chi \big)\, \Pi^0_{(A)} 
- \big( \partial_i\, \Pi^i_{(A)} \big)\,\chi \Big] \nonumber\\
&\equiv& \int d^{D-1} x \, \Big[\big(\partial_0 \chi\big) \, \Pi^0_{(A)} 
+ \big(\partial_i\, \chi \big)\, \Pi^i_{(A)} \Big ],  
\end{eqnarray}
which leads to the $U (1)$ gauge symmetry transformation: $\delta_g\, A_\mu = \partial_\mu\, \chi$
if we use the canonical equal-time commutators (10) appropriately. Second, we note that the gauge theories are always 
endowed with the first-class constraints in the terminology of Dirac's prescription for classification scheme of 
constraints (see, e.g. [5-8]). Third, the first-class constraints are responsible for the appropriate definitions of the 
gauge symmetry generators. In other words, the first-class constraints generate the gauge symmetry transformations which lead to the 
sacrosanct (and the most modern) definition of a {\it classical} gauge theory. 
Fourth, it is the presence
of the St$\ddot u$ckelberg-compensating field ($\phi$) that has been able to convert the second-class
constraints of the {\it original}  Proca Lagrangian density [${\cal L}_{(P)}$] into the first-class constraints of the
{\it modified}  Lagrangian density [${\cal L}_{(S)} $].  Fifth, it is clear from the P. C.  [cf. Eq. (3)] and S. C. [cf. Eq. (5)]  that (in
the {\it massless}  limit) we shall have, once again, the first-class constraints where $\Pi^0_{(A)} \approx 0$ 
as the P. C. and the S. C. [cf. Eq. (5)] reduces to:  $\partial_i\, E_i = \vec \nabla \cdot \vec E \approx 0$ (i.e. the Gauss law
of the source-free Maxwell's equations). Finally, we note that the {\it modified}  Proca theory and 
Maxwell's theory present themselves as the field-theoretic examples of the {\it simple}  gauge theories  as they respect the  infinitesimal, 
continuous and local {\it classical} gauge symmetries
that are generated by the first-class constraints.\\

\subsection{Noether Conserved Charge: First-Class Constraints}

The purpose of this subsection is to establish a connection between the Noether 
conserved charge and the first-class constraints of the St$\ddot u$ckelberg-modified massive Abelian 1-form 
(i.e. Proca) theory whose massless limit will be valid for Maxwell's theory. 
In this connection, we would like to point out that the local, continuous and infinitesimal 
{\it classical} gauge symmetry transformations $(\delta_g)$ of equation (6) lead to the derivation of the 
following Noether current $(J^\mu_{(A)})$ from the {\it modified} Lagrangian density  [${\cal L}_{(S)} $], namely;  
\begin{eqnarray} 
J^\mu_{(A)} = -\, F^{\mu\nu}\, (\partial_\nu\, \chi)  \pm m\, (\partial^\mu \, \phi \mp m\, A^\mu)\, \chi, 
\end{eqnarray}
where the superscript $(A)$ on the Noether current  denotes the fact that we are dealing with the {\it massive} 
Abelian 1-form ($A^{(1)} =  A_\mu \; d x^\mu$) theory where the {\it basic}
gauge field is $A_\mu$. The conservation law ($\partial_\mu\, J^\mu_{(A)} = 0$) can be proven by using 
the following EL-EoMs 
\begin{eqnarray} 
&&\partial_\mu\, F^{\mu\nu} + m^2\, A^\nu \mp m\, \partial^\mu\, \phi = 0, \nonumber\\
&&\partial_\mu\, [\partial^\mu\, \phi \mp {m}\, A^\mu] = 0 \; \; \Longrightarrow \; \; \Box\, \phi \mp m\, (\partial \cdot A) = 0,
\end{eqnarray}
that emerge out from ${\cal L}_{(S)} $ [cf. Eq. (2)] w.r.t. the {\it basic} fields $A_\mu$ and $\phi$ of our 
St$\ddot u$ckelberg-{\it modified} massive Abelian 1-form  theory,  respectively.

At this juncture,  following the sacrosanct principle of Noether's theorem, 
we define the conserved charge $Q_{(A)}$ (i.e. $Q_{(A)} = \int d^{D - 1}\,x \, J^0_{(A)}$)  as follows:
\begin{eqnarray} 
 \nonumber\\ 
 Q_{(A)} &\equiv& \int d^{D - 1}\,x \, \Big[ -\, F^{0\nu}\, \partial_\nu\, \chi \pm m\, (\partial^0\, \phi \mp m\ A^0)\, \chi   \Big]. 
\end{eqnarray}
Since our theory is endowed with the constraints, we have to be careful in expanding the r.h.s. of the above equation. 
Taking into account the primary constraint ($\Pi^0_{(A)} =  -\, F^{00} \approx 0$) and the secondary constraint 
($\partial_i\, \Pi^i_{(A)} \mp m\, \Pi_{\phi} \approx 0$), we obtain the following: 
\begin{eqnarray} 
Q_{(A)} &=&  \int d^{D - 1}\,x \, \Big[ -\, F^{00}\, \partial_0 \, \chi -\, F^{0i}\, \partial_{i} \, \chi 
\pm m\, \Pi_{(\phi)} \, \chi \Big] \nonumber\\ 
&\equiv&  \int d^{D - 1}\,x \, \Big[ \Pi^0_{(A)} \, (\partial_0 \, \chi) + \Pi^i_{(A)} \, (\partial_{i} \, \chi)
\pm m\, \Pi_{(\phi)} \, \chi \Big]. 
\end{eqnarray}
Using  Gauss's divergence theorem, the above equation can be recast into the following
\begin{eqnarray} 
Q_{(A)}  = \int d^{D - 1}\,x \, \Big[ \Pi^0_{(A)} \, (\partial_0 \, \chi) 
- \big( \partial_{i}\, \Pi^i_{(A)} \mp  m\, \Pi_{(\phi)} \big ) \, \chi \Big],
\end{eqnarray} 
where the  primary constraint ($\Pi^0_{(A)}  \approx 0$) and the secondary constraint 
($\partial_i\, \Pi^i_{(A)} \mp m\, \Pi_{\phi} \approx 0$) appear explicitly. A close look at 
the above equation demonstrates that we have already derived the generator $G$ [cf. Eq. (8)] of our previous subsection
as $Q_{(A)}$ which contains the primary and  secondary constraints of our {\it modified} massive Abelian 1-form theory.

We end this subsection with the following remarks. First of all, as pointed out earlier, it is the presence of the 
St$\ddot u$ckelberg-compensating pure scalar field $(\phi)$ that is responsible for the conversion of the 
second-class constraints of the {\it original} Proca  theory into the first-class constraints. 
It is obvious that, for the {\it massless} Abelian 1-form gauge theory (where $m = 0, \, \phi = 0$),
the constraints: $\Pi^0_{(A)} \approx 0$ and  $\partial_i \Pi^i_{(A)} \approx 0$ are the first-class in nature, too. 
Second, the standard form of the generators for the local, continuous and infinitesimal {\it classical} gauge symmetry transformations 
(see, e.g. [42, 43, 26, 8]) in the cases of the {\it modified} massive and massless theories are, in fact,  deeply connected with the Noether conserved charges which contain the first-class constraints 
in a precise manner. Third, we would like to lay emphasis that, because of the presence of the constraints, 
the expression for the conserved charge (derived from the conserved Noether current) should be expressed in a careful manner 
so that the constraints  appear at the precise places in it. In other words, the first-class constraints should 
{\it not} be {\it strongly} set equal to zero whenever they appear in the expression for the conserved charge.
Finally, we note that the $Q_{(A)}$ in (16) reduces  (with $m = 0, \, \phi  = 0$) to the Noether conserved charge for the Maxwell theory 
(i.e. $Q_{(A)} \longrightarrow Q_{(A)}^{(M)} = \int d^{D - 1} x\, [(\partial_0 \chi)\,\Pi^0_{(A)} - \chi\, (\partial_i \Pi^i _{(A)})]$)
which is equal to the generator $G^{(M)}$ in (11) where the superscript $(M)$ stands for the Maxwell theory. The {\it latter}
contains the first-class constraints (i.e.  $\Pi^0_{(A)} \approx 0$ and $ \partial_i\Pi^i_{(A)} \approx 0$) in an explicit manner.

\subsection{(Anti-)BRST Charges: First-Class Constraints}

The central objective of our present subsection is to establish the existence of the 
first-class constraints (i.e. $\Pi^0_{(A)}  \approx 0, \; \partial_i\, \Pi^i_{(A)} \mp m\, \Pi_{\phi} \approx 0$)
at the {\it quantum} level within the framework of Becchi-Rouet-Stora-Tyutin (BRST) formalism 
where we demonstrate that the physicality criteria ($Q_{(a)b}| phys > = 0$) on the physical states 
(i.e. $| phys > $) w.r.t. the conserved and off-shell nilpotent (anti-)BRST charges $[Q_{(a)b}]$ of our theory 
lead to the annihilation of the physical states (in the {\it total} Hilbert space of states) by the operator 
forms of the first-class constraints. Towards this goal in mind, we note that the following off-shell nilpotent 
[$s_{(a)b}^2 = 0$] and absolutely anticommuting $(s_b\, s_{ab} + s_{ab}\, s_b = 0)$ (anti-)BRST symmetry transformations [$s_{(a)b}$]
\begin{eqnarray}
&&s_{ab}\,A_{\mu} = \partial_\mu\,\bar C,\quad s_{ab}\,\bar C = 0, \quad s_{ab}\,C = -\,i\,B,\quad s_{ab}\,B = 0,
\quad s_{ab}\,\phi = \pm\, m\,\bar C,\nonumber\\
&&s_{b}\,A_{\mu} = \partial_\mu\, C,\qquad s_{b}\,C = 0, \quad s_{b}\,\bar C = i\,B,\qquad s_b\,B = 0,\qquad s_{b}\,\phi = \pm\, m\,C,
\end{eqnarray}
leave the following (anti-)BRST invariant Lagrangian density
\begin{eqnarray} 
{\cal L}_{B}&  = & {\cal L}_{(S)} + B\,(\partial \cdot A \pm m\,\phi) + \frac {B^2}{2}
 -\,i\,\partial_\mu\,\bar C\,\partial^\mu\,C + i\,m^2\,\bar C\,C\nonumber\\
& \equiv & -\,\frac {1}{4}\, F^{\mu\nu}\,F_{\mu\nu}  + \frac {m^2}{2}\,A_\mu\, A^\mu \mp m\, A^\mu \partial_\mu \phi + \frac {1}{2}\, \partial_\mu \phi\, \partial^\mu \phi + B\, (\partial\cdot A \pm m\, \phi)\nonumber\\ 
& + & \frac {B^2}{2} - i\, \partial_\mu \bar C	 \, \partial^\mu 	C + i\, m^2 \, \bar C\, C,
\end{eqnarray} 
quasi-invariant in the sense that we observe the following: 
\begin{eqnarray} 
s_{ab}\,{\cal L}_{B} = \partial_\mu \,[B\,\partial^\mu\,\bar C], \qquad\qquad 
s_b\,{\cal L}_{B } = \partial_\mu \,[B\,\partial^\mu\,C]. 
\end{eqnarray} 
In other words, the Lagrangian density (18) transforms to the {\it total} spacetime derivatives under
the (anti-)BRST symmetry transformations (17) thereby rendering the action integral $S = \int d^Dx\, {\cal L}_B$
invariant $[s_{(a)b}\, S = 0]$ under the (anti-)BRST symmetry transformations. Hence, the off-shell nilpotent, 
continuous and infinitesimal (anti-)BRST transformations are the {\it symmetry} 
transformations for the action integral of our {\it quantum} theory.

At this crucial juncture, a few key and crucial  comments are in order. First of all, the {\it quantum} (anti-)BRST symmetry transformations 
(17) are the generalizations of the {\it classical} gauge symmetry transformations (6) within the framework 
of BRST formalism. Second, the local and infinitesimal {\it classical} gauge symmetry transformation parameter $\chi \, (\vec x, t)$
has been traded with the fermionic  $(C^2 = \bar C^2 = 0, \, C\, \bar C + \bar C\, C = 0)$ (anti-)ghost 
$(\bar C) \, C$ fields which are invoked to maintain the unitarity of the {\it quantum} gauge 
(i.e. BRST and anti-BRST) symmetry invariant BRST-quantized  theory. 
Third, the St$\ddot u$ckelberg-modified Lagrangian density ${\cal L}_{(S)}$ [cf. Eq. (2)] has been generalized to the 
(anti-)BRST invariant Lagrangian density ${\cal L}_{B}$ at the {\it quantum} level that incorporates the 
gauge-fixing and Faddeev-Popov ghost terms. Fourth, the auxiliary field $B$ is the Nakanishi-Lautrup field which is invoked 
to linearize the gauge-fixing term. Finally, the covariant canonical quantization (i.e. BRST-quantization)
can be performed for  the Lagrangian density  ${\cal L}_{B}$ because {\it all} the dynamical 
fields in the theory  have their counterpart  canonical conjugate momenta.

As pointed out earlier, the (anti-)BRST symmetry transformations (17) are the infinitesimal, continuous, 
off-shell nilpotent and absolutely anticommuting {\it symmetry} transformations of the action integral. 
As a consequence, they lead  to the derivation of the Noether currents ($J^\mu_{(r)}, \, r = ab, \, b$)
due to the celebrated Noether theorem, namely; 
\begin{eqnarray}
J^{\mu}_{(ab)} = -\,F^{\mu\nu}\,\partial_\nu\,\bar C + B\, \partial^\mu\,\bar C \pm m\,\bar C\,\partial^\mu\,\phi - m^2\,\bar C\,A^\mu,\nonumber\\
J^{\mu}_{(b)} = -\,F^{\mu\nu}\,\partial_\nu\,C + B\, \partial^\mu\,C \pm m\,C\,\partial^\mu\,\phi - m^2\,C\,A^\mu,
\end{eqnarray}
which are found to be conserved ($\partial_\mu\, J^\mu_{(r)} = 0, \, r = ab, \, b$) due to the
following EL-EoMs that emerged out from the ${\cal L}_{B}$, namely; 
\begin{eqnarray}
&&\partial_\mu\,F^{\mu\nu} =  \partial^\nu\,B \pm m\,\partial^\nu\,\phi -\, m^2\,A^\nu,
\qquad  \Box\,\phi \mp m\,(\partial \cdot A) = \pm m\,B, \nonumber\\
&&(\Box + m^2)\,C = 0,  \qquad \qquad (\Box + m^2)\,\bar C = 0.  
\end{eqnarray}
According to the basic idea behind the Noether theorem, the above conserved currents in (20) lead to the definitions 
of the conserved charges $Q_{(r)} = \int d^{(D - 1)} x \, J^{0}_{(r)}$ (with $r = b, ab$). In other words,  we have 
the following explicit expressions for the charges 
\begin{eqnarray}
Q_{ab}   =  \int d^{D - 1}\,x \, \Big[- F^{0\,i}\,(\partial_{i}\,\bar C) 
+B\,\dot {\bar C} \pm m\, \bar C\,\dot \phi  - m^2\,A_0\,\bar C \big] \equiv \int d^{D - 1}\,x \,\big[B\,\dot {\bar C} - \dot B\,\bar C \big],\nonumber\\
Q_b   =  \int d^{D - 1}\,x \, \Big[- F^{0\,i}\,(\partial_{i}\,C) 
+B\,\dot C \pm m\, C\,\dot \phi  - m^2\,A_0\,C \big] \equiv \int d^{D - 1}\,x \,\big[B\,\dot C - \dot B\,C \big],
\end{eqnarray}
where we have used the {\it first} entry of the EL-EoMs that have been derived in (21) from the (anti-)BRST invariant 
Lagrangian density ${\cal L}_B$. In other words, we have used the EL-EoM: 
$\dot B = \partial_i \,F^{i0} \mp m\,\partial_0\,\phi +  m^2\,A_0$ which 
emerges out from the {\it first} entry of (21).

At this stage, we define the physical states  of our BRST-quantized theory where we demand that the physical states ($| phys>$),  
in the total Hilbert space of states,  are {\it those} that are annihilated by the conserved and off-shell nilpotent (anti-)BRST charges.
In this context, it is pertinent to point out that the (anti-)BRST charges are conserved [i.e. $\dot Q_{(a)b} \equiv \partial_0 Q_{(a)b} = 0 $]. 
To corroborate  this claim, we note that the applications of the EL-EoMs: $(\Box + m^2) \, C = 0,\; (\Box + m^2) \, \bar C = 0$
and $(\Box + m^2) \, B = 0$, and (ii) the Gauss divergence theorem lead to the expressions  for the (anti-)BRST charges to be {\it zero}
 (i.e. $\dot Q_{(a)b} = 0$) when we apply  a 
{\it direct} time derivative on $Q_{(a)b}$. Furthermore, we note that the
 off-shell nilpotency [$Q_{(a)b}^2 = 0$] of the 
(anti-)BRST charges [$Q_{(a)b}$] is also {\it true}\footnote{We shall see that the standard Noether (anti-)BRST charges are {\it not} off-shell nilpotent 
in the cases of the {\it modified} massive and massless Abelian 2-form theories because of the existence of the non-trivial CF-type restrictions. In the 
case of our D-dimensional Abelian 1-form theory, the CF-type restriction is {\it trivial}. This is why, we have 
a {\it single} (anti-)BRST invariant Lagrangian density ${\cal L}_B$  [cf. Eq. (18)] and the Noether theorem leads to the derivations of the 
{\it conserved} and off-shell {\it nilpotent} (anti-)BRST charges.}:
$s_{ab}\, Q_{ab} = -\,i\,\{Q_{ab}, Q_{ab}\} = 0 \; \Longrightarrow  \;  Q^{2}_{ab} = 0,\;
s_b\, Q_b   =   -\,i\,\{Q_b, Q_b\} = 0 \; \Longrightarrow \;Q^{2}_{b} = 0$ 
where we have used (i) the direct applications of the (anti-)BRST symmetry transformation (17) on the concise forms of the 
(anti-)BRST charges [cf. Eq. (22)], and (ii) the basic relationship between  the continuous symmetry transformations and their generators
 (as the conserved Noether charges). We now {\it impose} the physicality criteria on the physical states (i.e. $| \,phys> $) w.r.t. the conserved and off-shell 
 nilpotent (anti-)BRST charges, namely; 
 \begin{eqnarray}
Q_{(a)b}\,|\,phys> \,= 0 \quad \Longrightarrow \quad  B\, |\,phys> \,= 0, \qquad  \quad  \dot B \, |\,phys> \,= 0,
\end{eqnarray}
where we have taken onto account the non-physical nature of the (anti-)ghost fields. 
At this juncture, we note that the canonical conjugate momenta w.r.t. the basic fields $A_\mu$ and $\phi$ from the 
Lagrangian density ${\cal L}_B$ are:   
\begin{eqnarray} 
&&\Pi^\mu_{(A)}   = \frac{\partial\, {\cal L}_{(S)}}{\partial\, (\partial_0\, A_\mu)} = -\, F^{0\mu} + \eta^{\mu0}\, B
\; \Longrightarrow \;  \Pi^0_{(A)}  =  B,   \nonumber\\ 
&& \Pi_{(\phi)} =  \frac{\partial\, {\cal L}_{(S)}}{\partial\, (\partial_0\, \phi)} 
= \mp  \,m\, A^0 + \partial^0\, \phi \equiv \partial_0 \phi \mp \, m\, A_0. 
\end{eqnarray}
On the other hand, from the appropriate equations of motion (21), we obtain the following 
\begin{eqnarray} 
\dot B  & = & \partial_i E_i + m^2\, A_0 \mp m\, \phi \; \equiv  \; \partial_i \Pi^{i}_{(A)} \mp m\, \Pi_{\phi},
\end{eqnarray}
which is nothing but the secondary constraint of the {\it original} St{$\ddot u$}ckelberg-modified Lagrangian density [${\cal L}_{(S)}$].
It is interesting  to point out   that the Nakanishi-Lautrup field $B$ is nothing but the expression for 
$\Pi^{0}_{(A)}$ (which is the P. C. emerging out from   ${\cal L}_{(S)})$. Thus, it is perfectly obvious, from  the physicality 
criteria (23), that we have:  
 \begin{eqnarray}
&& B\, |\,phys> \,= 0 \quad \Longrightarrow \quad  \; \Pi^0 _{(A)}\, |\,phys> \,= 0, \nonumber\\
&& \dot B\, |\,phys> \,= 0 \quad \Longrightarrow \quad  \; ( \partial_i \Pi^{i}_{(A)} \mp m\, \Pi_{\phi})\, |\,phys> \,= 0.
\end{eqnarray}
The above observations show that the operator forms of the first-class constraints of the {\it original} St{$\ddot u$}ckelberg-modified
{\it massive} theory [described by ${\cal L}_{(S)}$]  annihilate the physical states. Thus, the physicality criteria [cf. Eq. (23)]
w.r.t. the conserved and off-shell nilpotent (anti-)BRST charges $[Q_{(a)b}]$ are consistent  with the Dirac quantization conditions
where the operator forms of the constraints are required to annihilate the physical states (of a system  endowed with any kind of constraints). 
It is interesting to point out that the concise forms of the (anti-)BRST charges [cf. Eq. (22)] remain the {\it same} for the {\it modified} 
massive and massless Abelian 1-form theories. However, in the case of the {\it latter} (with $m = 0, \phi = 0$), the second entry of (26) becomes:
$\partial_i \Pi^i_{(A)} |\,phys>\, = 0$ due to the fact that $\dot  B = \partial_i E_i \equiv \partial_i \Pi^i_{(A)}$ 
is the secondary constraint on the {\it massless} theory.

We end our Sec. 2 with the final remark that the first-class constraints are the characteristic and decisive features of 
a gauge theory at the {\it classical} as well as at the {\it quantum} level. In the {\it latter} case,
within the framework of BRST formalism,  they 
appear as operators in the Dirac quantization conditions through physicality criteria w.r.t. the conserved and off-shell nilpotent
versions of the (anti-)BRST charges. The nilpotency property of the (anti-)BRST charges is very {\it crucial} when we take 
into account the physicality criteria w.r.t the conserved (anti-)BRST charges within the purview of BRST formalism.

\section{First-Class Constraints and Generators: Modified Massive and Massless Abelian 2-Form Theories}

We begin with the following Lagrangian density
for the {\it massive} Abelian 2-form theory (with the {\it rest}  mass equal to $m$) in any arbitrary D-dimension of spacetime (see, e.g. [44])
\begin{eqnarray}
{\cal L}_{0}^{(B)} = \frac {1}{12}\,H^{\mu \nu \lambda}\,H_{\mu \nu \lambda} - \frac {m^2}{4} \, B_{\mu\nu}\,B^{\mu\nu},
\end{eqnarray}
where the 3-form  $H^{(3)} = d\,B^{(2)} \equiv  \frac{1}{3!}\,H_{\mu\nu\lambda}\; (d\,x^\mu \wedge d\,x^\nu \wedge d\,x^\lambda)$ defines the 
field-strength tensor $H_{\mu\nu\lambda} = \partial_\mu\,B_{\nu\lambda} + \partial_\nu\, B_{\lambda\mu} + \partial_\lambda \, B_{\mu\nu}$ 
for the Abelian 2-form  $B^{(2)} = \frac{1}{2!} \,B_{\mu\nu} \; (d\,x^\mu \wedge d\,x^\nu)$ 
antisymmetric $(B_{\mu\nu} = -\,B_{\nu\mu})$ tensor field $B_{\mu\nu}$.  
Here the operator $d$ (with $d^2 = 0)$ is the exterior derivative of differential geometry [45-49]. 
The above  Lagrangian density does {\it not} respect any {\it gauge} symmetry transformation because it 
is endowed with the second-class constraints in the terminology of Dirac's prescription for the  
classification scheme of constraints [5-8]. To restore the gauge symmetry transformations (of the {\it massless} 
Abelian 2-form theory),  we exploit the beauty and theoretical potential of the St{$\ddot u$}ckelberg-technique of the 
compensating field (e.g. a vector field $\phi_\mu$) in the following replacement (see, e.g. [44]) 
\begin{eqnarray}
B_{\mu\nu} \longrightarrow B_{\mu\nu} \mp \frac {1}{m}\,(\partial_\mu\,\phi_\nu - \partial_\nu\,\phi_\mu)
\quad \equiv \quad B_{\mu\nu} \mp \frac {1}{m}\, \Phi_{\mu\nu}, 
\end{eqnarray}
where the 2-form $\Phi^{(2)} = \frac{1}{2!}\, \Phi_{\mu\nu} \; (d\,x^\mu \wedge d\,x^\nu)$ defines the field-strength antisymmetric 
($\Phi_{\mu\nu} = -\,  \Phi_{\nu\mu}$) tensor field $\Phi_{\mu\nu} = \partial_\mu\,\phi_\nu - \partial_\nu\,\phi_\mu$
for the 1-form $\Phi^{(1)} = \phi_\mu \;  d\, x^\mu$ Lorentz vector St{$\ddot u$}ckelberg-compensating  field $(\phi_\mu)$.
The substitution (28) into  ${\cal L}_{0}^{(B)}$ leads to the  following St{$\ddot u$}ckelberg-modified 
Lagrangian density ${\cal L}_{(S)}^{(B)} $ for the antisymmetric tensor field, namely;  
\begin{eqnarray}
{\cal L}_{0}^{(B)} \longrightarrow {\cal L}_{(S)}^{(B)} =  \frac {1}{12}\,H^{\mu \nu \lambda}\,H_{\mu \nu \lambda} - \frac {m^2}{4} \, B_{\mu\nu}\,B^{\mu\nu}
\pm \frac{m}{2}\, B_{\mu\nu}\,\Phi^{\mu\nu} - \frac{1}{4}\, \Phi_{\mu\nu}\,\Phi^{\mu\nu},
\end{eqnarray}
which respects [i.e $\delta_g\, {\cal L}_{(S)}^{(B)} = 0$] the following gauge symmetry transformations
\begin{eqnarray}
&&\delta_g\,B_{\mu\nu} = (\partial_\mu\, \Lambda_\nu - \partial_\nu\,\Lambda_\mu), \qquad 
\delta_g\,\phi_{\mu} = \pm\,m\,\Lambda_\mu - \partial_\mu \Sigma,\nonumber\\
&&\delta_g \, H_{\mu \nu \lambda} = 0, \quad\qquad \qquad \qquad \;\; 
 \delta_g\,\Phi_{\mu\nu} = \pm\, m\,(\partial_\mu\, \Lambda_\nu - \partial_\nu\,\Lambda_\mu),
\end{eqnarray}
where $\Lambda_\mu$ and $\Sigma$ are the Lorentz vector and scalar gauge symmetry transformation parameters. 
It is an undeniable truth that the first-class constraints on a theory generate the 
infinitesimal  local  gauge symmetry transformations. In our present case, 
there are primary and secondary constraints on our theory.  The present theory  is {\it classically} described  by the Lagrangian 
density ${\cal L}^{(B)}_ {(S)}$, for which, the canonical conjugate momenta w.r.t. the {\it basic} fields $B_{\mu\nu}$
and $\phi_\mu$ are: 
\begin{eqnarray}
&&\Pi_{(B)}^{\mu\nu}  = \frac {\partial {\cal L}_{(S)}}{\partial (\partial_0 B_{\mu\nu})}
 = \frac {1}{2}\, H^{0\mu\nu} \quad  \Longrightarrow \quad
\Pi^{0i}_{(B)}  = \frac {1}{2}\, H^{00i} \approx 0, \nonumber\\
&&\Pi_{(\phi)}^{\mu}  = \frac {\partial {\cal L}_{(S)}}{\partial (\partial_0 \phi_\mu)} 
= \pm\, m\, B^{0\mu} - \Phi^{0\mu} \quad \Longrightarrow \quad
\Pi^{0}_{(\phi)}  = \pm\,m\, B^{00}\ - \Phi^{00}\approx 0,\nonumber\\
&&\Pi ^{ij}_{(B)} = \frac {1}{2}\, H^{0ij}, \qquad  \qquad \Pi_{(\phi)}^{i}  = \pm\,m\, B^{0i} - \Phi^{0i}.  
\end{eqnarray}
From the above equations, it is clear that $\Pi^{0i}_{(B)} \approx 0$
and $\Pi_{(\phi)}^{0} \approx 0$ are the primary constraints (PCs) on our theory where we have used the Dirac notation 
for the idea of {\it weakly} zero. Thus, we are allowed to take a {\it first-order} time derivative on the above PCs to obtain the 
secondary constraints (SCs) on our theory. Towards this goal in mind, we note that the following 
EL-EoMs, w.r.t. the basic fields $B_{\mu\nu}$ and $\phi _{\mu}$ from ${\cal L}_{(S)}^{(B)} $,  are: 
\begin{eqnarray}
&& \partial_\mu H^{\mu\nu\lambda}  + m^2\, B^{\nu\lambda}  \mp \, m\, \Phi^{\nu\lambda} = 0, \nonumber\\
&& \partial_\mu [\,\pm m\, B^{\mu\nu}  - \Phi^{\mu\nu}] \equiv \pm\,m\,\partial_\mu B^{\mu\nu} - \partial_\mu \Phi^{\mu\nu} = 0.  
\end{eqnarray} 
For our simple system, the Lagrangian formulation is good enough to deal with  the 
derivations of the SCs. In this context, we note that the very specific  choices: $\nu = 0, \; \lambda = j$ 
in (32) lead to the following explicit forms of the EoMs, namely; 
\begin{eqnarray}
&& \partial_0 H^{00j}  + \partial_i H^{i0j}  + m^2\, B^{0j}  \mp \, m\, \Phi^{0j} = 0, \nonumber\\
&& (\pm m\, \partial_0 \,B^{00}  - \, \partial_0 \, \Phi^{00}) + (\pm m\, \partial_i \,B^{i0}  - \, \partial_i \, \Phi^{i0})  = 0.  
\end{eqnarray} 
Taking the inputs from (31), it is clear that we have the following time-evolution invariance [24] of the PCs of our theory, namely;
\begin{eqnarray}
  \frac{\partial\, \Pi_{(B)}^{0j}}{\partial\, t} & = &  \partial_i\, \Pi^{ij}_{(B)} \mp \frac{m}{2}\, \Pi^j_{(\phi)} \approx 0, 
\nonumber\\
\frac{\partial\, \Pi_{(\phi)}^{0}}{\partial\, t} & = & \partial_i\, \Pi^{i}_{(\phi)} \approx 0.
\end{eqnarray} 
The above equations of motion lead to the derivations of SCs on our theory as: 
$\partial_i \Pi^{ij}_{(B)} \mp \frac {m}{2}\,\Pi^{j}_{(\phi)} \approx 0, \; \partial_i \Pi^{i}_{\phi}\approx 0$.
Since all the PCs and SCs are expressed in terms of the components of the conjugate momenta, 
it is clear that they commute among themselves. Hence, {\it these}  constraints are of the first-class
variety in the terminology of Dirac's prescription for the classification scheme of constraints [5-8].
There are {\it no} further constraints on the theory. Using the standard formula (see, e.g. [42, 43] for details) for the 
expression of the generator in terms of the first-class constraints, the generator $G_{(B)}$ for our St$\ddot u$ckelberg-modified  
{\it massive}  Abelian 2-form theory is    
\begin{eqnarray}
G_{(B)}  & = & \int d^{D-1} x \, \Big[ (\partial_0\, \Lambda_i)\, \Pi^{0i}_{(B)} +  (\partial_i\, \Lambda_0)\, \Pi^{i0}_{(B)} - 
\Lambda_j\, \big(\partial_i\, \Pi^{ij}_{(B)} \mp \frac{m}{2}\,\Pi^j_{(\phi)}\big) 
\nonumber\\
 & - &  \Lambda_i\, \big(\partial_j\, \Pi^{ji}_{(B)} \mp \frac{m}{2}\, \Pi^i_{(\phi)}\big) 
-\, \big(\partial_0\, \Sigma \mp m\, \Lambda_0 \big)\, \Pi^0_{(\phi)} + \Sigma\, \partial_i\, \Pi^i_{(\phi)}
\Big],
\end{eqnarray} 
where $(\Lambda_0, \Lambda_i)$ are the components of the Lorentz vector gauge transformation parameters 
$\Lambda_\mu$ and $\Sigma$ is the Lorentz scalar transformation parameter which are present in the gauge 
symmetry transformation of our theory in (30). The above generator can be re-expressed, in a more transparent and useful
form, as: 
\begin{eqnarray}
G_{(B)} & = & \int d^{D-1} x \, \Big[ \big(\partial_0\, \Lambda_i - \partial_i\, \Lambda_0 \big)\, \Pi^{0i}_{(B)} 
+ \big(\partial_i\, \Lambda_j - \partial_j\, \Lambda_i \big)\, \Pi^{ij}_{(B)}\nonumber\\
& - & \big(\partial_0\, \Sigma \mp m\, \Lambda_0 \big)\, \Pi^0_{(\phi)} - \big(\partial_i\, \Sigma \mp m\, \Lambda_i \big)\, \Pi^i_{(\phi)}
\Big].
\end{eqnarray}
The above form of the generator  leads to the derivation of the gauge transformations $(\delta_g)$ for the {\it basic} fields 
(i.e. the antisymmetric $B_{\mu\nu}$ field and vector field $\phi_\mu$) of our theory as
\begin{eqnarray}
&&\delta_g\, \phi_0\, (\vec x, t) = -\, i \, [\phi_0\, (\vec x, t), \, G_{(B)}] = -\, \partial_0 \Sigma \pm m\, \Lambda_0,
\nonumber\\
&&\delta_g\, \phi_i\, (\vec x, t) = -\, i \, [\phi_i\, (\vec x, t), \, G_{(B)}] = -\, \partial_i\, \Sigma \pm m\, \Lambda_i,
\nonumber\\
&&\delta_g\, B_{0i}\, (\vec x, t) = -\, i \, [B_{0i} \, (\vec x, t), \, G_{(B)}] = \partial_0\, \Lambda_i - \partial_i\, \Lambda_0, 
\nonumber\\
&&\delta_g\, B_{ij}\, (\vec x, t) = -\, i \, [B_{ij} \, (\vec x, t), \, G_{(B)}] = \partial_i\, \Lambda_j - \partial_j\, \Lambda_i,
\end{eqnarray} 
provided we take into account the following {\it non-trivial} equal-time canonical commutators
\begin{eqnarray}
&&[\phi_0\, (\vec x, t), \, \Pi^0_{(\phi)}\, (\vec y, t)] \; = \; i\, \delta^{(D-1)}\, (\vec x - \vec y), \nonumber\\
&&[\phi_i\, (\vec x, t), \, \Pi^j_{(\phi)}\, (\vec y, t)] \; = \; i\,\delta_i^j \, \delta^{(D-1)}\, (\vec x - \vec y),
\nonumber\\
&&[B_{0i} (\vec x, t), \, \Pi^{0j}_{(B)}\, (\vec y, t)] \; = \; i\,\delta_i^j \, \delta^{(D-1)}\, (\vec x - \vec y),
\nonumber\\
&&[B_{ij} (\vec x, t), \, \Pi^{kl}_{(B)}\, (\vec y, t)] \; = \; \frac{i}{2!}\,(\delta_i^k \, \delta_j^l
 - \delta_i^l \, \delta_j^k  )\, \delta^{(D-1)}\, (\vec x - \vec y),
\end{eqnarray}
and the {\it rest} of the brackets for our theory are {\it trivially   zero}.
We have taken the natural units ($\hbar = c = 1$) in the above definition of the brackets.  Ultimately, 
it is clear that we have derived the local gauge symmetry transformations in their {\it covariant} form as given in (30).
Under the above local and infinitesimal gauge symmetry transformations, it is straightforward to 
check that we have the gauge invariance (i.e. $\delta_g\,{\cal L}_{(S)}^{(B)} = 0$).

We wrap-up this subsection  with the following remarks. First, we have taken the notations: 
$\Pi_{(\phi)}$ and $\Pi_{(\phi)}^i$ for the conjugate momenta w.r.t. the St{$\ddot u$}ckelberg-compensating 
fields $\phi$ and $\phi_\mu$ in the context of {\it modified} Proca theory and massive  Abelian 2-form theory, respectively, 
where the subscripts are the {\it same}. However, the contravariant   indices  on these conjugate momenta take care of the 
{\it relevance} of these  definitions in their appropriate  contexts. Second, it is obvious that the generator  $G_{(B)}^{(m  = 0)} $ for the 
{\it massless} Abelian 2-form gauge theory can be obtained  from (36) by taking into account the conditions:
$m = 0$ and $\phi_\mu = 0$. In other words, for the {\it massless} Abelian 2-form theory, we have the following
\begin{eqnarray}
G_{(B)}^{(m = 0)}  &=& \int d^{D-1} x \, \Big[ (\partial_0\, \Lambda_i)\, \Pi^{0i}_{(B)} +  (\partial_i\, \Lambda_0)\, \Pi^{i0}_{(B)} 
 -  \Lambda_i\, (\partial_j\, \Pi^{ji}_{(B)}) - \Lambda_j\, (\partial_i\, \Pi^{ij}_{(B)}) \Big],
\nonumber\\
&\equiv&
 \int d^{D-1} x \, \Big[ \big(\partial_0\, \Lambda_i - \partial_i\, \Lambda_0 \big)\, \Pi^{0i}_{(B)} 
+ \big(\partial_i\, \Lambda_j - \partial_j\, \Lambda_i \big)\, \Pi^{ij}_{(B)}\Big],  
\end{eqnarray}
which is derived from $G_{(B)}$ [cf. Eqs. (35), (36)] with the substitutions: $m = 0, \; \Pi^i_{(\phi)} =  \Pi^0_{(\phi)} = 0$ (due to $\phi_\mu = 0$). 
Finally, using the equal-time commutators appropriately, it is straightforward to note that we obtain the 
gauge transformation:  $\delta_g\,B_{\mu\nu} = (\partial_\mu\, \Lambda_\nu - \partial_\nu\,\Lambda_\mu)$
in its covariant form from (39) when we exploit and use the theoretical strength of the appropriate commutators from
(37) with the replacement: $G_{(B)} \longrightarrow G_{(B)}^{(m = 0)}$. Under this transformation (i) the limiting case (i.e. $ m = 0)$ of the Lagrangian density (27) and/or (ii)
the limiting case (i.e. $m = 0, \; \phi_\mu = 0)$ of the Lagrangian density (29),  remains 
invariant because we observe that
the field-strength tensor is gauge invariant (i.e. $\delta_g \, H_{\mu \nu \lambda } = 0$).

\section{Noether Conserved Charges and  Constraints: Modified Massive and Massless Abelian 2-Form Theories}

According to the {\it basic} tenets of the Noether theorem, the existence of the  
infinitesimal, continuous and local gauge symmetry transformations (30) leads to the following Noether's current
for the {\it modified} massive Abelian 2-form theory [described  by (29)], namely; 
\begin{eqnarray*}
J^\mu_{(B)} &=& \frac{1}{2}\, H^{\mu \nu \lambda }\, (\partial_\nu\, \Lambda_\lambda - \partial_\lambda\,\Lambda_\nu)
+ (\pm m\, B^{\mu\nu} - \Phi^{\mu\nu})\ (\pm\, m\, \Lambda_\nu - \partial_\nu\, \Sigma) 
\end{eqnarray*}
\begin{eqnarray}
 &\equiv&  H^{\mu \nu \lambda }\, (\partial_\nu\,\Lambda_\lambda)
+ (\pm m\, B^{\mu\nu} - \Phi^{\mu\nu})\ (\pm\, m\, \Lambda_\nu - \partial_\nu\, \Sigma),   
\end{eqnarray}
where the totally antisymmetric property of $H_{\mu\nu\lambda}$ has been used. 
The conservation law ($\partial_\mu\, J^\mu_{(B)} = 0$) can be proven in a straightforward
fashion by using the EL-EoMs w.r.t the {\it basic} fields $B_{\mu\nu}$ and $\phi_\mu$ of our theory
[cf. Eq. (32)].  In addition, we have to keep in mind that the sum of the antisymmetric and symmetric 
indices {\it always} turns out to be {\it zero}. The conserved Noether charge [$Q_{(B)} = \int d^{D-1} x \, J^0_{(B)}$], derived from the 
above Noether conserved current [$J^\mu_{(B)}$], is as follows: 
\begin{eqnarray}
Q_{(B)}  \equiv  \int d^{D-1} x \, \Big[ \frac{1}{2}\, H^{0 \nu \lambda }\, (\partial_\nu\, \Lambda_\lambda - \partial_\lambda \,\Lambda_\nu)
+ (\pm m\, B^{0 \nu} - \Phi^{0 \nu})\ (\pm\, m\, \Lambda_\nu - \partial_\nu\, \Sigma)\Big].  
\end{eqnarray}
Since our theory is endowed with constraints [cf. Eqs. (31, (34)] we have to be careful in expanding the r.h.s. 
of (41) so that the  presence of constraints,  at appropriate places,  is taken care of. 
In other words, we have the following explicit expression for the Noether conserved charge
for our {\it modified} massive Abelian 2-form  theory, namely;   
\begin{eqnarray}
Q_{(B)} & = & \int d^{D - 1}x\,\Big[(\partial_0 \Lambda_i - \partial_i\Lambda_0)\,\Pi^{0i}_{(B)}  
+  (\partial_i \Lambda_j - \partial_j\Lambda_i)\,\Pi^{ij}_{(B)} \nonumber\\
& + & (\,\pm\,m\,\Lambda_0 - \partial_0\Sigma)\, \Pi^{0}_{(\phi)}
+ (\,\pm\,m\,\Lambda_i - \partial_i\Sigma)\, \Pi^{i}_{(\phi)}\Big].
\end{eqnarray}   
 Using (i) the theoretical strength of the Gauss divergence theorem, and (ii) a simple algebraic trick: 
$ \pm m \,  \Lambda_i\,\Pi^i_{(\phi)} \equiv \pm\, ({m}/{2})\,[\Lambda_i\,\Pi^i_{(\phi)} + \Lambda_j\,\Pi^j_{(\phi)}],$
 we can re-express the above conserved Noether charge [i.e. $Q_{(B)}$] as follows
\begin{eqnarray}
Q_{(B)} & = & \int d^{D - 1}x\,\Big[(\partial_0 \Lambda_i - \partial_i\Lambda_0)\,\Pi^{0i}_{(B)}  
- \Lambda_i \,\big(\partial_j \Pi^{ji}_{(B)} \mp \frac {m}{2}\,\Pi^i_{(\phi)}\big) \nonumber\\
& - & \Lambda_j \,\big(\partial_i \Pi^{ij}_{(B)} \mp \frac {m}{2}\,\Pi^j_{(\phi)}\big)
+ (\,\pm\,m\,\Lambda_0 - \partial_0\Sigma)\, \Pi^{0}_{(\phi)} - \Sigma\,(\partial_i \Pi^i_{(\phi)})\Big],
\end{eqnarray}
where the primary constraints $(\Pi^{0i}_{(B)} \approx 0, \, \Pi^{0}_{(\phi)} \approx 0$) and the secondary constraints of our theory 
$(\partial_i\Pi^{ij}_{(B)} \mp ({m}/{2})\, \Pi^j _{(\phi)} \approx 0, \, \partial_i\Pi^{i}_{(\phi)} \approx 0)$
appear in an explicit form. A close look at (43) demonstrates that, in reality, we have obtained the generators 
$G_{(B)}$ in terms of the constraints [cf. Eq. (35)] that has been  derived in our  previous section (cf. Sec. 3).

We end this section with the following interesting remarks. First, we have established a deep relationship 
between (i) the Noether conserved charge (derived from the infinitesimal, local and continuous gauge symmetry transformations), and (ii)
the standard formula for the generator of the infinitesimal 
gauge symmetry transformations (expressed in terms of the first-class constraints of the 
{\it original} St{$\ddot u$}ckelberg-modified massive {\it classical} gauge theory). Second, it 
is straightforward to note that in the limit: $m = 0, \, \phi_\mu = 0,$  we obtain the Noether conserved charge
[$Q_{(B)} ^{( m =0)}$] from the charges (42) and/or (43) as follows 
\begin{eqnarray}
Q_{(B)}^{(m = 0)} & = & \int d^{D - 1}x\,\Big[(\partial_0 \Lambda_i - \partial_i\Lambda_0)\,\Pi^{0i}_{(B)}  
-  (\partial_i \Pi^{ij}_{(B)})\,\Lambda_j - (\partial_j \Pi^{ji}_{(B)})\,\Lambda_i)\Big] \nonumber\\
& \equiv & 
\int d^{D - 1}x\,\Big[(\partial_0 \Lambda_i - \partial_i\Lambda_0)\,\Pi^{0i}_{(B)}  
+  (\partial_i \Lambda_j - \partial_j\Lambda_i)\,\Pi^{ij}_{(B)} \Big],
\end{eqnarray}
where the first-class constraints $(\Pi^{0i}_{(B)} \approx 0, \, \partial_i\, \Pi^{ij}_{(B)} \approx 0$) of the {\it massless} Abelian 2-form 
theory appear explicitly. Finally, due to the presence of constraints on our {\it modified} massive and 
massless Abelian 2-form {\it gauge} theories, one has to be careful in the derivations of the Noether conserved charges
 where the constraints can {\it not} to be  set {\it strongly} equal to {\it zero}.

\section{First-Class Constraints and Their Appearance  at the Quantum Level: BRST Approach}

This section is divided into {\it two} parts. In subsection 5.1, we consider the {\it massless}
Abelian 2-form theory within the framework of BRST formalism and show that the physicality criteria 
with the conserved and off-shell nilpotent (anti-)BRST charges lead to the annihilation of the physical states 
(present in the {\it total} quantum Hilbert space of states)
by the operator forms of the first-class constraints of the {\it original} classical Abelian 2-form gauge theory 
(with {\it zero} rest mass). This exercise is repeated for the St{$\ddot u$}ckelberg-modified {\it massive} Abelian 2-form 
theory in our subsection 5.2 to establish the existence of the first-class constraints,  within the purview of BRST formalism,  at
the {\it quantum}  level.\\

\subsection{Massless Abelian 2-Form Theory: (Anti-)BRST Charges}

We begin with the following off-shell nilpotent [$s_{(a)b}^2 = 0$] (anti-)BRST symmetries $s_{(a)b}$
[which are the generalizations of the {\it classical } infinitesimal and local  gauge symmetry transformations\footnote{We 
have taken purposely a minus sign in the gauge transformation because it is consistent with the (anti-)BRST symmetry transformations:
$s_{ab} \,B_{\mu\nu} = -\, (\partial_\mu\, \bar C_\nu - \partial_\nu\,\bar C_\mu)$ 
and $s_b \,B_{\mu\nu} = -\, (\partial_\mu\, C_\nu - \partial_\nu\,C_\mu)$ 
which have been {\it precisely} derived from the superfield approach to BRST formalism in the context of 
Abelian 2-form {\it gauge}  theory where the horizontality condition has played a decisive role (see, e.g. [29]).}: 
$\delta_g\,B_{\mu\nu} = -\, (\partial_\mu\, \Lambda_\nu - \partial_\nu\,\Lambda_\mu)$
under which the {\it classical} Lagrangian density ${\cal L}_0 = ({1}/{12})\, (H^{\mu \nu \lambda }\, H_{\mu \nu \lambda })$, 
for our {\it massless} theory,  remains invariant], namely; 
\begin{eqnarray}
&& s_{ab} B_{\mu\nu} = - (\partial_\mu {\bar C}_\nu - \partial_\nu {\bar C}_\mu), 
\quad s_{ab} {\bar C}_\mu = - \partial_\mu {\bar \beta}, \qquad s_{ab} C_\mu = {\bar B}_\mu, \nonumber\\
&& s_{ab} \phi = \rho, \quad s_{ab} \beta = - \lambda, \;\; s_{ab} B_\mu = \partial_\mu \rho, 
\quad s_{ab} \bigl [\rho, \lambda, {\bar\beta}, 
{\bar B}_\mu, H_{\mu\nu\kappa} \bigr ] = 0,\nonumber\\
&& s_b B_{\mu\nu} = - (\partial_\mu C_\nu - \partial_\nu C_\mu), 
\qquad s_b C_\mu = - \partial_\mu \beta, \qquad s_b \bar C_\mu = - B_\mu, \nonumber\\
&& s_b \phi = \lambda, \quad s_b \bar \beta = - \rho, \quad s_b \bar B_\mu  = -\, \partial_\mu \lambda,\quad 
s_b \bigl [\rho, \lambda, \beta,  B_\mu, H_{\mu\nu\kappa} \bigr ] = 0.
\end{eqnarray}
The above (anti-)BRST transformations are the {\it symmetries}  of the following coupled (but equivalent)
Lagrangian densities (see, e.g. [40] for details)
\begin{eqnarray}
{\cal L}_{B} &=& \frac {1}{12} \,H^{\mu \nu \kappa} \,H_{\mu\nu\kappa}
+  B^{\mu} \,\Bigl ( \partial^{\nu} B_{\nu\mu} - \partial_{\mu}\phi \Bigr)
+ B \cdot  B + \partial_{\mu}{\bar \beta }\, \partial^{\mu} \beta \nonumber\\
&+& (\partial_{\mu} {\bar C}_{\nu} - \partial_{\nu}{\bar C}_{\mu})\,(\partial^{\mu}C^{\nu})
+(\partial \cdot C - \lambda)\,\rho + (\partial \cdot {\bar C}+ \rho )\,\lambda ,
\end{eqnarray}
\begin{eqnarray}
{\cal L}_{\bar B} &=&  \frac {1}{12}\, H^{\mu \nu \kappa} \, H_{\mu\nu\kappa}
+  {\bar B}^{\mu} \, \Bigl (\partial^{\nu} \, B_{\nu\mu} + \partial_{\mu} \phi\Bigr)
+ {\bar B} \cdot  {\bar B} + \partial_{\mu}\, {\bar \beta}  \, \partial^{\mu}  \beta \nonumber\\
&+& (\partial_{\mu} {\bar C}_{\nu} - \partial_{\nu}{\bar C}_{\mu})\, (\partial^{\mu}C^{\nu})
+ (\partial \cdot C - \lambda)\, \rho + (\partial \cdot{\bar C}+ \rho )\, \lambda,
\end{eqnarray}
where the Nakanishi-Lautrup auxiliary fields ($B_{\mu}, \,\bar B_{\mu} $) are related to each-other by the CF-type restriction:
$B_\mu - \bar B_\mu - \partial_\mu \phi = 0$. Here $\phi$ is a pure scalar field  that is present 
in the  (anti-)BRST invariant Lagrangian densities [cf. Eqs. (46), (47)] of our theory. The fermionic (anti-)ghost fields 
$(\bar C_\mu)C_\mu$ are the generalization of the Lorentz vector gauge transformation parameter ($\Lambda_\mu$)
of the {\it classical} gauge symmetry transformation: $\delta_g\,B_{\mu\nu} = -\, (\partial_\mu\, \Lambda_\nu - \partial_\nu\,\Lambda_\mu)$
and they carry the ghost numbers $(- 1)+1$, respectively. The bosonic (anti-)ghost fields 
$(\bar\beta)\beta$ are endowed with the ghost numbers $(- 2)+2$, respectively, and the fermionic 
($\rho^2 = \lambda^2 = 0, \, \rho\, \lambda + \lambda\, \rho = 0$) auxiliary (anti-)ghost fields $(\rho)\, \lambda$
carry the ghost numbers $(- 1)+1$, respectively, because $\rho = -\,({1}/{2})\,(\partial\cdot\bar C)$ 
and $\lambda  = ({1}/{2})\,(\partial\cdot C)$. The conserved (anti-)BRST Noether currents,  
with the superscripts $(ab)$ and $(b)$, namely; 
\begin{eqnarray}
J^{\mu}_{(ab)}=\rho \,{\bar B}^{\mu} -(\partial^{\mu} {\bar C}^{\nu} - \partial^{\nu}{\bar C}^{\mu}) 
\,{\bar B}_{\nu} -(\partial^{\mu} C^{\nu} - \partial^{\nu} C^{\mu}) \, \partial_{\nu}{\bar\beta} 
- \lambda \;\partial^{\mu}{\bar\beta} - H^{\mu\nu\kappa}\; (\partial_\nu \bar C_\kappa),\nonumber\\
J^{\mu}_{(b)}=(\partial^{\mu} {\bar C}^{\nu} - \partial^{\nu}{\bar C}^{\mu}) \;\partial_{\nu}\beta 
-(\partial^{\mu} C^{\nu} - \partial^{\nu} C^{\mu}) \, B_{\nu} -\lambda \; B^{\mu}
-\rho \;\partial^{\mu}\beta -  H^{\mu\nu\kappa}\; (\partial_\nu C_\kappa),
\end{eqnarray}
lead to the definitions of the conserved (anti-)BRST charges: 
\begin{eqnarray}
Q_{ab} = {\displaystyle \int} d^{D-1} x \, \Bigr[\rho \,{\bar B}^{0} -(\partial^{0} {\bar C}^{i} - \partial^{i}
{\bar C}^{0}) \,{\bar B}_{i} -(\partial^{0} C^{i} - \partial^{i} C^{0}) \, \partial_{i}{\bar\beta}
- \lambda \;\partial^{0}{\bar{\beta}}- H^{0ij}\; (\partial_i \bar C_j) \Bigl ],\nonumber\\
Q_b = {\displaystyle \int}
d^{D-1} x \, \Bigr [(\partial^{0} {\bar C}^{i} - \partial^{i}{\bar C}^{0}) 
\,\partial_{i}\beta -(\partial^{0} C^{i} - \partial^{i} C^{0}) B_{i} -\lambda \; B^{0}
-\rho \; \partial^{0}\beta - H^{0ij}\; (\partial_i C_j ) \Bigl ]. 
\end{eqnarray}
It turns out, however, that the above charges are {\it not} nilpotent [$Q_{(a)b}^2 \ne 0$]
of order {\it two}. We would like 
to point out that (unlike the cases of the {\it modified} massive and massless Abelian 1-form theories where the CF-type restriction
is {\it trivial}) for our present {\it massless}  Abelian 2-form theory,  there exits a {\it non-trivial} (anti-)BRST invariant 
(i.e. $s_{(a)b}\,\{B_\mu - \bar B_\mu - \partial_\mu \phi\} = 0$) CF-type restriction where the Nakanishi-Lautrup  auxiliary fields 
$(B_\mu, \, \bar B_\mu)$ and the scalar field $(\phi)$ are restricted to obey: $B_\mu - \bar B_\mu - \partial_\mu \phi = 0$.
This is primarily the reason behind the observation that the  standard {\it conserved} Noether (anti-)BRST charges $Q_{(a)b}$ are non-nilpotent 
(i.e. $Q_{(a)b}^ 2 \neq 0$). This is {\it not} true for the {\it modified} massive and massless Abelian 1-form theories within the framework 
of BRST formalism (cf. Subsec. 2.3).

We have developed [40] a systematic theoretical procedure to obtain the off-shell {\it nilpotent} (anti-)BRST charges from the 
{\it non-nilpotent} standard  Noether (anti-)BRST charges where we have exploited  (i) the appropriate EL-EoMs that emerge out from the specific Lagrangian 
density, and (ii) the Gauss divergence theorem. The off-shell nilpotent versions of the (anti-)BRST charges [$Q_{(a)b} ^{(1)}$] for 
our {\it massless} Abelian 2-form gauge theory are  [40] 
\begin{eqnarray}
Q_{ab} \longrightarrow Q_{ab} ^{(1)} & = & {\displaystyle \int}\;
d^{D - 1} x \Bigr [(\partial^0 \bar B^i - \partial ^i \bar B^0)\, \bar C_i 
 +  (\partial^{0} {C}^{i} - \partial^{i}{C}^{0}) \, \partial_i \bar \beta\nonumber\\ 
 & - & (\partial^{0} \bar C^{i} - \partial^{i} \bar C^{0}) \, \bar B_{i}  + \rho \, \bar B^{0} 
 - 2\, \bar \beta\, \dot\lambda  - \lambda\, \dot{\bar\beta}  \Bigl],\nonumber\\
 Q_b \longrightarrow Q_b ^{(1)} & = & {\displaystyle \int}\;
d^{D - 1} x \; \Bigr [(\partial^0 B^i - \partial ^i B^0)\, C_i - (\partial^{0} {\bar C}^{i} - \partial^{i}{\bar C}^{0}) \;\partial_{i}\beta \nonumber\\
& + & 2\, \beta\, \dot\rho - (\partial^{0} C^{i}
  - \partial^{i} C^{0}) B_{i} -\lambda B^{0} -\rho\, \dot\beta  \Bigl ], 
\end{eqnarray} 
where it can be explicitly checked that: 
\begin{eqnarray}
s_b \, Q_b ^{(1)}  & = & -\, i\, \{Q_b ^{(1)}, \; Q_b ^{(1)}\}  = 0 \quad \Longrightarrow \quad [Q_b ^{(1)}]^2 = 0,\nonumber\\
s_{ab} \, Q_{ab} ^{(1)}  & = & -\, i\, \{Q_{ab} ^{(1)}, \; Q_{ab} ^{(1)}\}  = 0 \quad \Longrightarrow \quad [Q_{ab} ^{(1)}]^2 = 0.
\end{eqnarray}
We are now in the position to impose  the physicality   criteria: 
$Q_{(a)b}^{(1)}\,|phys>\, = 0$ on the physical states that are present in the total Hilbert space of states. Since the (anti-)ghost fields 
are {\it not} the  physical fields, we note that the {\it above} criteria lead to the following: 
\begin{eqnarray}
Q_b^{(1)}\,|\,phys>\, = 0 \quad \Longrightarrow  \quad B_i\,|\,phys>\, = 0, \;\;\; \quad  (\partial_0 B_i - \partial_i B_0)\, |\,phys>\, = 0,\nonumber\\
Q_{ab}^{(1)}\,|\,phys>\, = 0\quad \Longrightarrow \quad \bar B_i\,|\,phys>\, = 0, \;\;\; \quad (\partial_0 \bar B_i - \partial_i \bar B_0)\, |\,phys>\, = 0.
\end{eqnarray} 
To explicitly see the appearance of the first-class constraints, we focus on the Lagrangian density ${\cal L}_B$ and ${\cal L}_{\bar B}$
and define the conjugate momenta w.r.t. the gauge field $(B_{\mu\nu})$: 
\begin{eqnarray}
&&\Pi^{\mu\nu}_{(B)} = \frac {\partial {\cal L}_B}{\partial (\partial_0 B_{\mu\nu})} = \frac {1}{2}\, H^{0\mu\nu}\, + \frac {1}{2}\,
(\eta^{0\mu}\, B^{\nu} - \eta^{0\nu}\, B^\mu), \nonumber\\
&&\Pi^{\mu\nu}_{(\bar B)} = \frac {\partial {\cal L}_{\bar B}}{\partial (\partial_0 B_{\mu\nu})} = \frac {1}{2}\, H^{0\mu\nu}\, + \frac {1}{2}\,
(\eta^{0\mu}\, \bar B^{\nu} - \eta^{0\nu}\, \bar B^\mu).
\end{eqnarray}  
The above conjugate momenta have the following components: 
\begin{eqnarray}
&&\Pi^{0i}_{(B)} = \frac {1}{2}\,B^i \equiv -\, \frac {1}{2}\,B_i, \qquad \qquad \qquad  \Pi^{ij}_{(B)} = \frac {1}{2}\,H^{0ij},\nonumber\\
&&\Pi^{0i}_{(\bar B)} = \frac {1}{2}\,\bar B^i \equiv -\, \frac {1}{2}\,\bar B_i, \qquad \qquad \qquad  \Pi^{ij}_{(\bar B)} = \frac {1}{2}\,H^{0ij}.
\end{eqnarray} 
 Thus, we note that the primary constraints $\Pi_{(B)}^{0i} \approx 0$ and/or $\Pi_{(\bar B)}^{0i} \approx 0$ 
are connected with the Nakanishi-Lautrup auxiliary fields $(B_i$ and/or $\bar B_i)$ as far as the Lagrangian densities
${\cal L}_B$ and ${\cal L}_{\bar B}$ are concerned. Furthermore, from the following EL-EoMs that are derived
 from ${\cal L}_B$ and ${\cal L}_{\bar B}$, respectively,  namely;
\begin{eqnarray}
\partial_\mu H^{\mu\nu\lambda} + (\partial^\nu B^\lambda - \partial^\lambda B^\nu) = 0, \quad \qquad 
\partial_\mu H^{\mu\nu\lambda} + (\partial^\nu \bar B^\lambda - \partial^\lambda \bar B^\nu) = 0,
\end{eqnarray}   
we find that,  for the choices: $\nu = 0, \,\lambda  = j$, we have the following:    
\begin{eqnarray}
\partial_i H^{0ij}  =  (\partial^0 B^j - \partial^j B^0), \qquad  \qquad \partial_i H^{0ij}  =  (\partial^0 \bar B^j - \partial^j \bar B^0).
\end{eqnarray} 
Using the definitions of the components of the conjugate momenta from (54), we note that we have obtained the following
relationship between the first-class constraints and the specific combination 
(e.g. $\partial_0 B_i - \partial_0 B_0, \; \partial_0 \bar B_i - \partial_0 \bar B_0$) of the derivatives  on the Nakanishi-Lautrup axillary 
fields $(B_i, \bar B_i)$ of the Lagrangian densities ${\cal L}_B$ and ${\cal L}_{\bar B}$: 
\begin{eqnarray}
 2\, \partial_i \Pi^{ij}_{(B)} = -\,(\partial_0 B_j - \partial_j B_0), \quad\qquad 2\, \partial_i \Pi^{ij}_{(\bar B)} =
 -\,(\partial_0 \bar B_j - \partial_j \bar B_0). 
\end{eqnarray} 
A close look at the equations (54) and (57) demonstrate that we have obtained the {\it same} kinds of quantization conditions 
(i.e. restrictions on the physical states) from the conserved and nilpotent
 versions of the (anti-)BRST charges (with $B_i \longleftrightarrow \bar B_i$). Ultimately, 
we claim, however, that we have obtained the conditions where the operator forms of the first-class constraints annihilate 
the physical states of our theory. In other words, mathematically, we have obtained the following
conditions on the physical states (from {\it both}  the charges):       
\begin{eqnarray}
 B_i\, |phys>\, = 0\qquad  &\Longrightarrow & \qquad   \Pi^{0i}_{(B)} \, |phys>\, = 0,\nonumber\\
(\partial_0 B_i - \partial_i B_0)\, |phys>\, = 0 \qquad &\Longrightarrow &\qquad   \partial_i\Pi^{ij}_{(B)} \, |phys>\, = 0.
\end{eqnarray}    
We conclude that, within the framework of BRST formalism, the operator forms of the first-class constraints 
(i.e. $\Pi^{0i}_{(B)} \approx  0$ and $\partial_i \Pi^{ij}_{(B)} \approx  0)$ of the {\it massless} Abelian 2-form theory 
annihilate the physical states of the {\it quantum} theory which is consistent with the requirements of the 
Dirac-quantization conditions (for the physical systems that are endowed with 
the constraints of any kind/variety [5-8] in Dirac's terminology).

We end this subsection with a few remarks. First, the Noether theorem does {\it not} lead to the off-shell nilpotent 
(anti-)BRST charges whenever the {\it non-trivial} CF-type restriction(s) exist on a BRST-quantized theory (see, e.g. [40]). 
Second, to obtain the conserved and off-shell nilpotent (anti-)BRST charges from the {\it standard} Noether {\it non-nilpotent} 
(anti-)BRST charges, one has to exploit  $(i)$ the appropriate EL-EoMs from the appropriate Lagrangian density, 
and $(ii)$ the beauty and strength of the Gauss divergence theorem. 
Finally, it is the physicality criteria with the conserved and off-shell nilpotent versions of the  (anti-)BRST charges that trace  the trajectory of the elevation  of the first-class constraints
from the {\it classical} level to the  {\it quantum} level within the framework of of BRST formalism. \\

\subsection{Modified Massive Abelian 2-Form Theory: BRST Formalism}

We begin with the following (anti-)BRST invariant coupled (but equivalent) Lagrangian densities (${\cal L}_b $ and ${\cal L}_{\bar b} $)
for the St$\ddot u$ckelberg-modified {\it massive}  Abelian 2-form theory\footnote{We have taken a specific sign with a specific 
term of the Lagrangian densities because these signs  
satisfy all the requirements that are essential for an (anti-)BRST invariant theory. The {\it uniqueness} of these 
Lagrangian densities has been proven in our very recent work (see, e.g. [50] for details).} in any arbitrary D-dimension of spacetime
(see, e.g. [40] for details) 
\begin{eqnarray}
 {\cal L}_b & = & \frac{1}{12} \, H_{\mu\nu\eta} H^{\mu\nu\eta}
-\frac{1}{4}\,m^2\, B_{\mu\nu} B^{\mu\nu} - \frac{1}{4}\, \Phi_{\mu\nu} \Phi^{\mu\nu} +\frac{1}{2}\, m\, B_{\mu\nu} \Phi^{\mu\nu} - B^2 \nonumber\\ 
& - & B \left(\partial\cdot\phi + m \,\varphi \right) + B^\mu B_\mu + B^\mu \left(\partial^\nu B_{\nu\mu} - \partial_\mu \varphi + m\,  \phi_\mu \right)
- m^2\, \bar \beta \beta \nonumber\\ 
& + & \left(\partial_\mu \bar C_\nu - \partial_\nu \bar C_\mu \right) \left(\partial^\mu C^\nu \right)
- \left(\partial_\mu \bar C - m \, \bar C_\mu \right) \left(\partial^\mu C - m \,C^\mu \right) + \partial_\mu \bar \beta \,\partial^\mu \beta  \nonumber\\ 
 & + & \left(\partial\cdot \bar C + \rho +  m \, \bar C\right) \lambda + \left(\partial\cdot C - \lambda
 +  m \, C \right) \rho,
\end{eqnarray}  
\begin{eqnarray}
 {\cal L}_{\bar b} & = & \frac{1}{12} \, H_{\mu\nu\eta}H^{\mu\nu\eta}
-\frac{1}{4}\,m^2\, B_{\mu\nu}B^{\mu\nu} -\frac{1}{4}\, \Phi_{\mu\nu} \Phi^{\mu\nu} + \frac{1}{2}\, m\, B_{\mu\nu} \Phi^{\mu\nu} - {\bar B}^2 \nonumber\\ 
& + & \bar B \left(\partial\cdot \phi - m \,\varphi\right) + \bar B_\mu \bar B^\mu 
+ \bar B^\mu \left(\partial^\nu B_{\nu\mu} + \partial_\mu \varphi + m\, \phi_\mu \right) - m^2\, \bar \beta \beta \nonumber\\ 
 & + & \left(\partial_\mu \bar C_\nu - \partial_\nu \bar C_\mu \right)(\partial^\mu C^\nu)
-  \left(\partial_\mu \bar C - m \, \bar C_\mu \right) \left(\partial^\mu C - m \, C^\mu \right)  + \partial_\mu \bar \beta\, \partial^\mu \beta \nonumber\\ 
& + & \left(\partial\cdot \bar C + \rho +  m \, \bar C\right) \lambda + \left(\partial\cdot C - \lambda + m \, C\right) \rho,
\end{eqnarray} 
where the fermionic (anti-)ghost fields $(\bar C_\mu)C_\mu$ and $(\bar C)C$ carry the ghost numbers $(-\, 1)+1$, respectively, 
and the bosonic (anti-)ghost fields $(\bar\beta)\beta$ are endowed with the ghost numbers $(-\, 2)+2$, respectively.
In addition, there are {\it fermionic} auxiliary fields $(\rho)\lambda$ with the ghost numbers (-1)+1, respectively.
All the other symbols are {\it same} as in the subsection 5.1 {\it but} for the additional Nakanishi-Lautrup type auxiliary fields
$(\bar B)B$ which are invoked  to linearize the gauge-fixing terms for  the St$\ddot u$ckelberg-compensating 
vector field $\phi_\mu$. We  have the rest  mass for {\it all} the basics fields, with well-defined   gauge-fixing terms, in 
our {\it  modified} massive Abelian 2-form theory which is denoted by the symbol $m$. The above coupled (but equivalent)
Lagrangian densities respect the following (anti-)BRST symmetry transformations (see, e.g. [40, 50] for details): 
\begin{eqnarray}
&&s_{ab} B_{\mu\nu} = - \,(\partial_\mu \bar C_\nu - \partial_\nu \bar C_\mu), \qquad 
s_{ab} \bar C_\mu  = - \,\partial_\mu \bar \beta, \qquad s_{ab} \phi_\mu = \partial_\mu \bar C - m\, \bar C_\mu, \nonumber\\ 
&& s_{ab}  C_\mu =  \bar B_\mu, \quad s_{ab} \beta = - \,\lambda, \quad s_{ab} \bar C = -\, m\, \bar \beta, \quad  s_{ab}  C = \bar B, 
\quad s_{ab} B = - \, m\, \rho, \nonumber\\ 
&& s_{ab} \varphi = \rho, \qquad s_{ab} B_\mu =   \partial_\mu \rho, \qquad 
s_{ab} [\bar B, \rho, \lambda, \bar \beta, \bar B_\mu,  H_{\mu\nu\kappa}] = 0,\nonumber\\  
&&s_b B_{\mu\nu} = - \,(\partial_\mu C_\nu - \partial_\nu C_\mu), \qquad 
s_b C_\mu  = - \, \partial_\mu \beta,  \qquad s_b \phi_\mu = \partial_\mu C - \, m\, C_\mu, \nonumber\\ 
&& s_b \bar C_\mu = -\,B_\mu, \quad s_b \bar \beta = - \,\rho, \quad 
s_b C = - \, m\,\beta, \quad s_b \bar C =  B, \quad s_b \bar B = -\, m \,\lambda,  \nonumber\\ 
&& s_b \varphi = \lambda, \qquad s_b \bar B_\mu =  - \,\partial_\mu \lambda, \qquad s_b [B, \rho, \lambda, \beta, B_\mu, H_{\mu\nu\kappa}] = 0.
\end{eqnarray} 
It is interesting to point out that the  above symmetry transformations are off-shell nilpotent of order two (i.e. $s_{(a)b}^2 = 0$)
and {\it their} absolute anticommutativity property (i.e. $\{s_b, s_{ab}\} = 0$) is also satisfied {\it except}  in the two specific cases which are as follows: 
\begin{eqnarray}
&& \{s_b, \, s_{ab}\}\, B_{\mu\nu} =  \partial_\mu(B_\nu - \bar B_\nu) - \partial_\nu(B_\mu - \bar B_\mu),\nonumber\\ 
&& \{s_b, \, s_{ab}\}\, \Phi_\mu =  \partial_\mu(B + \bar B) + m\, (B_\mu - \bar B_\mu).
\end{eqnarray}
However, we have the following (anti-)BRST invariant CF-type restrictions
\begin{eqnarray}
B_\mu - \bar B_\mu -\partial_\mu \phi = 0, \qquad B + \bar B  + m\, \phi = 0, 
\end{eqnarray}  
on our {\it modified}  massive Abelian 2-form theory within the framework of BRST formalism. 
The above restrictions have been derived from the superfield approach to BRST formalism (see, e.g. [29]). 
It can be checked that we have: $\{s_b, s_{ab}\}\, B_{\mu\nu} = 0$ and $\{s_b, s_{ab}\}\, \phi_\mu = 0$ in equation (62) due to (63).
Thus, the (anti-)BRST symmetry transformations, listed in (61),  are off-shell nilpotent and absolutely anticommuting 
in nature. Taking the precise forms of the off-shell nilpotent (anti-)BRST symmetry transformations for the appropriate fields, 
it is straightforward to check that we have the following: 
\begin{eqnarray}
s_{(a)b}\,[B_\mu - \bar B_\mu -\partial_\mu \phi] = 0, \qquad s_{(a)b}\,[B + \bar B  + m\, \phi] = 0. 
\end{eqnarray}
Thus,  the CF-type restrictions in (63), on our St$\ddot u$ckelberg-modified  {\it massive}
Abelian 2-form theory, are {\it physical} in the sense that {\it both} of them are (anti-)BRST invariant.

According to Noether's theorem, the infinitesimal, {\it continuous}, off-shell nilpotent and absolutely anticommuting 
(anti-)BRST symmetry transformations lead to the derivations of the following (anti-)BRST charges 
(see, e.g. [40] for details) 
\begin{eqnarray}
Q_{ab}  & = &  \int d^{D-1}x \Big[- H^{0ij} \,\big(\partial_i \bar C_j) 
-\big(\partial^0 \bar C^i - \partial^i \bar C^0\big)\, \bar B_i  +  \big(\partial^0 \bar C - m \,\bar C^0\big)\, \bar B \nonumber\\
& - & \big(\partial_i \bar C - m \, \bar C_i\big) \, \big(\Phi^{0i} - m \, B^{0i}\big) + m \, \big(\partial^0 C - m\, C^0\big)\, \bar \beta\nonumber\\
& - & \big(\partial^0 C^i - \partial^i C^0\big) (\partial_i \bar \beta) - \lambda \,\partial^0 \bar \beta + \rho\, \bar B^0 \Big],\nonumber\\ 
Q_b & = & \int d^{D-1}x \Big[- H^{0ij} \,\big(\partial_i  C_j)
- \big(\partial^0 C^i - \partial^i C^0\big)\, B_i  - \big(\partial^0 C - m \,C^0\big)\ B\nonumber\\
& - & \big(\Phi^{0i} - m  \,B^{0i}\big)\, \big(\partial_i C - m \, C_i\big) - m \,  \big(\partial^0 \bar C - m \,\bar C^0\big)\, \beta \nonumber\\
& + & \big(\partial^0 \bar C^i - \partial^i \bar C^0\big) (\partial_i \beta) - \rho \,\partial^0 \beta - \lambda\, B^0 \Big],
\end{eqnarray}
which are the generators of the (anti-)BRST symmetry transformations. However, these conserved 
charges are found to be {\it non-nilpotent} [i.e. $Q_{(a)b}^2 \neq 0$]. A systematic theoretical methodology has been developed
in our earlier work (see, e.g. [40]) where we have exploited  the theoretical strength of (i) the Gauss divergence theorem, and (ii)
appropriate EL-EoMs from the suitable Lagrangian density, to  obtain the conserved and nilpotent (anti-)BRST
charges $[Q_{(a)b}^{(2)}]$ from the standard {\it but} non-nilpotent  versions of the conserved (anti-)BRST charges $[Q_{(a)b}]$  
that have been derived from the basic principle behind the Noether theorem. The conserved and off-shell nilpotent versions of the 
(anti-)BRST charges are [40]: 
\begin{eqnarray}
Q_b & \longrightarrow & Q_{b}^{(2)}  =  \int d^{D - 1} x\, \Big [(\partial^0 B^i - \partial^i B^0)\, C_i + 
(\dot B + m\,B^0)\, C + m\, (\dot {\bar C} - m\, \bar C^0)\, \beta\nonumber\\
& - & \,(\partial^0 \bar C^i - \partial^i \bar C^0)\, \partial_i \beta 
 \,+\, 2\, \dot\rho\,\beta - \rho\,\dot{\beta} - \lambda\,  B^0 - (\partial^0  C^i - \partial^i  C^0)\, B_i   
- (\dot {C} - m\, C^0)\, B \Big],  \nonumber\\
Q_{ab} & \longrightarrow & Q_{ab}^{(2)}   =  \int d^{D - 1} x\, \Big [(\partial^0 \bar B^i - \partial^i \bar B^0)\, \bar C_i - 
(\bar B^0 - m\,\bar B^0)\, \bar C + (\partial^0 C^i - \partial^i C^0)\, \partial_i\bar \beta \nonumber\\
& - & m\, (\dot C - m C^0)\, \bar\beta + 2\, \dot\lambda\bar\beta - \lambda\,\dot{\bar\beta} + \rho\, \bar B^0 
+ (\dot {\bar C} - m \bar C^0)\, \bar B - (\partial^0 \bar C^i - \partial^i \bar C^0)\, \bar B_i \Big],  
\end{eqnarray}
where the superscript $(2)$   has been taken to distinguish these charges from our subsection 5.1 where the conserved and nilpotent 
charges [cf. Eq. (50)] have been denoted by the superscript $(1)$. 
It is an elementary exercise to check that we have 
\begin{eqnarray}
s_b \, Q_b ^{(2)}  & = & -\, i\, \{Q_b ^{(2)}, \; Q_b ^{(2)}\}  = 0 \quad \Longrightarrow \quad [Q_b ^{(2)}]^2 = 0,\nonumber\\
s_{ab} \, Q_{ab} ^{(2)}  & = & -\, i\, \{Q_{ab} ^{(2)}, \; Q_{ab} ^{(2)}\}  = 0 \quad \Longrightarrow \quad [Q_{ab} ^{(2)}]^2 = 0,
\end{eqnarray}
where the l.h.s. can be readily computed by using the (anti-)BRST symmetry transformations (61) and the  explicit expressions for 
the nilpotent (anti-)BRST charges  from (66).

The stage is set now to exploit the beauty and strength  of the physicality criteria with the help of conserved and off-shell nilpotent 
(anti-)BRST charges  $Q_{(a)b}^{(2)}$, namely; 
\begin{eqnarray}
Q_b^{(2)} \, |phys>\, = 0  \qquad &\Longrightarrow & \qquad  B_i\, |phys>\, = 0, \nonumber\\ 
\qquad &\Longrightarrow & \qquad  B\, |phys>\, = 0,  \nonumber\\
\qquad  &\Longrightarrow & \qquad   (\partial^0 B^i - \partial^i B^0)\, |phys>\, = 0, \nonumber\\
\qquad &\Longrightarrow &\qquad  (\dot B + m\,B^0)\, |phys>\, = 0.
\end{eqnarray}
In the above, we have taken into consideration {\it only} the BRST charge $[Q_b^{(2)}]$ which is off-shell nilpotent and conserved because
we {\it shall} get the {\it same} conditions from the anti-BRST charge $[Q_{ab}^{(2)}]$ except that the 
Nakanishi-Lautrup auxiliary fields $(B, \,  B_i)$ will be replaced by $(\bar B, \,  \bar B_i)$.
However, these notational changes  will {\it not} alter any physical consequences as far as the constraint analysis,  at the
{\it quantum} level,  is concerned. To observe the physical consequences from the conditions (68), 
it is essential that we focus on the following conjugate momenta (derived from the 
perfectly BRST invariant Lagrangian density ${\cal L}_b$), namely: 
\begin{eqnarray}
&&\Pi^{\mu\nu}_{(B)} = \frac {\partial {\cal L}_b}{\partial (\partial_0 B_{\mu\nu})} = \frac {1}{2}\, H^{0\mu\nu}\, + \frac {1}{2}\,
(\eta^{0\mu}\, B^{\nu} - \eta^{0\nu}\, B^\mu), \nonumber\\
&& \Pi^{\mu}_{(\phi)} =  \frac {\partial {\cal L}_b}{\partial (\partial_0 \phi_{\mu})} =  + m\, B^{0\mu} - \Phi^{0\mu} - \eta^{0\mu}\, B, 
\end{eqnarray}
which lead to the derivations  of the following components of the conjugate momenta:   
\begin{eqnarray}
&&\Pi^{0i}_{(B)} = -\, \frac{1}{2}\, B_i,  \qquad \qquad \Pi^{ij}_{(B)} = -\, \frac{1}{2}\, H^{0ij}, \nonumber\\
&&\Pi^{0}_{(\phi)} = -\,  B_i,  \qquad \qquad  \Pi^{i}_{(\phi)} = + m\, B^{0i} - \Phi^{0i}.  
\end{eqnarray}
The above expressions demonstrate that, within the framework of BRST formalism, {\it all} the components of momenta are 
{\it non-zero} and, hence, the covariant canonical quantization (i.e. BRST-quantization) can be performed in a straightforward manner.  
In addition to (70), we require the following EL-EoMs, w.r.t the basic fields $B_{\mu\nu}$ and $\phi_{\mu}$, namely; 
\begin{eqnarray}
&&\partial_\mu H^{\mu\nu\lambda} + m^2\, B^{\nu\lambda} - m\, \Phi^{\nu \lambda }
 + (\partial^\nu B^\lambda - \partial^\lambda B^\nu) = 0, \nonumber\\ 
&&\partial_\mu \Phi^{\mu\nu} - m\, \partial_\mu\, \Phi^{\mu\nu} + \partial^\nu\, B + m\, B^\nu = 0,
\end{eqnarray}
that are derived from the {\it perfectly} BRST-invariant Lagrangian density 
${\cal L}_{b}$ [cf. Eq. (59)]. Taking the  choices: $\nu = 0, \lambda = j$ in  (70), we obtain the following 
\begin{eqnarray}
&&(\partial^0\, B^i - \partial^i\, B^0) = 2\, \big[ \partial_j\, \Pi^{ji}_{(B)} - \frac{m}{2}\, \Pi^i_{(\phi)}\big], \nonumber\\
&& (\dot B + m\, B^0)  = -\, \big[\partial_i\, \Pi^i_{(\phi)}\big], 
\end{eqnarray}
where we have used the notations from (70) for the components of the conjugate momenta. 
It is crystal clear now that we have the following consequences from (68) in terms of explicit forms of 
the first-class constraints:  
\begin{eqnarray}
B\, | \, phys > \, = 0 \qquad &\Longrightarrow& \qquad    \Pi^0_{(\phi)}\, | \, phys > \, = 0, \nonumber\\
B_i \, | \, phys > \, = 0 \qquad &\Longrightarrow& \qquad   \Pi^{0i}_{(B)}\, | \, phys > \, = 0, \nonumber\\
(\dot B + m\, B^0) \, | \, phys > \, = 0 \qquad &\Longrightarrow& \qquad  \partial_i \,  \Pi^i_{(\phi)}\, | \, phys > \, = 0, \nonumber\\
(\partial^0\, B^i - \partial^i\, B^0) \, | \, phys > \, = 0 \qquad &\Longrightarrow& \qquad  
 \big[ \partial_j\, \Pi^{ji}_{(B)} - \frac{m}{2}\,  \Pi^i_{(\phi)} \big] \, | \, phys > \, = 0. 
\end{eqnarray}
Ultimately, we conclude that the operator forms of the first-class constraints 
[$\Pi^{0i}_{(B)} \approx 0, \; \Pi^{0}_{(\phi)} \approx 0, \;  
(\partial_j\,  \Pi^{ji}_{(B)} - \frac{m}{2}\,  \Pi^i_{(\phi)}) \approx 0, \; \partial_i\, \Pi^i_{(\phi)} \approx 0$]
of the St$\ddot u$ckelberg-modified {\it massive}  Abelian 2-form theory 
[{\it classically} described  by the Lagrangian density (29)] annihilate the physical states 
(in the {\it total} quantum Hilbert of states) within the framework of BRST formalism {\it through} 
the celebrated physicality criteria w.r.t. the off-shell nilpotent and conserved (anti-)BRST charges
for our {\it modified} massive Abelian 2-form theory.  
A noteworthy point, at this stage, is the observation that we have {\it not} taken into account any condition on the 
physical state $(| phys >)$ from the term: $-\, \lambda\, B^0$ from the BRST charge $Q_b^{(2)}$. This due to the 
fact that (i) the auxiliary ghost field $\lambda$ is {\it not} the basic
[i.e. $\lambda = (\frac{1}{2})\, (\partial\cdot C + m\, C)$] field in the theory, and (ii) the canonical conjugate 
momentum  $[\Pi_{(\phi)}]$,  w.r.t. the $\phi$ field for the Lagrangian density (59), is: 
$\Pi_{(\phi)} = -\, B^0$ which is {\it not} a constraint. In other words, the Nakanishi-Lautrup auxiliary field  $B^0$ 
is {\it not} connected with any kinds of constraints (that are present in our {\it modified} massive Abelian 2-form theory at the
classical level).  To be precise, the scalar field $\phi$ appears only at the {\it quantum} level in our theory.

\section{Modifications in St$\ddot u$ckelberg-Technique: 2D Proca  and 4D Massive Abelian 2-Form Theories}

In this section, we discuss the {\it modifications} in the St$\ddot u$ckelberg-technique of the 
compensating fields in the context of (i) the  2D {\it massive}  Abelian 1-form (i.e. Proca) theory, and (ii) the 
4D {\it massive}  Abelian 2-form theory. Thus, our present section is divided into {\it two} subsections to 
capture the {\it above} appropriate  modifications {\it separately}.  For this section to be self-contained, we 
have taken some theoretical materials from our earlier works [40, 50].

\subsection{2D  Massive Abelian 1-Form Theory: Special Features}

We begin with the following Lagrangian density  for the  two (1 + 1)-dimensional (2D)
 Abelian  1-form (i.e. Proca) theory (see, e.g. [41, 38, 39])
\begin{eqnarray}
{\cal L}_{(P)}^{(2D)} = -\frac {1}{4} F_{\mu\nu} F^{\mu\nu} +  \frac {m^2}{2}  A_\mu \, A^\mu
 \equiv +  \frac {1}{2}  E^2   +  \frac {m^2}{2}  A_\mu \, A^\mu, 
\end{eqnarray}
where, in the {\it specific} two (1 + 1)-dimensions of spacetime, the field-strength tensor $F_{\mu\nu} =  \partial_\mu\, A_\nu - \partial_\nu\, A_\mu$
has only {\it one} independent and existing component (i.e. $F_{01} = \partial_0 \, A_1 - \partial_1\, A_0  = E$) 
which is nothing but a pseudo-scalar (electric) field that changes sign under the parity operation.  In  any arbitrary 
dimension of spacetime, the {\it usual}  St$\ddot u$ckelberg-technique, can be written in the following 
form in the language of the differential 1-forms, namely; 
\begin{eqnarray}
A^{(1)} \longrightarrow A^{(1)} \mp \frac{1}{m}\, d\, \Phi^{(0)}, 
\end{eqnarray}
where the 0-form $\Phi^{(0)} = \phi$ is a pure scalar field without any tensorial index  and the operator $d$
(with $d^2 = 0$) is the exterior derivative of the differential geometry [45-49]. 
Taking into account $d = \partial_\mu \; d\,x^\mu$ and $  A^{(1)}  =   A_\mu \; d\,x^\mu$, we find that the above 
relationship (75) is nothing but the {\it usual} St$\ddot u$ckelberg-technique for the replacement:
 $A_\mu \longrightarrow A_\mu \mp \frac {1}{m}\, \partial_\mu \phi$ which leads to the derivation of 
the St$\ddot u$ckelberg-modified Lagrangian density $[{\cal L}_{(S)}^{(2D)}]$ from ${\cal L}_{(P)}^{(2D)}$  as
\begin{eqnarray}
{\cal L}_{(P)}^{(2D)} \longrightarrow {\cal L}_{(S)}^{(2D)} = \frac {1}{2}\, E^2
+ \frac {m^2}{2} A_\mu\, A^ \mu  \mp m \,A_\mu \,\partial^\mu\, \phi + \frac {1}{2}\, \partial_\mu\, \phi\, \partial^\mu\, \phi,
\end{eqnarray}
which respects the following gauge symmetry transformations for 2D theory, namely; 
\begin{eqnarray}
\delta_g\, A_{\mu} = \partial_\mu\,\Sigma,\qquad \delta_g\,\phi = \pm\,m\,\Sigma,\qquad \delta_g\, E = 0,
\end{eqnarray}
because of the presence of the first-class constraints on the St$\ddot u$ckelberg-modified 
Lagrangian density for the 2D Proca theory in (76) where the infinitesimal  and local gauge symmetry transformation
parameter is $\Sigma (\vec x, t)$. It is pertinent to point out that the second-class constraints of ${\cal L}_{(P)}^{(2D)}$
have been converted  into the first-class constraints\footnote{For our 2D theory, the 
constraints are: $\Pi^0_{(A)} \approx 0$ and $\partial_i\, \Pi^{(i = 1)}_{(A)} \mp m\, \Pi_{(\phi)} \equiv 
\partial_1\, E \mp m\, \Pi_{(\phi)} \approx 0$ where $\Pi^{\mu}_{(A)} = -\, \varepsilon^{0\mu}\, E $ defines
 $\Pi^0_{(A)} \approx 0$ and  $\Pi^{(i = 1)}_{(A)} = E$  and we  also have $\Pi_{(\phi)} = \dot  \phi \mp m\, A_0$
as the conjugate momentum w.r.t. the pure scalar field ($\phi$). 
These first-class constraints consist of the primary constraint  ($\Pi^0_{(A)} \approx 0$) and the secondary constraint 
($\partial_1\, E \mp m\, \Pi_{(\phi)} \approx 0$) which commute with each-other. } 
due to the presence of the 
St$\ddot u$ckelberg-compensating field $(\phi)$ which happens to be a pure scalar field.

The two (1 + 1)-dimensional (2D) spacetime is very {\it special} for the massive Abelian 1-form theory because 
the {\it usual} St$\ddot u$ckelberg-technique (with a pure scalar field $\phi$) gets modified and
 it can incorporate a pseudo-scalar field ($\tilde \phi$), too. In other words, we find
that (75) changes to the following form (in the {\it special} 2D case), namely; 
\begin{eqnarray}
A^{(1)} \longrightarrow A^{(1)} \mp \frac{1}{m}\, d\, \Phi^{(0)} \pm \frac{1}{m}\, *\, d\,\tilde{\Phi}^{(0)},
\end{eqnarray}
where the 0-form $\tilde \Phi^{(0)} = \tilde \phi$ is a pseudo-scalar field (which changes  sign under the parity operation)
and $*$ is the Hodge duality operation so that the parity symmetry is maintained in our theory. 
In the language  of the vector field $(A_\mu)$, we have the following replacement:   
\begin{eqnarray}
A_\mu \longrightarrow A_\mu \mp \frac {1}{m}\, \partial_\mu \phi \mp \frac{1}{m}\,\varepsilon _{\mu\nu}\,\partial^{\nu}\,\tilde{\phi}.
\end{eqnarray}
Here $\varepsilon _{\mu\nu}$ is the 2D Levi-Civita tensor with the choices: 
$\varepsilon _{01} = +1 = \varepsilon ^{10}, \,\, \varepsilon _{10} = \varepsilon ^{01} = -1$ and $\varepsilon _{\mu\nu}\, \varepsilon ^{\mu\nu} = -\,2!,\,
\varepsilon _{\mu\nu}\,\varepsilon ^{\mu\lambda} = -\,\delta^\lambda _\nu,$ etc. As a side 
remark, we point out that the 2D field-strength tensor $F_{\mu\nu}$ (with {\it only} $F_{01} = E$ as the independent and existing 
component) can {\it also} be expressed in terms of the Levi-Civita tensor 
$(\varepsilon _{\mu\nu})$ as: $F_{01} = E \equiv  -\, \varepsilon ^{\mu\nu}\,\partial_\mu A_\nu.$
The {\it modified}  St$\ddot u$ckelberg-technique (with a pure scalar field $\phi$ and a pseudo-scalar field $\tilde {\phi}$) remains
invariant under the following discrete duality  symmetry transformations: 
\begin{eqnarray}
&&A_\mu \to \mp\, i\, \varepsilon_{\mu\nu} A^\nu, \qquad \qquad \; \phi \to \mp\, i\, \tilde \phi, 
\qquad \qquad \tilde \phi \to \mp \,i \, \phi. 
\end{eqnarray} 
We call the above transformations as  the ``duality" symmetry transformations because the origin of the {\it basic} symmetry transformation
for the gauge field (i.e. $A_\mu \longrightarrow \mp\,i\,\varepsilon _{\mu\nu}A^\nu$) owes its origin to the {\it self-duality} condition
on the 1-form field $A^{(1)}$ in 2D spacetime. In other words, we have 
the following explicit form: 
\begin{eqnarray}
*\, A^{(1)} \; \equiv\;  A_\mu\; \bigl [*\, (dx^\mu) \bigr] \equiv\; \tilde A_\mu\; d x^{\mu} \quad 
\;\Longrightarrow\; \quad  \tilde A_\mu  = -\, \varepsilon _{\mu\nu} A^\nu.
\end{eqnarray} 
Thus, it is clear that the self-duality condition on the {\it basic}  1-form $(A^{(1)} = dx^\mu A_\mu)$ 
field $A_\mu$ is the root-cause behind the transformations: $A_\mu \longrightarrow \mp\,i\,\varepsilon _{\mu\nu} \,A^\nu$ 
(modulo some signs and $`i$' factors). Under the replacement (79), we have the following explicit replacements: 
\begin{eqnarray}
A_0 & \longrightarrow & A_0 \mp \frac {1}{m}\,(\partial_0 \phi - \partial_1 \tilde \phi),\nonumber\\
A_1 & \longrightarrow & A_1 \mp \frac {1}{m}\,(\partial_1 \phi - \partial_0 \tilde \phi),\nonumber\\
F_{01} & \longrightarrow & F_{01} \pm \frac {1}{m}\, \Box\, \tilde \phi. 
\end{eqnarray} 
The last entry implies that: $E \longrightarrow E \pm \frac {1}{m}\, \Box \, \tilde \phi$. However, we note that, in the 
St{$\ddot u$}ckelberg-modified Lagrangian density (76), we have the kinetic term for the gauge field as: 
$(1/2)\, E^2$ in the 2D spacetime. 
Thus, for the 2D {\it modified} massive Proca theory, we shall have higher derivative  terms in the following change [due to (79)], namely; 
\begin{eqnarray}
\frac {1}{2}\, E^2 \longrightarrow \frac {1}{2}\,( E \pm \frac {1}{m}\, \Box \, \tilde \phi)^2, 
\end{eqnarray} 
which will be problematic for a renormalizable and consistent theory in the two (1 + 1)-dimensions of spacetime.
We can get rid of this problem, if we take the help of the Klein-Gorden equation of motion: 
$(\Box + m^2)\, \tilde \phi = 0  \Longrightarrow \Box \,\tilde \phi  = -\, m^2\, \tilde \phi$ in an ad-hoc and
arbitrary manner at this stage. However,  from the appropriately  
defined Lagrangian density for the {\it modified} 2D Proca theory, ultimately,  we should obtain the EL-EoM: 
$(\Box + m^2)\, \tilde \phi = 0 $ so that the sanctity of {\it this} substitution can be explicitly corroborated. 
Taking the input  of the on-shell condition: 
$ \Box \,\tilde \phi  = -\, m^2\, \tilde \phi$, we observe that we have the following transformation  for  the kinetic term 
of the 2D St{$\ddot u$}ckelberg-{\it modified}  Proca theory (where $\phi$ and $\tilde\phi$ are present as the compensating fields), namely; 
\begin{eqnarray}
\frac {1}{2}\, E^2 \longrightarrow \frac {1}{2}\,( E \mp \,m\,\tilde \phi)^2, 
\end{eqnarray} 
which is a renormalizable term in the 2D spacetime. Finally, we have the following gauge-fixed Lagrangian density 
for the {\it modified} 2D Proca theory [with the modified form of the St{$\ddot u$}ckelberg-technique in (79)], namely; 
\begin{eqnarray}
{\cal L}_{(2D)}  &=& \frac {1}{2}\, {(E \mp m\,\tilde\phi)}^2 \pm m\,  E\,\tilde\phi - 
\frac {1}{2}\,\partial_\mu \,\tilde\phi\,\,\partial^\mu\,\tilde\phi 
+ \frac {m^2}{2} A_\mu\, A^ \mu \nonumber\\ &\mp& m \,A_\mu \,\partial^\mu\, \phi + \frac {1}{2}\, \partial_\mu\, \phi\, \partial^\mu\, \phi - \frac{1}{2}
\,(\partial\,\cdot A \pm m\,\phi)^2,
\end{eqnarray}
where the {\it last} term is nothing but the gauge-fixing term in the `t Hooft gauge (see, e.g. [41]). At this stage, we lay emphasis on 
{\it three} crucial points connected with (85). First, it can be checked that the gauge-fixed Lagrangian density respects
the discrete duality symmetry transformations: $A_\mu \to \mp\, i\, \varepsilon_{\mu\nu} A^\nu, \;  \phi \to \mp\, i\, \tilde \phi, 
\;  \tilde \phi \to \mp \,i \, \phi$ along with $(\partial\,\cdot A) \to \pm i\, E,  \;  E \to \pm i\, (\partial\,\cdot A)$. 
The {\it latter} two transformations are due to: $A_\mu \to \mp\, i\, \varepsilon_{\mu\nu} A^\nu$. 
Second, we can check explicitly that the EL-EoMs\footnote{This shows that, in going from (83) to (84), whatever we have chosen
in an ad-hoc and arbitrary manner  is the {\it correct} way to get rid of the higher order derivative terms for our 2D theory. This is due to the
fact that the Lagrangian density (85) produces  $(\Box +  m^2)\, \tilde \phi  = 0$ as the EL-EoM.}: 
$(\Box + m^2)\,  \phi = 0$ and $(\Box + m^2)\,\tilde \phi = 0$ are 
satisfied from the gauge-fixed Lagrangian density. 
Finally, we note that the kinetic term for the pseudo-scalar field $\tilde\phi$ carries a {\it negative} sign 
in comparison to the scalar field $\phi$.  Such fields with {\it negative}  kinetic terms (but {\it with}  well-defined rest mass) are {\it exotic} fields 
which are relevant in the   context of the cyclic, bouncing and self-accelerated cosmological models of the 
Universe [51-53] and they are also a set of possible candidates  for the dark matter
[54, 55] in the modern literature on {\it this} challenging topic of research activity.

Within the framework of BRST formalism, the above gauge-fixed Lagrangian density can be generalized to incorporate the 
Faddeev-Popov (FP) ghost terms as follows 
\begin{eqnarray}
{\cal L}_{(B)}^{(2D)} &=& \frac {1}{2}\, {(E \mp m\,\tilde\phi)}^2 \pm m\,  E\,\tilde\phi - 
\frac {1}{2}\,\partial_\mu \,\tilde\phi\,\,\partial^\mu\,\tilde\phi 
+ \frac {m^2}{2} A_\mu\, A^ \mu  \mp m \,A_\mu \,\partial^\mu\, \phi  \nonumber\\ 
&+& \frac {1}{2}\, \partial_\mu\, \phi\, \partial^\mu\, \phi - \frac{1}{2} \,[(\partial\,\cdot A) \pm m\,\phi]^2
-\, i\, \partial_\mu\, \bar C\, \partial^\mu\, C + i\, m^2\, \bar C\, C,  
\end{eqnarray}
where the (anti-)ghost fields $(\bar C)C$ are fermionic (i.e. $ C^2 = \bar C^2 = 0, \; \bar C\, C + C\, \bar C = 0$)
in nature and they are invoked to maintain the {\it unitarity} in the theory. 
The subscript $(B)$ on the Lagrangian density stands for its (anti-)BRST and (anti-)co-BRST invariant form. 
We can linearize the quadratic  kinetic and gauge-fixing terms by  invoking a couple of  Nakanishi-Lautrup type auxiliary fields (${\cal B}, B$) as 
\begin{eqnarray}
{\cal L}_{(B, \, {\cal B})}^{(2D)} &=& {\cal B}\, {(E \mp m\,\tilde\phi)} - \frac{1}{2}\,{\cal B}^2 \pm m\,  E\,\tilde\phi - 
\frac {1}{2}\,\partial_\mu \,\tilde\phi\,\,\partial^\mu\,\tilde\phi 
+ \frac {m^2}{2} A_\mu\, A^ \mu + \frac {1}{2}\, \partial_\mu\, \phi\, \partial^\mu\, \phi \nonumber\\ &\mp& m \,A_\mu \,\partial^\mu\, \phi
+ B\,(\partial\cdot A \pm m\,\phi) + \frac{1}{2}\,{B}^2  -\,i\,\partial_\mu\,\bar C\, \partial^\mu\,C + i\, m^2\,\bar C\,C,
\end{eqnarray}
which respects the generalized version of the {\it classical} discrete duality symmetry transformations (80) in the 
following form at the {\it quantum} level: 
\begin{eqnarray}
&&\phi \to \mp\, i\, \tilde \phi, 
\qquad \tilde \phi \to \mp \,i \, \phi, \qquad 
A_\mu \to \mp\, i\, \varepsilon_{\mu\nu} A^\nu,  \qquad B\to \mp i\, {\cal B},\quad {\cal B} \to \mp i\,B, \nonumber\\
&&C \to \pm i\,\bar C, \quad \qquad \;\; \bar C \to \pm i\, C, \quad \quad \;(\partial\,\cdot A) \to \pm i\, E \qquad E \to \pm i\, (\partial\,\cdot A).
\end{eqnarray}
It is very interesting to point out that the Lagrangian density (87) respects, in addition to a couple of discrete 
duality symmetry transformations in (88), a set of {\it six} continuous symmetry transformations that incorporate
{\it four} fermionic symmetry transformations and {\it  two} bosonic. In precise terms, the {\it four} fermionic 
(i.e. nilpotent) symmetry transformations are the BRST, anti-BRST, co-BRST and anti-co-BRST symmetry transformations. 
On the other hand, there exists a {\it unique} bosonic symmetry transformation that turns up as a  suitable anticommutator of the 
(anti-)BRST and (anti-)co-BRST transformations [38, 39].
Finally, there exists  a ghost-scale symmetry transformation that is also  {\it bosonic} in nature. These total 
(i.e. discrete and continuous) symmetries entail upon the 2D {\it modified} Proca theory to become a field-theoretic model 
for the Hodge theory (see, e.g. [39] for details).

As far as our present investigation is concerned, we mention here only {\it two} fermionic
(i.e. nilpotent) BRST and co-BRST (i.e. dual BRST) symmetry transformations which are denoted by $s_b$
and $s_d$,  respectively. These are assimilated   as follows [38, 39]
\begin{eqnarray}
&&s_{b}\,A_\mu = \partial_\mu\,C,\qquad \;\; s_{b}\,C = 0,\qquad s_{b}\,\bar C = +\,i\,B, \qquad s_{b}\,\phi = \pm m\,C, \nonumber\\
&&s_{b}\,B = 0,\qquad\qquad \;\;s_{b}\,E = 0,\qquad \; s_{b}\,{\cal B} = 0,\qquad \qquad s_{b}\,{\tilde \phi} = 0,  \nonumber\\
&& s_{d} \,A_\mu = - \varepsilon_{\mu\nu} \partial^\nu \,\bar C, \quad   s_{d} \, \bar C = 0, \quad s_{d}\, C = -\, i\, {\cal B},
 \quad \quad s_{d}\, {\cal B} = 0,\qquad s_{d} \,\phi =  0,\nonumber\\ 
&& s_{d}\, B =  0,\qquad \qquad s_{d} \,\tilde \phi = \mp\, m\, \bar C,\qquad s_{d} \,(\partial \cdot A) = 0,\qquad s_{d}\, E = \Box \,\bar C, 
\end{eqnarray}
which demonstrate that (i) the electric field (i.e. $E$), owing its origin to the 
exterior derivative, remains invariant under $s_b$, and (ii) the gauge-fixing term (i.e. $\partial \cdot A$), 
tracing back its origin to the co-exterior derivative, remains invariant under the dual-BRST (i.e. co-BRST) symmetry 
transformations. Hence, the nomenclatures of the BRST and co-BRST symmetry transformations are appropriate because 
of their {\it origin} (and connection with the exterior and co-exterior derivatives). 
The infinitesimal, continuous and off-shell nilpotent symmetry transformations lead, according to the Noether theorem, 
to the derivations of the conserved current and corresponding conserved charges. These {\it latter} charges 
(which are off-shell nilpotent of order two) are as follows (see, e.g. [38, 39] for details): 
\begin{eqnarray}
Q_b = \int dx\, [B\, \dot C - \dot B\, C], \qquad\qquad Q_d = \int dx\, [{\cal B}\, \dot {\bar C} - \dot {\cal B}\, \bar C],
\end{eqnarray}
where the pair  $(B, {\cal B})$ are the Nakanishi-Lautrup auxiliary fields and $(\bar C)C$ are the fermionic (anti-)ghost 
fields of the Lagrangian density (87). It is clear,  from the {\it latter},  that we have: ${\cal B} =  (E \mp m\,\tilde\phi)$
and $B = -\,(\partial\cdot A \pm m\,\phi)$. Further, we note that the conjugate momenta w.r.t. the {\it massive}
gauge field $(A_\mu)$,  from (87),  is as follows 
\begin{eqnarray}
\Pi ^\mu _{(A)} = \frac {\partial {\cal L}_{(B)}}{\partial (\partial_0 A_\mu)} 
= -\, \varepsilon ^{0\mu}\, {\cal B} + \eta^{0\mu}\, B \mp m\,\varepsilon ^{0\mu}\, \tilde\phi,
\end{eqnarray}
where we have taken the help of the covariant form of the electric field: $E = -\,\varepsilon ^{\mu\nu}\, \partial_\mu A_\nu$
 and the subscript $(A)$ on the conjugate momenta denotes that the field $A_\mu$ is invoked (in the canonical definition of the conjugate momenta).
 It is clear that we have the following components of the momenta in the 2D Minkowskian spacetime, namely;
 \begin{eqnarray}
\Pi ^0 _{(A)}  = B \equiv  -\,(\partial\cdot A \pm m\,\phi), \qquad\qquad \Pi^{(i = 1)}_{(A)} = {\cal B} \pm m\, \tilde\phi  \equiv  E. 
\end{eqnarray}
In addition, we have the following EL-EoM w.r.t the massive gauge field $A_\mu$, namely; 
\begin{eqnarray}
\varepsilon^{\mu\nu}\, \partial_\nu\, {\cal B} + \partial^\mu\, B \pm  m\, \varepsilon^{\mu\nu}\, \partial_\nu\, \tilde\phi 
= m^2\, A^\mu \mp m\, \partial^\mu\, \phi, 
\end{eqnarray}
which, for the choice: $\mu = 0$, leads to 
\begin{eqnarray}
\dot B = \partial_1\, ({\cal B} \pm m\, \tilde\phi) \mp m\, (\dot \phi \mp m\, A_0)  \quad \equiv \quad  \partial_1\, E \mp m\, \Pi_{(\phi)}, 
\end{eqnarray}
where $\Pi_{(\phi)} = \dot \phi \mp m\, A_0$ is the canonical conjugate momentum  w.r.t. the pure scalar field $\phi$
and we have also used ${\cal B} = E \mp m\, \tilde\phi$ that emerges out from (87).

We are in the position now to {\it impose}  the physicality criterion w.r.t. the off-shell nilpotent and conserved  BRST 
charge $Q_{(b)}$ [cf. Eq. (90)]. Since the (anti-)ghost fields are {\it not} physically important, we note that we have the following: 
\begin{eqnarray}
Q_{(b)}\, | \, phys> = 0  \qquad \Longrightarrow \qquad B\, |phys> = 0, \; \qquad \; \;  \dot B \, |phys > = 0. 
\end{eqnarray}
A close and careful look at (92) and (94) shows that, in the language of the first-class constraints of the 
St$\ddot u$ckelberg-modified massive 2D Proca theory [{\it classically} described by the Lagrangian density (76)], we have the following: 
\begin{eqnarray}
 B\, |phys> = 0  \qquad &\Longrightarrow& \qquad  \Pi^0_{(A)} \, |phys> = 0  \nonumber\\
\qquad &\Longrightarrow& \qquad    -\, (\partial \cdot A \pm m\, \phi)\, |phys> = 0,   \nonumber\\ 
\dot B \, |phys > = 0   \qquad &\Longrightarrow& \qquad (\partial_1\, E \mp m\, \Pi_{(\phi)})\, |phys> = 0  \nonumber\\
\qquad &\Longrightarrow& \qquad 
 -\, \partial_0(\partial \cdot A \pm m\, \phi)\, |phys> = 0.  
\end{eqnarray}
Hence,  the physical states (in the {\it total} quantum Hilbert space of states) are {\it those} that are
annihilated by the conserved and off-shell nilpotent BRST charge in any arbitrary dimension of 
spacetime (where the special 2D spacetime is {\it also} included). This automatically implies 
that the operator forms of the first-class constraints annihilate the physical states
(existing in the {\it total} quantum Hilbert space of states) at the {\it quantum} level.

We end this subsection with a crucial remark that is connected with the St$\ddot u$ckelberg-modified
Proca theory in the two (1+1)-dimension of spacetime which has already been proven 
(see, e.g. [56, 39] for details) to be a massive model of Hodge theory where the Hodge decomposition theorem
(see, e.g. [39, 56] for details) can be performed  in the quantum Hilbert space of states. 
The {\it most} symmetric  state,  in the above decomposition,  is the {\it harmonic} state that is annihilated by {\it both} the {\it basic} 
conserved and nilpotent charges of our theory (which are nothing but the conserved and off-shell nilpotent BRST and co-BRST charges). Thus, the 
physical state ($| phys> $),  in the total Hilbert space of states, is the harmonic state for a field-theoretic model of Hodge theory. 
Hence, the physicality criterion w.r.t. the off-shell nilpotent and conserved co-BRST charge leads to the following conditions, namely;  
\begin{eqnarray}
Q_d\, |phys> = 0 \qquad \Longrightarrow \qquad {\cal B} \, |phys> = 0, \quad  \qquad  \dot {\cal B} \, |phys> = 0, 
\end{eqnarray}
which can be equivalently  expressed  as follows   
\begin{eqnarray}
&&{\cal B}\, |phys > = 0 \qquad \Longrightarrow \qquad (E \mp m\, \tilde\phi)\, |phys> = 0, \nonumber\\
&&\dot {\cal B} \, |phys > = 0 \qquad \Longrightarrow \qquad \partial_0\, (E \mp m\, \tilde\phi)\, |phys> = 0. 
\end{eqnarray}
At this stage, first of all,  we note that $Q_b$ and $Q_d$
are related to each-other (i.e. $Q_b \longleftrightarrow Q_d$) by the discrete duality symmetry transformations: 
$B \longrightarrow \mp i\, {\cal B}, \; {\cal B} \longrightarrow  \mp i\, B, \; 
C \longrightarrow \pm i\, \bar C, \; \bar C \longrightarrow \pm i\, C$ that are present in the full set of discrete duality  symmetry 
transformations (88). Thus, we conclude that $Q_b\, |phys> = 0$ implies that the operator form of the first-class constraints {\it must}  annihilate the 
physical states.  On the other hand, the physicality criterion w.r.t. the conserved and off-shell nilpotent co-BRST charge $Q_d$ (i.e. $Q_d\, |phys> = 0$)
implies that the {\it dual}  versions of the operator form of the first-class constraints must
annihilate the physical states of a 2D field-theoretic model of Hodge theory, too. Mathematically, we capture these statements as follows 
\begin{eqnarray}
&&  B\, |phys> = 0  \quad \Longrightarrow \quad   -\, (\partial \cdot A \pm m\, \phi)\, |phys> = 0 \qquad  \Longleftrightarrow \nonumber\\ 
&&{\cal B} \,|phys > = 0 \quad \Longrightarrow \quad  (E \mp m\, \tilde\phi)\, |phys> = 0, \nonumber\\
&&\dot B \, |phys > = 0  \quad \Longrightarrow \quad   -\, \partial_0(\partial \cdot A \pm m\, \phi)\, |phys> = 0 \quad \,    \Longleftrightarrow \nonumber\\ 
&&\dot \,{\cal B} |phys > = 0 \quad \Longrightarrow \quad  \partial_0\, (E \mp m\, \tilde\phi)\, |phys> = 0,  
\end{eqnarray}
where we have used the duality symmetry transformations: 
$ (\partial \cdot A) \longrightarrow \pm i\, E, \; E \longrightarrow \pm i\, (\partial \cdot A), \;      
B \longrightarrow \mp i\, {\cal B}, \; {\cal B} \longrightarrow 
 \mp i\, B, \; \phi \longrightarrow \mp i\, \tilde\phi, \; \tilde\phi \longrightarrow \mp i\, \phi$  
which are a part of the  full set of discrete {\it duality} symmetry transformations that have been  written explicitly  in (88). 
Lastly, we would like to comment that for $\mu = 1$, we obtain from (93), the expression for the time derivative on the auxiliary field 
${\cal B}$ as: $\dot{\cal B} = -\, [\partial_1 \, (\partial \cdot A) \pm m\, (\dot {\tilde\phi} \pm m\, A_1)]$ which is the 
{\it dual} of the secondary first-class constraint: $\partial_1\, E \mp m\, \Pi_{\phi} \approx 0$ that annihilates the physical state due to 
$\dot B \, | phys > = 0$ in Eq. (96). 
It is the observations in (99) that have been responsible to prove that the {\it massless} 2D Abelian 1-form gauge theory
(and the 2D non-Abelian 1-form gauge theory) provide 
(i) a {\it new} class of topological field theories (TFTs)
which capture  a few key aspects of Witten-type TFTs [31] and  some salient features of Schwarz-type TFTs [32], 
and (ii) the perfect field-theoretic  examples of Hodge theory. 
The topological nature of the 2D {\it massless} Abelian 1-form theory 
happens because both the degrees of freedom of the 2D photon (in the Abelian case) can be gauged away due to the presence 
of BRST and co-BRST symmetries (see, e.g. [51, 33] for details).

\subsection{4D Massive Abelian 2-Form Theory: Special Observations}

Against the backdrop of the Subsec. 6.1, we would like to discuss the modification in the {\it standard}   St{$\ddot u$}ckelberg-technique 
of the compensating field(s) in the context of a {\it massive} Abelian 2-form 
theory in a very {\it specific}  four (3 + 1)-dimension of spacetime. 
We begin with the Lagrangian density ${\cal L}^{(B)}_{0}$ [cf. Eq. (27)] and express the standard  St{$\ddot u$}ckelberg-technique (28), valid
in any arbitrary D-dimension of spacetime, in the differential-form language (in terms of the 2-forms) as follows
\begin{eqnarray}
B^{(2)}\longrightarrow  B^{(2)} \mp \frac{1}{m}\,d\, \Phi^{(1)},
\end{eqnarray}
where the 1-form $\Phi^{(1)} = \phi_{\mu}\; d\,x^\mu$ defines the St{$\ddot u$}ckelberg-compensating field $(\phi_\mu)$ which turns out to be 
a Lorentz vector. In the {\it specific} four (3 + 1)-dimensional  (4D) Minkowskian flat spacetime, 
one can generalize (100) by incorporating an axial-vector field  
$\tilde \phi_\mu$ which is defined through the {\it pseudo} 1-form $\tilde\Phi^{(1)} =  \tilde \phi_\mu \; d\,x^\mu$ and using the {\it basic}
 concepts behind the exterior derivative  
and the Hodge duality $*$ operation on the 4D spacetime,  as follows: 
\begin{eqnarray}
B^{(2)}\longrightarrow  B^{(2)} \mp \frac{1}{m}\,d\, \Phi^{(1)} \mp \frac{1}{m}\,*\, d\, \tilde \Phi^{(1)}.
\end{eqnarray} 
We point out that the above St{$\ddot u$}ckelberg-modification has been made such that the 
parity symmetry is respected and maintained in our theory (unlike the theory of weak interactions where the parity 
symmetry is violated).  
In terms of the antisymmetric ($B_{\mu\nu} = -\,B_{\nu\mu}$) tensor field, we can re-express (101) as  follows 
\begin{eqnarray}
B_{\mu\nu} &\longrightarrow & B_{\mu\nu} \mp \frac {1}{m}\,(\partial_\mu\,\phi_\nu - \partial_\nu\,\phi_\mu) 
\mp \frac{1}{m}\, \varepsilon_{\mu \nu \lambda \xi }\,\partial^\lambda\,\tilde \phi^\xi,
\end{eqnarray}
which respects the following set of discrete duality symmetry transformations: 
\begin{eqnarray}
B_{\mu\nu} \longrightarrow  \mp \frac{i}{2!}\,\varepsilon_{\mu \nu \lambda \xi }\,B^{\lambda \xi},\qquad
\phi_{\mu}\longrightarrow  \pm i\,\tilde {\phi_\mu},\qquad \tilde {\phi_\mu}\longrightarrow  \mp i\, \phi_{\mu}.
\end{eqnarray}
This observation is exactly {\it similar} to our discussion in Subsec. 6.1 where we have discussed the ``self-duality"
in the context of the {\it modified} Abelian 1-form (i.e. Proca) theory in 2D  [cf. Eqs. (80), (81].
Using the notation from the  {\it ordinary} 2-form $\Phi^{(2)}  = \frac{1}{2!}\, \Phi_{\mu\nu}\; (dx^\mu \wedge dx^\nu)$ where
 $\Phi_{\mu\nu} = \partial_\mu \phi_\mu - \partial_\nu\phi_\mu$ and taking the symbol 
 $\tilde f_{\mu\nu} = \varepsilon_{\mu \nu \lambda \xi }\,\partial^\lambda\,\tilde \phi^\xi$, the {\it modified}  St{$\ddot u$}ckelberg technique of the
replacement (102) can be re-expressed in {\it two} different ways  as 
\begin{eqnarray}
B_{\mu\nu} &\longrightarrow & B_{\mu\nu} \mp \frac {1}{m}(\,\Phi_{\mu\nu} \mp \tilde f_{\mu\nu}) \equiv B_{\mu\nu} \mp \frac{1}{m}\,\Big(\Phi_{\mu\nu} 
+ \frac{1}{2} \, \varepsilon_{\mu\nu\lambda\xi}\, \tilde\Phi^{\lambda\xi}\Big),
\end{eqnarray} 
where $\tilde \Phi_{\mu \nu } = \partial_\mu  \tilde \phi_\nu - \partial_\nu \tilde \phi_\mu$. 
The {\it latter}  is defined through the {\it pseudo}  
2-form: $\tilde \Phi^{(2)}  = \frac{1}{2!}\,\tilde \Phi_{\mu\nu} \; (d\,x^\mu \wedge d\,x^\nu)$ which 
is nothing but an  antisymmetric {\it pseudo} tensor field of rank-2 and it is the {\it origin} of the kinetic term 
[$+ (1/4)\, \tilde \Phi^{\mu \nu }\, \tilde \Phi_{\mu \nu }$] for the axial-vector field $\tilde\phi_{\mu}$. 
With the input from (104), 
it is clear that we have the following transformations for the {\it mass} 
term {\it and} the field-strength tensor of the starting Lagrangian density ${\cal L}^{(B)}_{0}$ [cf. Eq. (27)]  
\begin{eqnarray}
-\,\frac{m^2}{4}\, B_{\mu\nu}\,B^{\mu\nu} \longrightarrow &-&\,\frac{m^2}{4}\, B_{\mu\nu}\,B^{\mu\nu} \pm \frac{m}{2}\,B_{\mu\nu}\,(\Phi^{\mu\nu} 
+ \frac{1}{2}\varepsilon^{\mu \nu \rho \sigma}\,\tilde\Phi_{\rho \sigma}) \nonumber\\
&-& \,\frac{1}{4}\,\Phi_{\mu\nu}\,\Phi^{\mu\nu} + \frac{1}{4}\tilde \Phi_{\mu\nu}\,\tilde \Phi^{\mu\nu},
\end{eqnarray}
\begin{eqnarray}
H_{\mu\nu\lambda}  \longrightarrow  H_{\mu\nu\lambda}  & \mp & \frac{1}{m}\,(\partial_\mu\,\Phi_{\nu\lambda} + \partial_\nu\,\Phi_{\lambda\mu}
+\partial_\lambda\,\Phi_{\mu\nu}) \nonumber\\
& \mp & \frac{1}{m}\,(\partial_\mu\,\tilde f_{\nu\lambda} + \partial_\nu\,\tilde f_{\lambda\mu}
+\partial_\lambda\, \tilde f_{\mu\nu}),~~~~~~~~~~~
\end{eqnarray}
where we note that the {\it second} term of (106) is equal to {\it zero} due to
: $\Phi_{\mu\nu} = \partial_\mu\phi_\nu - \partial_\nu\phi_\mu.$ We also lay emphasis on the observation  that there is a 
relative sign difference  between the kinetic terms of field $\phi_\mu$ and $\tilde\phi_\mu$, respectively, in (105). 
We shall come back to this point later in our Sec. 7 when we shall make some concluding remarks.

With the inputs from the equations (105) and (106), we note that the standard  kinetic term 
of the Lagrangian density ${\cal L}_{0}^{(B)}$ [cf. Eq. (27)] transforms as follows
\begin{eqnarray}
\frac{1}{12}\,H_{\mu\nu\lambda}\,H^{\mu\nu\lambda} \; \longrightarrow  \;  \frac{1}{12}\,H_{\mu\nu\lambda}\,H^{\mu\nu\lambda} \mp \frac{1}{6\,m}\,
H_{\mu\nu\lambda}\,\Sigma^{\mu\nu\lambda} + \frac{1}{12\,m^2}\,\Sigma_{\mu\nu\lambda}\,\Sigma^{\mu\nu\lambda},
\end{eqnarray} 
where we have introduced a new notation: $\Sigma_{\mu\nu\lambda}   = \partial_\mu\,\tilde f_{\nu\lambda} + \partial_\nu\,\tilde f_{\lambda\mu}
+\partial_\lambda\, \tilde f_{\mu\nu} $ which makes the transformation (106) as: 
$H_{\mu\nu\lambda} \longrightarrow H_{\mu\nu\lambda} \mp (1/m)\, \Sigma _{\mu\nu\lambda}$. 
The {\it latter} has been used in the  derivation of the transformation (107).
With the help of (105) and (107),  we note that the starting Lagrangian density ${\cal L}_0^{(B)}$ [cf. Eq. (27)]
for the massive Abelian 2-form theory transforms to the following St{$\ddot u$}ckelberg-modified massive Abelian 2-form theory
\begin{eqnarray}
{\cal L}_0^{(B)}   \longrightarrow  {\cal L}_S^{(m)}   = 
\frac{1}{12}\,H_{\mu\nu\lambda}\,H^{\mu\nu\lambda} \mp \frac{1}{6\,m}\,
H_{\mu\nu\lambda}\,\Sigma^{\mu\nu\lambda} + \frac{1}{12\,m^2}\,\Sigma_{\mu\nu\lambda}\,\Sigma^{\mu\nu\lambda}
&-& \frac{m^2}{4} B_{\mu\nu}\,B^{\mu\nu} \nonumber\\
\pm \frac{m}{2}\,B_{\mu\nu}\,[\Phi^{\mu\nu} 
+ \frac{1}{2}\varepsilon^{\mu \nu \rho \sigma}\,\tilde\Phi_{\rho \sigma}] 
- \,\frac{1}{4}\,\Phi_{\mu\nu}\,\Phi^{\mu\nu} + \frac{1}{4}\tilde \Phi_{\mu\nu}\,\tilde \Phi^{\mu\nu}, 
\end{eqnarray} 
where there are higher derivative terms in the {\it second} and {\it third} terms 
of the 4D Lagrangian density ${\cal L}_S^{(m)} $ and the superscript $(m)$ denotes that we have obtained (108) with the {\it modified} version of the 
St$\ddot u$ckelberg-technique of the compensating fields (102) in 4D.

Our present discussion will center 
around the Lagrangian density (108) where {\it all} its terms will be well-defined, consistent and renormalizable
in the specific {\it four} (3 + 1)-dimensions of Minkowskian flat spacetime using the on-shell condition for the 
{\it basic} field $\tilde\phi_\mu$. This systematic  exercise has  been performed in a very recent 
paper by us (see e.g. [50] for details). Thus, we skip many theoretical and mathematical 
steps because it will be only an academic exercise if we repeat those steps here. Finally, we state that we    
 have derived the coupled  (but equivalent)
(anti-)BRST and (anti-)co-BRST invariant Lagrangian densities in the four (3 + 1)-dimensions of 
spacetime where the generalizations of (i) the discrete duality symmetry  transformation (103), and (ii) the
continuous (dual-)gauge symmetry transformations have been taken into account (see e.g. [50, 30] for details). 
However, the numerical factors associated with the gauge-fixing terms and Faddeev-Popove ghost terms 
are quite different from our present discussion of the D-dimensional {\it modified} massive Abelian 2-form theory (cf. Subsec. 5.2 for details). 
To be consistent with 
notations adopted in Subsec. 5.2, we concentrate on the following (co-)BRST invariant Lagrangian density (${\cal L}_{\cal B}$) 
with the appropriate generalization of (108) that  incorporates into it the proper gauge-fixing and 
FP-ghost terms, namely; 
\begin{eqnarray*}
{\cal L}_{\cal B} &=&  {\cal B}_\mu\,{\cal B}^\mu 
+ {\cal B}^\mu \,\Big[\frac{1}{2}\,\varepsilon_{\mu \nu \lambda \xi }\,\partial^\nu\,B^{\lambda\xi} + m\,\tilde\phi_\mu 
- \partial_\mu \tilde\phi\Big] -\,\frac{m^2}{4}\, B_{\mu\nu}\,B^{\mu\nu}
\end{eqnarray*}
\begin{eqnarray}
&-& \frac{1}{4}\,\Phi_{\mu\nu}\,\Phi^{\mu\nu} + \frac{1}{4}\tilde \Phi_{\mu\nu}\,\tilde \Phi^{\mu\nu}
+ \frac{m}{2}\, B_{\mu\nu}\,\Big[\Phi^{\mu\nu}+ \frac{1}{2} \varepsilon^{\mu\nu\rho\sigma}\,\tilde\Phi_{\rho\sigma}\Big]\nonumber\\
&+& B^\mu\, \Big[ \partial^\nu\,B_{\nu\mu}  + m\, \phi_\mu - \partial_\mu\,\phi \Big]\,
+ B^\mu\,B_\mu - B\,\Big(\partial \cdot \phi+ m\,\phi \Big) \nonumber\\
&- & B^2 +  {\cal B}\, \Big(\partial \cdot \tilde\phi+ m\,\tilde\phi\Big) +  {\cal B}^2
+ \partial_\mu\,\bar\beta\, \partial^\mu\,\beta - \, m^2\, \bar\beta\,\beta  \nonumber\\ 
&+& (\partial_\mu\,\bar C_\nu - \partial_\nu\,\bar C_\mu)\,(\partial^\mu\,C^\nu)
- (\partial_\mu\,\bar C - m\, \bar C_\mu)\,(\partial^\mu\, C - m\, C^\mu) \nonumber\\
&+&  \Big(\partial \cdot \bar C + m\, \bar C + \rho\Big)\,\lambda +  \Big(\partial \cdot C + m\, C - \lambda\Big)\,\rho,
\end{eqnarray}
where {\it bosonic} (anti-)ghost fields $(\bar\beta)\beta$ carry the ghost numbers (-2)+2, respectively, and 
the {\it fermionic} (anti-)ghost fields $(\bar C_\mu)C_\mu$ are endowed with the ghost numbers (-1)+1, respectively. 
We have also linearized the kinetic term and gauge-fixing term for $B_{\mu\nu}$ and an axial-vector field 
$\tilde\phi_\mu$ by invoking Nakanishi-Lautrup type auxiliary fields $({\cal B}_\mu, {\cal B})$. 
It will be noted that we  also have a pseudo-scalar field {$\tilde\phi$} in our theory\footnote{A close look at (109) 
shows that it is the generalization of the {\it perfectly} BRST invariant Lagrangian 
density (${\cal L}_b$) that has been quoted in (59) of our Subsec. 5.2. For the sake of  simplicity, we have not bothered about the 
numerical factors that are present in various terms of the Lagrangian density that has been taken into account in 
our earlier works [30, 50] where the discrete and continuous symmetries are precisely   respected at every stage.
For our present Lagrangian density (109), we have considered only the infinitesimal, continuous and off-shell nilpotent 
(co-)BRST  symmetries $(s_d)s_b$.} (see, e.g. [50, 40, 30] for details).  
Our (co-)BRST invariant Lagrangian density (109) also contains additional (anti-) ghost fields $(\bar C)C$ which are fermionic 
$(C^2 = \bar C^2 = 0, C\,\bar C + \bar C\, C = 0)$ in nature with the ghost numbers (-1)+1, respectively. Finally, the FP ghost 
terms contain a set of  {\it auxiliary} (anti-)ghost fields $(\rho)\,\lambda$ with ghost numbers (-1)+1, respectively. The Lagrangian density
(109) respects the following (co-)BRST symmetry transformations $(s_d)s_b$
\begin{eqnarray}
&&s_d\, B_{\mu\nu} = -\, \varepsilon_{\mu \nu \lambda \xi }\,\partial^\lambda\,\bar C^\xi, \quad
s_d\,\bar C_\mu = -\, \partial_\mu\, \bar\beta, \quad s_d\,C_\mu = {\cal B}_\mu, \nonumber\\
&& s_d\,\beta = -\, \lambda,\quad s_d\,\tilde\phi_\mu = + (\partial_\mu\, \bar C - m\, \bar C_\mu),\quad
s_d\,C = {\cal B}, \quad s_d\, \bar C = - m\, \bar \beta, \nonumber\\
&& s_d\,\tilde\phi = - \rho, \quad s_d\,[\partial^\nu\,B_{\nu\mu},\, {\cal B}_\mu,\,B_\mu,\, {\cal B},\, 
\phi_\mu,\, \Phi_{\mu\nu}, \, \phi, \, \bar\beta,\, \lambda,\, \rho] = 0, \nonumber\\
&&s_b\, B_{\mu\nu} = -\, (\partial_\mu\, C_\nu - \partial_\nu\, C_\mu), \quad
s_b\, C_\mu = -\, \partial_\mu\, \beta,  \quad s_b \bar C_\mu = -\, B_\mu, \nonumber\\
&& s_b\, \bar\beta = -\, \rho, \quad s_b\, \phi_\mu = + (\partial_\mu C - m\, C_\mu), \quad
s_b\, \bar C =  B, \quad s_b\, \phi = + \lambda, \nonumber\\
&& s_b\, C = -\, m\, \beta, \quad
s_b\,[H_{\mu\nu\lambda},\, B,\,\lambda,\, \rho,\, B_{\mu},\, {\cal B}_\mu, \, \beta,\, 
{\cal B},\, \tilde\phi_\mu,\, \tilde\phi_{\mu\nu}, \, \tilde\phi] = 0,
\end{eqnarray}
which are characterized by the observations that (i) the field strength tensor $(H_{\mu\nu\lambda})$,  owing its origin to
the exterior derivative,  remains invariant under $s_b$, and (ii) the gauge-fixing term $(\partial^\nu B_{\nu\mu})$, tracing 
back its origin to the co-exterior derivative of differential geometry [45-49], remains invariant under $s_d$. The above (co-)BRST transformations are the 
{\it symmetry} transformations for the action integral $S = \int d^4 x\, {\cal L}_{\cal B}$ because we observe that ${\cal L}_{\cal B}$ transform 
under $(s_d)s_b$ to the total spacetime derivatives as:  
\begin{eqnarray}
s_d\,{\cal L}_{\cal B}&=& \partial_\mu\, \Big[ - m\, \varepsilon^{\mu \nu \lambda \xi }\,\phi_\nu\,\partial_\lambda\, \bar C_\xi
- \, (\partial^\mu\, \bar C^\nu - \partial^\nu\, \bar C^\mu)\,{\cal B}_\nu   + {\cal B}^\mu \, \rho \nonumber\\
&+& (\partial^\mu \, \bar C - m\, \bar C^\mu) \,{\cal B} -  \lambda\,(\partial^\mu\, \bar\beta)  \Big], \nonumber\\
s_b\, {\cal L}_{\cal B} &=& -\, \partial_\mu\, \Big[m\, \varepsilon^{\mu \nu \lambda \xi }\,\tilde\phi_\nu\,\partial_\lambda\, C_\xi
+ (\partial^\mu\, C^\nu - \partial^\nu\, C^\mu)\,B_\nu  +  \lambda \, B^\mu \nonumber\\
&+& (\partial^\mu \, C - m\, C^\mu) \, B +  \rho \,(\partial^\mu\, \beta) \Big].
\end{eqnarray}
According to the basic principles  behind Noether's theorem,  the existence of the continuous, infinitesimal and off-shell nilpotent 
$(s_d^2 = 0, \, s_b^2 = 0)$ (co-)BRST symmetry transformations $(s_d)s_b$ lead to the derivations of the conserved (co-)BRST currents 
and corresponding conserved charges. It has been shown, however, in our earlier work (see, e.g. [40]) that the conserved Noether BRST charge is 
{\it not} off-shell nilpotent. A systematic method has been developed to obtain the off-shell nilpotent BRST  charge [$Q_b^{(2)}$] from the non-nilpotent 
$(Q_b^2 \neq 0)$ Noether BRST charge $(Q_b)$. The conserved and off-shell nilpotent version 
of the BRST charge has been taken into consideration [cf. Eq. (66)] in our Subsec. 5.2 and a detailed analysis of the constraints from the physicality 
criterion  (w.r.t. the conserved and off-shell nilpotent BRST charge) has been performed [cf. Eq. (73)] where we observe that the operator forms of the 
first-class constraints of the {\it original} St$\ddot u$ckelberg-modified massive Abelian 2-form theory
(described by the Lagrangian density ${\cal L}_S^{(B)}$ [cf. Eq. (29)]) annihilate [cf. Eq. (68)] the physical states 
(in the {\it  total} Hilbert space of quantum states).

To obtain the off-shell nilpotent and conserved co-BRST charge $(Q_d^{(2)})$, first of all, 
we derive the standard Noether conserved current and corresponding conserved charge ($Q_d$). 
We wish to perform this exercise in explicit fashion because the co-BRST and anti-co-BRST charges 
have {\it not} been discussed in our earlier work [40] where we have developed a systematic theoretical method  to obtain the 
nilpotent (anti-)BRST charges from the non-nilpotent {\it standard} Noether (anti-)BRST charges. 
In this context, first of all, we note that the Noether  conserved co-BRST current ($J_{(d)}^{\mu}$) is 
\begin{eqnarray}  
J^\mu _{d}& = & (s_d B_{\alpha\beta}) \, \frac {\partial {\cal L}_{\cal B}}{\partial(\partial_\mu B_{\alpha\beta})} + 
(s_d \tilde\phi_\alpha) \, \frac {\partial {\cal L}_{\cal B}}{\partial(\partial_\mu \tilde\phi_\alpha)} +  
(s_d C_\alpha) \, \frac {\partial {\cal L}_{\cal B}}{\partial(\partial_\mu C_\alpha)} +  
(s_d \bar C_\alpha) \, \frac {\partial {\cal L}_{\cal B}}{\partial(\partial_\mu \bar C_\alpha)} \nonumber\\
& + & 
(s_d \tilde\phi) \, \frac {\partial {\cal L}_{\cal B}}{\partial(\partial_\mu \tilde\phi)} + 
(s_d \bar C) \, \frac {\partial {\cal L}_{\cal B}}{\partial(\partial_\mu \bar C)}  +
(s_d C) \, \frac {\partial {\cal L}_{\cal B}}{\partial(\partial_\mu C)} + 
(s_d \beta ) \, \frac {\partial {\cal L}_{\cal B}}{\partial(\partial_\mu \beta)}\nonumber\\
& + & 
(\partial^\mu \bar C^\nu - \partial^\nu \bar C^\mu)\, {\cal B}_\nu + \lambda \, \partial^\mu \bar\beta + 
m\, \varepsilon^{\mu \nu \lambda \xi }\, \phi_\nu\, \partial_\lambda\bar C_{\xi} - \rho \, {\cal B}^\mu - (\partial^\mu \bar C - m\, \bar C^\mu)\, {\cal B}, 
\end{eqnarray} 
which, ultimately, turns out to be 
\begin{eqnarray} 
J^\mu _{d}& = & \varepsilon^{\mu \nu \lambda \xi }\, \Big[(m\, \phi_\nu - B_\nu)\, (\partial_\lambda \bar C_\xi) 
+ \frac {m}{2}\, (\partial_\nu \bar C - m\, \bar C_\nu)\, B_{\lambda\xi}\Big] + m\, (\partial^\mu C - m\,C^\mu)\, \bar \beta \nonumber\\
&-& (\partial^\mu C^\nu -\partial^\nu C^\mu)\, (\partial_\nu\bar\beta) + \tilde\Phi^{\mu\nu}\, (\partial_\nu \bar C - m\, \bar C_\nu) + (\partial^\mu \bar C - m\, \bar C^\mu)\, {\cal B}  \nonumber\\
&  - &   (\partial^\mu \bar C^\nu - \partial^\nu \bar C^\mu)\, {\cal B}_\nu + \rho\, {\cal B}^\mu - \lambda \,\partial^\mu \bar\beta. 
\end{eqnarray} 
The conservation law $(\partial_\mu J^{\mu}_{(d)} = 0)$ can be proven by using the 
following EL-EoMs w.r.t. the fields: $B_{\mu\nu}, \tilde\phi_\mu, C_\mu, \bar C_\mu, C, \bar C, \beta, \tilde\phi$ 
from the  Lagrangian density (109), namely; 
\begin{eqnarray} 
&&\varepsilon^{\mu \nu \lambda \xi }\, \partial_\lambda {\cal B}_\xi - m^2\, B^{\mu\nu}  + m\, (\Phi^{\mu\nu} + \frac {1}{2}\, \varepsilon^{\mu \nu \rho\sigma}\, \tilde\Phi_{\rho\sigma}) - (\partial^\mu B^\nu - \partial^\nu B^\mu) = 0, \nonumber\\
&& \partial_\mu \big(\tilde\Phi^{\mu\nu} - \frac {m}{2}\, \varepsilon^{\mu \nu \lambda \xi }\, B_{\lambda\xi}\big) = m\, {\cal B}^\nu - \partial^\nu {\cal B}, \nonumber\\ 
&& \partial_\mu (\partial^\mu C^\nu - \partial^\nu C^\mu) + \partial^\nu \lambda
= m\, (\partial^\nu C - m\, C^\nu), \nonumber\\
&& \partial_\mu (\partial^\mu \bar C^\nu - \partial^\nu \bar C^\mu) - \partial^\nu \rho  = m\, (\partial^\nu \bar C - m\, \bar C^\nu),\nonumber\\
&& \partial_\mu\, (\partial^\mu\, \bar C - m\, \bar C^\mu) = m\, \rho, \quad 
 \partial_\mu\, (\partial^\mu\, C  - m\,  C^\mu) = -\,m\, \lambda, \nonumber\\
&& (\Box + m^2)\, \bar\beta = 0, \quad \partial\cdot {\cal B} = -\, m\, {\cal B}.
\end{eqnarray} 
Some of the additional EoMs that emerge out from the above are:
\begin{eqnarray} 
&& \Box\cdot {\cal B} - m\, (\partial\cdot {\cal B}) = 0, \qquad \qquad (\Box  
+ m^2)\, {\cal B} = 0, \nonumber\\ 
&&\varepsilon^{\mu \nu \rho\sigma}\, \partial_\rho
\big(m\,\phi_\sigma - B_\sigma\big)  
 - (\partial^\mu {\cal B}^\nu - \partial^\nu {\cal B}^\mu) - m\, (\tilde\Phi^{\mu\nu} + \frac {m}{2}\, \varepsilon^{\mu \nu \rho\sigma} B_{\rho\sigma}) = 0.
\end{eqnarray} 
The {\it last} entry, in the above equation, is nothing but the {\it first} entry of 
(114) which is very useful in the proof of the conservation law 
$(\partial_\mu\, J^\mu_{d} = 0)$.

From the conserved current (113), one can compute the expression for the conserved 
co-BRST charge $(Q_d = \int d^3x\ J^0_d )$ as follows: 
\begin{eqnarray} 
Q_d &=&  \int d^3 x \Big[ m\, (\partial^0\, C - m\, C^0)\, \bar\beta
 - (\partial^0\, C^i - \partial^i \, C^0)\, \partial_i\, \bar\beta  + \Phi^{0i}\, (\partial_i\, \bar C - m\, \bar C_1) \nonumber\\
 &+&  (\partial^0\, \bar C - m\, \bar C^0)\, {\cal B} - (\partial^0\, \bar C^i - \partial^i\, \bar C^0 )\, {\cal B}_i 
 + \rho\, {\cal B}^0 - \lambda \, \partial^0\, \bar\beta \nonumber\\
& + & \varepsilon^{0ijk} \big\{(m\, \phi_i - B_i)\, \partial_j\, \bar C_k  + \frac{m}{2}\, (\partial_i\, \bar C - m\, \bar C_i)\, B_{jk}\big\}  \Big]. 
\end{eqnarray} 
Using the basic principle behind the continuous symmetries and their generators as the conserved Noether charges, it is straightforward to check that the following is true, namely; 
\begin{eqnarray}
s_d \, Q_d = -\, i\, \{Q_d, \, Q_d\} \neq 0 \quad \Longrightarrow  \quad Q_d^2 \neq 0, 
\end{eqnarray} 
where the l.h.s can be computed from the direct application of the co-BRST symmetry transformations ($s_d$)
on the conserved charge ($Q_d$) which lead to the following: 
\begin{eqnarray} 
s_d\, Q_d = \int d^3 x \Big [ m \, (\partial^0\, {\cal B} - m\, {\cal B}^0)\, \bar\beta 
- (\partial^0\, {\cal B}^i - \partial^i\, {\cal B}^0)\, \partial_i\, \bar\beta   \Big ] \neq 0.  
\end{eqnarray} 
Hence, we conclude that the standard Noether conserved co-BRST charge $(Q_d)$ is {\it not} 
off-shell nilpotent $(Q_d^2 \neq 0)$ of order two. To obtain the off-shell nilpotent (i.e. $[Q_d^{(2)}]^2 = 0)$ version of the co-BRST charge 
$(Q_d^{(2)} = 0)$, we follow our proposal (see, e.g.  [40] for details)
where the elegant combinations of (i) the Gauss divergence theorem, (ii) the suitable equations  of motion,  
and (iii) the off-shell nilpotent co-BRST symmetry transformations, will play the decisive roles. In this context, first of all,
we focus on the following {\it two} terms (which are present in the expression for $Q_d$) and apply the Gauss divergence theorem
\begin{eqnarray} 
\int d^3 x\,  \big(\tilde\Phi^{0i} + \frac {m}{2}\, \varepsilon^{0ijk} B_{jk} \big)\, \partial_i\, \bar C 
\equiv \int d^3 x\,   \partial_i\, \Big[ \big(\tilde\Phi^{0i} + \frac {m}{2}\, \varepsilon^{0ijk} B_{jk} \big)\,\bar C \Big] \nonumber\\
-   \int d^3 x\,  \partial_i\, \big(\tilde\Phi^{0i} + \frac {m}{2}\, \varepsilon^{0ijk} B_{jk} \big)\,  \bar C, 
\end{eqnarray} 
\begin{eqnarray} 
\int d^3 x\,  \varepsilon^{0ijk} \big(m\, \phi_i - B_i \big)\, \partial_j\, \bar C_k 
\equiv   \int d^3 x\, \partial_j\, \big[  \varepsilon^{0ijk} \big(m\, \phi_i - B_i \big)\,  \bar C_k   \big] \nonumber\\
-   \int d^3 x\, \varepsilon^{0ijk} \partial_j \, \big(m\, \phi_i - B_i \big)\,  \bar C_k,
\end{eqnarray} 
where the total space  derivative terms are dropped for the physically well-defined 
fields that vanish off as $x \rightarrow \pm \infty$.
A close look at (119) shows that its l.h.s. has been taken as the sum of appropriate terms from the last entries of the 
{\it first}  and {\it third} lines of $Q_d$ [cf. Eq. (116)].
 At this stage, we apply the following EL-EoMs 
\begin{eqnarray} 
&&\partial_i\,  \big(\tilde\Phi^{0i} + \frac {m}{2}\, \varepsilon^{0ijk} B_{jk} \big)\, \bar C 
=   (\partial^0 {\cal B} - m\,  {\cal B}^0)\, \bar C \equiv - \, (m\, {\cal B}^0 -\partial^0 {\cal B})\, \bar C,\nonumber\\
 &&\varepsilon^{0ijk} \partial_i \, \big(m\, \phi_j - B_j \big)\,  \bar C_k  =   (\partial^0 {\cal B}^i - \partial^i {\cal B}^0)\, \bar C_i + m\, (\tilde\Phi^{0i} + \frac {m}{2}\, \varepsilon^{0ijk} B_{jk})\, \bar C_i, 
\end{eqnarray} 
which are derived from the appropriate equations of (114) and (115), respectively. 
With the above inputs, the equations (119) and (120) can be re-written 
as follows (modulo total space derivative terms), namely;  
\begin{eqnarray} 
 -\, \int d^3 x\, \partial_i\,  \big(\tilde\Phi^{0i} + \frac {m}{2}\, \varepsilon^{0ijk} B_{jk} \big)\, \bar C 
& \equiv & \int d^3 x\,(m\,  {\cal B}^0 - \partial^0 {\cal B})\, \bar C, \nonumber\\
\int d^3 x\, \varepsilon^{0ijk} \partial_i \, \big(m\, \phi_j - B_j \big)\,  \bar C_k & \equiv &  \int d^3 x\, \Big[(\partial^0 {\cal B}^i - \partial^i {\cal B}^0)\, \bar C_i \nonumber\\
& + &  m\, (\tilde\Phi^{0i} + \frac {m}{2}\, \varepsilon^{0ijk} B_{jk})\, \bar C_i\Big].
\end{eqnarray}
At this stage, as per the theoretical method proposed in our earlier work [40], we carefully check whether there is any cancellation and/or addition between the r.h.s. of  (122) and  the {\it left-over} terms of $Q_d$. It turns out that the {\it second} term in the bottom-line  on the r.h.s. of (121) cancels with the remaining parts of the {\it last} terms of (i) the top-line, and
(ii) the {\it third} line of (116). Thus, the existing terms (that will be a part in the expression 
for the nilpotent version of the co-BRST charge) are written {\it together}  as follows 
\begin{eqnarray}
\int d^3\, x\, \Big [(\partial^0 {\cal B}^i - \partial^i {\cal B}^0)\, \bar C_i + (m\,  {\cal B}^0 - \partial^0 {\cal B})\, \bar C\Big]
\end{eqnarray}
If we apply the co-BRST symmetry transformations $(s_d)$ on  (123), we obtain
\begin{eqnarray}
\int d^3\, x\, \Big [-\, (\partial^0 {\cal B}^i - \partial^i {\cal B}^0)\, \partial_i\, \bar\beta -\, m \,(m\,  {\cal B}^0 - \partial^0 {\cal B})\, \bar \beta \Big].
\end{eqnarray}
which has to cancel out from the modified versions of the appropriate terms of the expression for $Q_d$.
In this context, we observe that we can modify two terms [i.e. $m\, (\partial^0\, C - m\, C^0)\, \bar\beta - (\partial^0\, C^i 
- \partial^i\, C^0)\, \partial_i  \bar \beta $] of $Q_d$ [cf. Eq. (116)] and utilize the interplay of (i) the Gauss divergence theorem, (ii) the
appropriate EL-EoMs, and (iii) application of the co-BRST symmetry transformations at appropriate places so that 
the resulting expression cancels (124). Using these type of tricks (see, e.g. [40] for details), we obtain the {\it final} 
expression for the off-shell nilpotent version of the conserved co-BRST charge as: 
\begin{eqnarray}
Q_d ^{(2)}  & = & \int d^3 x\, \Big [(\partial^0 {\cal B}^i - \partial^i {\cal B}^0)\, \bar C_i 
 - (\partial^0 {\cal B} -\, m\,{\cal B}^0)\,\bar C+ 
(\partial^0 C^i - \partial^i C^0)\, \partial_i \bar\beta  \nonumber\\
  & - &  m\,(\partial^0 C - m\, C^0)\, \bar\beta
  + 2\,\bar\beta\,\partial^0 \lambda - \lambda \, \partial^0 \bar\beta + \rho\, {\cal B}^0 -
  (\partial^0 \bar C - m\, \bar C^0)\, {\cal B}\nonumber\\
  & -& (\partial^0 \bar C^i - \partial^i \bar C^0)\, {\cal B}_i\Big].
\end{eqnarray}
It is straightforward to check that    
\begin{eqnarray}
s_d \,Q_d^{(2)}  = -\,i\, \{Q_d^{(2)}, Q_d^{(2)}\} = 0\quad \Longrightarrow [Q_d^{(2)}]^2 = 0,
\end{eqnarray}
where the l.h.s. of the above equation can  be explicitly computed by applying {\it directly} 
$s_d$ [cf. Eq. (110)] on the final expression for $Q_d^{(2)}$ [cf. Eq. (125)]. Thus, we have obtained the off-shell nilpotent version of 
the co-BRST charge $Q_d^{(2)}$ from the  
standard conserved (but non-nilpotent) Noether charge $Q_d$ [cf. Eq. (116].

The presence of the (co-)BRST symmetries allows us to perform the Hodge decomposition theorem (see, e.g. [57]) in the total Hilbert space of states where the 
most symmetric state turns out to be the {\it harmonic} state which can be chosen to be the physical state (i.e. $| phys > $). 
The {\it latter} is annihilated by the off-shell nilpotent and conserved versions of the
BRST and co-BRST charges {\it together} in an independent manner. 
The physicality criterion with the nilpotent co-BRST charge leads to: 
\begin{eqnarray} 
 Q_d^{(2)}\, | phys > = 0  \qquad \Longrightarrow  \qquad {\cal B} \,| phys > &=& 0,   \nonumber\\
 {\cal B}_i \,| phys > &=& 0, \nonumber\\
 (\dot {\cal B} - m \, {\cal B}^0)\, |\, phys > &=& 0,  \nonumber\\
 (\partial^0\, {\cal B}^i - \partial^i\, {\cal B}^0)\, |\, phys > &=& 0.  
\end{eqnarray} 
A noteworthy point, at this stage, is the observation that we have {\it not} taken into consideration  any 
physical condition on the physical state ($| phys > $) due to the presence of the term: $ -\, \rho\, {\cal B}^0$. 
This is due to the fact that the auxiliary anti-ghost field $\rho$ is {\it not}
a {\it basic} field of our theory [i.e.  $\rho = -(1/2)\, (\partial\cdot \bar C + m\, \bar C)$]. 
Thus, we do {\it not} have {\it separately} the condition: ${\cal B}^0 \,| phys > = 0$
from the physicality criterion w.r.t. the off-shell nilpotent and conserved co-BRST charge . 
Moreover, it is clear, from the non-ghost sector of the (co-)BRST invariant Lagrangian density (109),  that we 
have: 
\begin{eqnarray} 
\Pi_{(\tilde\phi)} = \frac{\partial\, {\cal L}_{\cal B}}{\partial\, (\partial_0\, \tilde\phi)} = \frac{1}{2}\, {\cal B}^0, 
\end{eqnarray} 
which is {\it not} a constraint (of any variety) on our theory. Hence, the physical constraint condition: 
${\cal B}^0 \, | phys >  = 0$ is {\it not} allowed and it is {\it not} even required from the 
basic principles that are connected with the quantization of theories  with constraints.

A close and careful look at the conditions (73) and (127) on the physical states (i.e. $| phys > $)
show that the four (3 + 1)-dimensional (4D) {\it modified} massive Abelian 2-form theory is very special  in the sense that the operator
forms of the first-class constraints as well as their dual versions (in the operator form) must annihilate the physical states. 
It is due to these {\it special} properties of the 4D theory that we have been able to establish that the {\it properly} St$\ddot u$ckelberg-modified 
4D Abelian 2-form theory is (i) a field-theoretic example of Hodge theory (see, e.g. [28, 30] for details)
where the symmetries and conserved charges provide the physical realizations of the de Rham cohomological operators of the 
differential geometry [45-49], and (ii) a quasi-topological field-theoretic (quasi-TFT) example 
of the formal quantum field theory where the proper topological invariants 
and their proper recursion  relations have been shown to exist (see, e.g. [34] for details).

\section{Conclusions}

In our present concise review, we have focused on the D-dimensional St$\ddot u$ckelberg-modified {\it massive} and 
{\it massless} Abelian 1-form gauge theories (without any interaction with matter fields) and shown that there is a 
deep connection between the standard generators (that are expressed in terms of the first-class constraints) and the 
conserved Noether charges  (that are derived by using the infinitesimal, local and 
continuous gauge symmetry transformations) at the {\it classical} level. This exercise has been repeated for the D-dimensional  
St$\ddot u$ckelberg-modified {\it massive} and {\it massless} Abelian 2-form gauge theories at the {\it classical} level. 
There is precise similarity between both {\it these} theories at the {\it classical} level. 

When we discuss the above two types of 1-form and 2-form theories (of the {\it modified} massive and massless varieties) 
at the {\it quantum} level, within the framework of BRST formalism, we find decisive differences. These differences 
owe their {\it basic} origin to the existence of the {\it trivial} and {\it non-trivial} CF-type of restriction(s) 
in the cases of the 1-form and 2-form modified  massive and  massless gauge theories, respectively. In the case of the 1-form 
theories, we find that the conserved (anti-)BRST charges 
(derived by using the standard Noether theorem) are off-shell nilpotent and conserved.
As a consequence, when we demand that the physical states (in the {\it total} quantum Hilbert space of states) are {\it those}
that are annihilated by the conserved and off-shell nilpotent (anti-)BRST charges, we observe that the operator forms 
of the first-class constraints (of the {\it classical} gauge theories) annihilate the physical states. This observation
is consistent  with the Dirac-quantization conditions for the systems that are endowed with any kinds of constraints [5-8]. This is {\it not}
the situation with the St$\ddot u$ckelberg-modified {\it massive}  and 
{\it massless} Abelian 2-form theories where the standard Noether theorem leads to the derivations of the
conserved (anti-)BRST charges  which are {\it not} found [40] to be off-shell nilpotent. Hence, they do {\it not} lead to the 
appearance of the first-class constraints 
(in their operator form) through the physicality criteria.

Against the backdrop of the above paragraphs, we focus on the 2-form theories. 
The {\it massless} Abelian 2-form theory is endowed with the  Curci-Ferrari (CF) type 
restriction (i.e. $B_\mu - \bar B_\mu -\partial_\mu \varphi = 0$)
at the {\it quantum} level when it is discussed within the framework of BRST formalism (see, Subsec. 5.1). As a consequence, when we apply the 
celebrated Noether theorem, we obtain the  conserved (anti-)BRST charges. However, these conserved Noether   
(anti-)BRST charges are found to be non-nilpotent [i.e. $Q_{(a)b}^2 \ne 0$] due to the presence of the  non-trivial CF-type 
restriction (i.e. $B_\mu - \bar B_\mu - \partial_\mu\, \phi = 0$). We have developed a systematic method to
obtain the conserved and off-shell nilpotent versions of the (anti-)BRST charges in our earlier work [40]. With 
{\it these} off-shell nilpotent versions of the (anti-)BRST charges, we obtain the presence of the operator forms of the 
first-class constraints (cf. Subsec. 5.1).

In the case of the St$\ddot u$ckelberg-modified massive Abelian 2-form theory, we have two {\it non-trivial} CF-type 
restrictions (i.e. $B_\mu - \bar B_\mu - \partial_\mu\phi= 0, \; B + \bar B + m\, \phi = 0)$ within the 
framework of BRST formalism (cf. Subsec. 5.2).  Hence, once again, the Noether theorem does not lead to
the derivation of conserved and off-shell nilpotent (anti-)BRST charges. We have obtained the off-shell 
nilpotent version [cf. Eq. (66)] of the conserved BRST charge [$Q_b^{(2)}$] in our earlier work [40]
which has been taken into account in our Subsec. 5.2 and we have shown that the physicality criterion with the nilpotent BRST charge 
(i.e. $Q_b^{(2)}\, | phys > = 0$) leads to the appearance of the operator forms of the first-class constraints (cf. Subsec. 5.2)
where they annihilate the physical states of the theory under consideration. 

Finally, we have pointed out the {\it modifications} in the St$\ddot u$ckelberg-technique in  
the cases of (i) the 2D massive Abelian 1-form 
(Proca) theory, and (ii) the  4D massive  Abelian 2-form theory, where the off-shell nilpotent and 
conserved (anti-)BRST and (anti-)co-BRST charges lead to the annihilation of the physical states by 
(i) the operator forms of the first-class constraints, and (ii) the {\it dual} versions of the 
first-class constraints. This is why  these $p$-form ($p = 1, 2$) {\it modified}  massive theories in $D = 2p$
(i.e. $D  = 2, 4$) dimensions of spacetime have some {\it special} features.
For  instance, the 2D {\it modified} Proca theory and {\it massless} Abelian 1-form gauge theory 
turn out be a set of prefect field-theoretic models of Hodge theory 
and the 2D  {\it massless} Abelian 1-form theory belongs to a {\it  new} class of topological field theory 
(see, e.g. [33] for details). On the other hand, the 4D {\it  massless} and modified {\it massive} 
Abelian 2-form theories are found to be the  perfect field-theoretic examples of Hodge theory 
(see, e.g. [27-30] for details). In addition, the 4D {\it massless} Abelian 2-form theory has {\it also}
been shown to be  a field-theoretic model of the quasi-topological field theory (quasi-TFT) with proper topological invariants and 
their precise recursion relations (see, e.g. [34] and references therein). 

One of the novel features of our  present endeavour is the systematic derivation of the Noether conserved co-BRST charge for the 
St$\ddot u$ckeberg-modified 4D massive Abelian 2-form theory which is found to be non-nilpotent 
of order two (i.e. $Q_d^2 \neq 0$). In our earlier work [40], we have {\it not} considered the off-shell nilpotency of the co-BRST and anti-co-BRST charges.
Instead, we have focused {\it only} on the derivation of the off-shell nilpotent versions of the (anti-)BRST charges from 
the standard non-nilpotent Noether (anti-)BRST charges. 
 In our present endeavor, however, we have very 
systematically derived the off-shell nilpotent version of the co-BRST charge from the non-nilpotent {\it standard} 
Noether conserved co-BRST charge. It is gratifying to state that our theoretical proposal 
(in the context of the (anti-)BRST charges [40]) is valid in the context of the co-BRST symmetry
and corresponding derivation of the off-shell nilpotent co-BRST charge, too.

We would like to briefly mention the importance of the new fields that have been introduced  in the cases of (i) the 2D {\it modified}
massive Abelian 1-form, and (ii) the 4D {\it modified} massive Abelian 2-form theories where the modifications in the   
St$\ddot u$ckeberg-technique have been taken into account (cf. Subsecs. 6.1 and 6.2). It turns out that, in the case of the 2D {\it modified} 
Proca theory the pseudo-scalar field $\tilde\phi$ is endowed with the negative kinetic term [cf. Eq. (85)]. However, it 
satisfies the usual Klein-Gorden equation $(\Box  + m^2) = 0$. This observation shows that the pseudo-scalar field is a relativistic 
field/particle with a well-defined rest mass equal  to  $m$ but it carries a negative kinetic term. Exactly similar kinds of statements 
can be made about the axial-vector field $\tilde\phi_\mu$ and the pseudo-scalar field $\tilde\phi$ for the 4D Lagrangian density (109)
where we observe that these {\it new} exotic fields (that have been introduced in the theory due to the {\it modification} in the 
St$\ddot u$ckeberg-technique) have {\it negative}  kinetic terms but they carry  the well-defined rest mass. 
Such kinds of fields/particles are the candidates of dark matter (see, e.g. [54, 55]) and these have been called 
a ``phantom" and/or ``ghost" fields in the context of the cyclic, bouncing and self-accelerated cosmological models of 
Universe (see, e.g. [51-53] and references therein). In our studies, these fields appear on the basis of symmetry transformations 
of our 2D and 4D theories.

In a recent set of interesting works [58-60], the St$\ddot u$ckelberg-modified (SUSY) quantum electrodynamics and some other aspects of 
(non-)interacting Abelian gauge theories have been considered and an ultralight dark matter candidate has been proposed.
 It will be nice to apply the BRST approach to these examples in our future endeavors. Moreover,
we plan to capture various aspects of the (non-)Abelian $p = 4, 5$ form gauge theories from various theoretical angels in our future endeavors 
(especially in the $D = 8, 10$ dimensions of spacetime [61] where {\it these} 
(non-)Abelian theories might turn out to be the field-theoretic examples of Hodge theory within the framework of BRST formalism.\\

\noindent
{\bf Acknowledgments}\\

\noindent
One of us (AKR) thankfully acknowledges the financial support from the 
Institution of Eminence (IoE) {\it Research Incentive Grant} of PFMS (Scheme No. 3254-World Class Institutions)
 to which Banaras Hindu University, Varanasi, belongs. 
 All the authors dedicate their present work, very humbly and respectfully, to the memory of Prof. G. Rajasekaran 
who was one of the very influential and prominent mentors of the theoretical high energy physics group at BHU, Varanasi, and 
who passed away on May 29, 2023. Fruitful comments by our esteemed {\it Reviewer} are gratefully acknowledged, too.\\

\noindent
{\bf Conflicts of Interest}\\

\noindent
The authors declare that there are no conflicts of interest. \\

\noindent
{\bf Data Availability}\\

\noindent
No data were used to support this study.

\begin{center}
{\bf Appendix A: On the Derivation of (85) in Covariant Notation}\\
\end{center}

\noindent
In Subsec. 6.1, we have derived the well-defined form of the 2D Lagrangian density for the {\it modified} Proca theory with the 
gauge-fixing term in a non-covariant manner because we have focused on: $F_{01} = \partial_0\, A_1 - \partial_1\, A_0$
and applied the {\it modified} form of St$\ddot u$ckelberg-technique [cf. Eq (79)]. The purpose of 
our present Appendix is to derive (85) in a covariant fashion as we have derived in the context of the 
{\it modified} massive Abelian 2-form theory (see, e.g. [50]). We begin with the  Lagrangian density of the 2D Proca theory 
(74) and focus on the individual terms. It turns out that, under (79),  the kinetic term of 
(74) transforms to 
\[
-\, \frac{1}{4}\,F^{\mu\nu}\, F_{\mu\nu} \longrightarrow -\, \frac{1}{4}\,F^{\mu\nu}\, F_{\mu\nu} 
\mp \frac{1}{2\,m}\, F^{\mu\nu}\,\Sigma_{\mu\nu}
- \, \frac{1}{4\,m^2}\, \Sigma^{\mu\nu}\,\Sigma_{\mu\nu}, 
\eqno (A.1)
\]
where the {\it new}  notation $\Sigma_{\mu\nu} $ is an antisymmetric $(\Sigma_{\mu\nu} = -\, \Sigma_{\nu\mu} )$ tensor
that is defined in terms of the derivatives on the pseudo-scalar field $(\tilde\phi)$ as follows 
\[
\Sigma_{\mu\nu} = (\varepsilon_{\mu\rho}\,\partial_\nu - \varepsilon_{\nu\rho}\,\partial_\mu)\,(\partial^\rho\,\tilde\phi),  
\eqno (A.2)
\]
where $\varepsilon_{\mu\nu}$ is the antisymmetric ($\varepsilon_{\mu\nu} = -\, \varepsilon_{\nu\mu}$) 2D Levi-Civita 
tensor and  $\tilde\phi$ is the pseudo-scalar field (that is present in the {\it modified} form of 
the St$\ddot u$ckelberg-technique [cf. Eq (79)]). For the 2D Proca theory, it is clear that 
\[
-\, \frac{1}{4}\,F_{\mu\nu}\, F^{\mu\nu} =  \frac{1}{2}\, E^2 
\equiv \frac{1}{2}\, (-\, \varepsilon^{\mu\nu}\, \partial_\mu\, A_\nu)\, (-\, \varepsilon^{\alpha \beta }\, \partial_\alpha \, A_\beta), 
\eqno (A.3)
\]
where $E =  \partial_0\, A_1 - \partial_1\, A_0$ has been expressed in its covariant form as: $E = -\, \varepsilon^{\mu\nu}\, \partial_\mu\, A_\nu$. 
The {\it second} term of (A.1) is a term that contains higher derivatives (e.g. {\it three} derivatives for our 2D massive Abelian 1-form theory). 
To get rid of {\it one}  higher derivative,  we see that the {\it second}  term can be expressed as 
 \[
\pm \frac{1}{m}\,F^{\mu\nu}\, (\varepsilon_{\nu\lambda }\, \partial_\mu\, \partial^\lambda\, \tilde\phi)
\equiv \mp \frac{1}{m}\, (\partial_\mu\, F^{\mu\nu)}) \, \varepsilon_{\nu\lambda } \partial^\lambda\, \tilde\phi,  
\eqno (A.4)
\]
where we have dropped a total spacetime derivative and used the antisymmetric properties of $F^{\mu\nu}$
and $\Sigma_{\mu\nu}$. Using the on-shell condition: $\partial_\mu\, F^{\mu\nu} + m^2\, A^\nu = 0$
(which is equivalent to $(\Box + m^2 )\, A_\mu = 0$ provided we take into account the subsidiary condition:
$\partial \cdot A = 0 $ for $m^2 \ne 0$), we can re-express (A.4) as:  
\[
\pm m\, A^\nu\, \varepsilon_{\nu\lambda}\, \partial^\lambda\, \tilde\phi 
= \mp m \, (\varepsilon^{\nu\lambda}\, \partial_\lambda\, A_\nu)\, \tilde\phi
\equiv \mp m\, (- \, \varepsilon^{\mu\nu}\, \partial_\mu\, A_\nu)\, \tilde\phi. 
\eqno (A.5)
\]
To derive (A.5),  we have used the on-shell conditions: $\partial_\mu\, F^{\mu\nu} + m^2\ A^\nu = 0$
and $(\Box + m^2 )\, A_\mu = 0$ in an ad-hoc and arbitrary fashion. However, we shall see that the appropriately 
defined Lagrangian density (with an appropriate gauge-fixing term) for the {\it modified} 2D Proca  theory will produce these 
{\it equivalent} equations as the  EL-EoMs.

At this stage, let us focus on the {\it third}  term of (A.1) which contains {\it four} derivatives. We can write  explicitly 
this term as: 
\[
- \, \frac{1}{4\,m^2}\, \Sigma^{\mu\nu}\,\Sigma_{\mu\nu} = 
- \, \frac{1}{4\,m^2}\,\Big[(\varepsilon^{\nu \lambda }\,\partial^\mu - \varepsilon^{\mu \lambda }\,\partial^\nu) \Big](\partial_\lambda \,\tilde\phi)\, 
\Big[(\varepsilon_{\nu\rho}\,\partial_\mu - \varepsilon_{\mu\rho}\,\partial_\nu)\Big](\partial^\rho\,\tilde\phi),  
\eqno (A.6)
\]
which will lead to {\it four} terms. However, it turns out that {\it two} of them are 
total spacetime derivatives which can be automatically ignored as they  are a part of the Lagrangian density. Only {\it two} terms contribute {\it equally} 
which can be added together to produce:                                        
\[
- \, \frac{1}{2\,m^2}\,(\varepsilon^{\nu \lambda }\,\partial^\mu \partial_\lambda \,\tilde\phi)
\,(\varepsilon_{\nu \rho }\,\partial_\mu \partial^\rho \,\tilde\phi)  
\equiv  \frac{1}{2\,m^2}\,(\Box \, \partial_\lambda \,\tilde\phi)\, (\partial^\lambda \,\tilde\phi), 
\eqno (A.7)
\]
where we have dropped a total spacetime derivative and have taken: 
$\varepsilon^{\nu \lambda}\, \varepsilon_{\nu \rho} = -\, \delta^\lambda_\rho$. 
To get rid of the  higher derivative in the above equation, we take the help of the on-shell condition: 
$(\Box + m^2 )\, \tilde\phi = 0 \Longrightarrow  (\Box + m^2 )\, \partial_\mu\, \tilde\phi = 0$. 
Finally, we obtain,  from (A.7), the following
\[
\frac{1}{2}\,\partial_\mu\, \tilde\phi\, \partial^\mu\, \tilde\phi  = -\, \frac{1}{2} \tilde\phi\, \Box\, \tilde\phi
\;\; \equiv \;\;+ \frac{1}{2}\, m^2\,\tilde\phi^2,
\eqno (A.8)
\]
where we have used: $(\Box + m^2 )\, \tilde\phi = 0$ in  an ad-hoc and arbitrary fashion. 
However, we shall show its sanctity from the 
appropriately defined 2D Lagrangian density.

Taking into account the equations (A.3), (A.5) and (A.8), we observe that (A.1) can be expressed 
in the language of the renormalizable terms for our 2D theory as: 
\[
-\, \frac{1}{4}\,F^{\mu\nu}\, F_{\mu\nu} \longrightarrow 
\frac{1}{2}\, \Big[ (-\, \varepsilon^{\mu\nu}\, \partial_\mu\, A_\nu)\, (-\, \varepsilon^{\alpha \beta }\, \partial_\alpha \, A_\beta)
\mp 2\, m\, (- \, \varepsilon^{\mu\nu}\, \partial_\mu\, A_\nu)\, \tilde\phi +  m^2\,\tilde\phi^2  \Big]
\]
\[
=  \frac{1}{2}\,  \Big[ (-\, \varepsilon^{\mu\nu}\, \partial_\mu\, A_\nu) \mp m\, \tilde\phi  \Big]^2 
\equiv \frac{1}{2}\,\Big[ E  \mp m\, \tilde\phi \Big]^2.  
\eqno (A.9)
\]
Thus, we have obtained a well-defined transformations (A.9) from the transformations (A.1)   as far as our {\it modified} 2D Proca 
theory is concerned. It should be noted that the {\it latter} contained the higher derivatives
in the {\it second} and {\it third} terms. 
 We now focus on the transformation of the mass term of (74) under the {\it modified} form of the 
St$\ddot u$ckelberg-technique in (79), namely; 
\[
\frac {m^2}{2}\, A_\mu\,A^\mu \longrightarrow \frac {m^2}{2}\, A_\mu\,A^\mu \mp m \,A_\mu \,\partial^\mu\, \phi
+ \frac {1}{2}\, \partial_\mu\, \phi\, \partial^\mu\, \phi
- \frac {1}{2}\,\partial_\mu \,\tilde\phi\,\,\partial^\mu\,\tilde\phi
\mp m\, \varepsilon^{\mu \rho } \, A_\mu\, \partial_\rho \tilde\phi
\]
\[
\equiv 
\frac {m^2}{2}\, A_\mu\,A^\mu \mp m \,A_\mu \,\partial^\mu\, \phi
+ \frac {1}{2}\, \partial_\mu\, \phi\, \partial^\mu\, \phi
- \frac {1}{2}\,\partial_\mu \,\tilde\phi\,\,\partial^\mu\,\tilde\phi
\pm m\, (-\, \varepsilon^{\mu\nu}\, \partial_\mu\, A_\nu) \,\tilde\phi
\eqno (A.10)
\]
where we have dropped a total spacetime derivative term. Taking into account: 
$E =  -\, \varepsilon^{\mu\nu}\, \partial_\mu\, A_\nu$, we observe that the {\it modified} form of the 
2D Proca Lagrangian density (74), along with a gauge-fixing term in the 't Hooft gauge 
[i.e. $-\, (1/2)\, (\partial \cdot A \pm m\, \phi)^2$] is nothing but the sum of (A.9), (A.10) and the 
appropriate  gauge-fixing term that turns out to be equal to (85) which we have derived in a different manner in subsection 6.1.
It is straightforward to check that (85) produces the on-shell conditions:
$(\Box + m^2 )\, A_\mu = 0, \; (\Box + m^2 )\, \phi = 0$ and $(\Box + m^2 )\, \tilde\phi = 0$
which have been taken into considerations to get rid of the higher derivative terms (emerging out from the {\it modified} form of the 
St$\ddot u$ckelberg-technique [79]).

\begin{center}
{\bf Appendix B: On the CF-Type Restrictions}\\
\end{center}

\noindent
The purpose of this Appendix is to show (very concisely) that the Lagrangian density (109) 
respects the anti-BRST symmetry transformations, too, provided the symmetry consideration is discussed on a sub-manifold in the 
quantum Hilbert space of fields where the CF-type restrictions: $B_\mu - \bar B_\mu - \partial_\mu\, \phi = 0$
and $B + \bar B + m\, \phi = 0$ are satisfied. In this context, first of all, we list here the infinitesimal and continuous anti-BRST 
symmetry transformations ($ s_{ab}$) from our earlier work (see, e.g. [40] for details).  
\[
s_{ab} B_{\mu\nu} = - \,(\partial_\mu \bar C_\nu - \partial_\nu \bar C_\mu), \quad 
s_{ab} \bar C_\mu  = - \,\partial_\mu \bar \beta, \quad s_{ab} \phi_\mu = \partial_\mu \bar C - m\, \bar C_\mu, \quad s_{ab} \phi = \rho,\]
\[s_{ab}  C_\mu =  \bar B_\mu, \quad s_{ab} \beta = - \,\lambda, \quad s_{ab} \bar C = -\, m\, \bar \beta, \quad  s_{ab}  C = \bar B, 
\quad s_{ab} B = - \, m\, \rho, \]
\[s_{ab} B_\mu =   \partial_\mu \rho, \qquad 
s_{ab} [\rho, \, \lambda, \, \bar \beta, \, \bar B, \, \, {\cal B}, \, \bar {\cal B}, \, \tilde \phi, \, \bar B_\mu, \, {\cal B}_\mu, \, 
  \tilde\phi_\mu,   H_{\mu\nu\kappa}] = 0,
\eqno (B.1)
\]
which are found to be off-shell nilpotent $(s_{ab}^2 = 0)$ and absolutely anticommuting 
$(s_b\, s_{ab} + s_{ab}\, s_b = 0)$ with the off-shell nilpotent $(s_{b}^2 = 0)$  BRST symmetry transformations 
(110) provided the CF-type restriction: $B_\mu - \bar B_\mu - \partial_\mu\, \phi = 0$, 
$B + \bar B + m\, \phi = 0$ are invoked for the proof of the {\it latter} (see, e.g. [29, 30, 50]). 
Mathematically, this can be expressed as: 
\[~~~~
\{s_b, \, s_{ab}\}B_{\mu\nu} =  \partial_\mu(B_\nu - \bar B_\nu) - \partial_\nu(B_\mu - \bar B_\mu),\]
\[\{s_b, \, s_{ab}\}\Phi_\mu =  \partial_\mu(B + \bar B) + m\, (B_\mu - \bar B_\mu).  
\eqno (B.2) 
\]
In other words, we note that $\{ s_b, \, s_{ab} \} = 0$ in (B.2) provided we invoke the validity of the {\it above} 
CF-typer restrictions. It is straightforward to note that when we apply the anti-BRST symmetry transformations on ${\cal L}_{{\cal B}}$
[cf. Eq. (109)], we obtain the following:  
\[
s_{ab} {\cal L}_{{\cal B}} =  \partial_\mu \bigg[-\, m \, \varepsilon^{\mu\nu\eta\kappa} 
\tilde \phi_\nu \big(\partial_\eta \bar C_\kappa \big) 
+ B^{\mu\nu}\, \partial_\nu\, \rho + (\bar B^\mu + m\, \phi^\mu)\, \rho 
- \lambda\,  \big(\partial^\mu \bar \beta \big)
\]
\[
- \big(\partial^\mu \bar C^\nu - \partial^\nu \bar C^\mu  \big)\, B_\nu   
- \big(\partial^\mu \bar C - m \bar C^\mu\big)\, B \bigg] 
+  \big[B_\mu - \bar B_\mu - \partial_\mu \varphi \big] \big(\partial^\mu \rho \big) 
\]
\[
~~~~~~+  \big(\partial^\mu \bar C^\nu - \partial^\nu \bar C^\mu  \big)\, \partial_\mu\big[B_\nu - \bar B_\nu - \partial_\nu \varphi  \big]
+ m\,  \big(\partial^\mu \bar C - m \bar C^\mu  \big)\,  \big[B_\mu - \bar B_\mu - \partial_\mu \phi \big] 
\]
\[
+ m\, \big[B + \bar B + m \phi \big] \rho
+ \big(\partial^\mu \bar C - m \bar C^\mu  \big)\,\partial_\mu \big[B + \bar B + m \phi \big], ~~~~~~~~~~~~~~~~
\eqno (B.3)
\]
A close look at the above expression shows that the Lagrangian density ${\cal L}_{\cal B}$ [cf. Eq. (109)] respects the anti-BRST
symmetry transformations, too, provided we take into account the validity of the CF-type restrictions: 
$B_\mu - \bar B_\mu - \partial_\mu \phi = 0$ and $B + \bar B + m\, \phi  = 0$. In other words, if we impose the 
CF-type restrictions from {\it outside}, we find that ${\cal L}_{{\cal B}}$ {\it also} 
respects anti-BRST symmetry transformations because this Lagrangian density transforms to: 
\[
s_{ab} {\cal L}_{{\cal B}} =  \partial_\mu \bigg[-\, m \, \varepsilon^{\mu\nu\eta\kappa} 
\tilde \phi_\nu \big(\partial_\eta \bar C_\kappa \big) 
+ B^{\mu\nu}\, \partial_\nu\, \rho + (\bar B^\mu + m\, \phi^\mu)\, \rho 
- \lambda\,  \big(\partial^\mu \bar \beta \big)
\]
\[
- \big(\partial^\mu \bar C^\nu - \partial^\nu \bar C^\mu  \big)\, B_\nu   
- \big(\partial^\mu \bar C - m \bar C^\mu\big)\, B \bigg]. 
\eqno (B.4)
\]
Hence, the action integral: $S = \int d^4 x \, {\cal L}_{{\cal B}} $ respects $(s_{ab}\, S = 0)$
the anti-BRST symmetry transformations (B.1) on the sub-manifold of the quantum Hilbert space of fields where 
the CF-type restrictions (i.e. $B_\mu - \bar B_\mu - \partial_\mu\, \phi = 0$, 
$B + \bar B + m\, \phi = 0$) are satisfied.

We wrap-up this Appendix with the remark that for the {\it massless} Abelian 2-form theory, the Lagrangian density 
(109) reduces to the following form: 
\[
{\cal L}_{\cal B}^{(m = 0)} =  {\cal B}_\mu\,{\cal B}^\mu 
+ {\cal B}^\mu \,\Big[\frac{1}{2}\,\varepsilon_{\mu \nu \lambda \xi }\,\partial^\nu\,B^{\lambda\xi} 
- \partial_\mu \tilde\phi\Big]
+ B^\mu\, \Big[ \partial^\nu\,B_{\nu\mu} - \partial_\mu\,\phi \Big]\,
+ B^\mu\,B_\mu 
\]
\[
+  \partial_\mu\,\bar\beta\, \partial^\mu\,\beta 
+ (\partial_\mu\,\bar C_\nu 
- \partial_\nu\,\bar C_\mu)\,(\partial^\mu\,C^\nu)
+ \Big(\partial \cdot \bar C 
+ \rho\Big)\,\lambda + \Big(\partial \cdot C
- \lambda\Big)\,\rho,
\eqno (B.5)
\]
which transforms under the following anti-BRST symmetry transformations: 
\[
s_{ab} B_{\mu\nu} = - \,(\partial_\mu \bar C_\nu - \partial_\nu \bar C_\mu), \,\,
s_{ab} \bar C_\mu  = - \,\partial_\mu \bar \beta, \,\,
s_{ab}  C_\mu =  \bar B_\mu, \quad s_{ab} \beta = - \,\lambda,  
\]
\[
s_{ab}  C = \bar B, \quad s_{ab} \phi = \rho, \quad s_{ab} B_\mu =   \partial_\mu \rho, \quad 
s_{ab} [  \rho, \, \lambda, \,  \bar \beta, \,\tilde\phi,\,  {\cal B}_\mu, \, \bar B_\mu, \,  H_{\mu\nu\kappa}] = 0,
\eqno (B.6)
\]
to the following total spacetime derivative {\it plus} terms that  vanish due to CF-type restriction: 
\[
s_{ab}\, {\cal L}_{\cal B}^{(m = 0)} =  \partial_\mu\, \Big[   
\bar B^\mu \rho + B^{\mu\nu}\, \partial_\nu\, \rho
- \lambda\, \partial^\mu\, \bar\beta - (\partial^\mu\, \bar C^\nu - \partial^\nu\, \bar C^\mu)\, B_\nu \Big]
\]
\[
~~~~~~~~~  ~~~~~~~~~ +  (B_\mu - \bar B_\mu - \partial_\mu \phi)\, (\partial^\mu\, \rho)
+  (\partial^\mu\, \bar C^\nu - \partial^\nu\, \bar C^\mu)\,\partial_\mu\, (B_\nu - \bar B_\nu - \partial_\nu \phi)
\Big].
\eqno (B.7)
\]
In other words, if we invoke the validity of the CF-type restrictions: 
$B_\mu - \bar B_\mu - \partial_\mu \phi = 0$, we obtain the following
\[
s_{ab}\, {\cal L}_{\cal B}^{(m = 0)} = \partial_\mu\, \Big[   
\bar B^\mu \rho + B^{\mu\nu}\, \partial_\nu\, \rho
- \lambda\, \partial^\mu\, \bar\beta - (\partial^\mu\, \bar C^\nu - \partial^\nu\, \bar C^\mu)\, B_\nu \Big],
\eqno (B.8)
\]
which shows that the action integral: $S = \int d^4 x \, {\cal L}_{{\cal B}}^{(m = 0)}$ respects 
$(s_{ab} \, S = 0)$ the anti-BRST symmetry transformations (B.6). Thus, as far as the symmetry considerations are 
concerned, we have shown that the CF-type restrictions are to be respected if we wish to have 
the BRST and anti-BRST symmetry {\it together} for the Lagrangian density (109) for the 
modified {\it massive} as well as {\it massless} cases of the Abelian 2-form theories. 
The existence of the CF-type restriction(s) is as fundamental at the {\it quantum} level as the existence of the first-class
 constraints for a {\it classical} gauge theory when the {\it latter} is discussed within the framework of BRST formalism. 
The CF-type restriction(s) of the higher Abelian $p$-form ($p = 2, 3$) gauge theories  are connected with the geometrical objects 
called gerbes [62, 63].


\begin{thebibliography}{99}
\bibitem{RPM1}   C. N. Yang,   Einstein's Impact on Theoretical Physics, {\it Physics Today} {\bf 33},  42 (1980)
\bibitem{RPM2}   E. P. Wigner, Symmetry and Conservation Laws,  {\it Physics Today} {\bf 17}, 34 (1964)  
\bibitem{RPM3}   C. N. Yang, Symmetry and Physics, \\ {\it Proceedings of the American Philosophical Society},  {\bf 140},  267 (1996)
\bibitem{RPM4}   R. Jackiw, N. S. Manton,  Symmetries and Conservation Laws in Gauge Theories, \\ {\it Ann. Phys.} {\bf 127}, 257  (1980)            
\bibitem{RPM5}      P. A. M. Dirac, {\it Lectures on Quantum Mechanics} (Belfer Graduate
                    School of Science), Yeshiva University Press, New York (1964)
\bibitem{RPM6}      K. Sundermeyer, {\it Constraint Dynamics, Lecture Notes in Physics}, \\Vol. 169, 
                    Springer-Verlag, Berlin (1982)
\bibitem{RPM7}      E. C. G. Sudarshan, N. Mukunda, {\it Classical Dynamics: A Modern Perspective},\\
                    Wiley, New York (1972)
\bibitem{RPM8}      D. M. Gitman, I. V. Tyutin, {\it Quantization of Fields with Constraints},\\
                    Springer-Verlag, Berlin Heidelberg (1990) 
\bibitem{RPM13}     S. Weinberg, The Making of the Standard Model,  {\it Eur. Phys. J.}  C {\bf 34}, 5 
                     (2004)
\bibitem{RPM9}      Mary K. Gaillard, Paul D. Grannis, Frank J. Sciulli, The Standard Model of Particle Physics,  
                    {\it Rev. Mod. Phys.} {\bf 71}, S96  (1999)             
\bibitem{RPM10}     G. Rajasekaran, {\it Building-up the Standard Model of High Energy Physics}, \\ in ``Gravitation, Gauge Theories and 
                    Early Universe", Eds. B. R. Iyer, {\it etal.} \\ Kluwer Academic Publications, 185 (1989)
\bibitem{RPM11}     T. W. B.  Kibble, The Standard Model of Particle Physics, \\ {\it European Review} {\bf 23}, 36 (2015)
\bibitem{RPM12}     R. Mann, {\it An Introduction to Particle Physics and the Standard Model}, (1st ed.) \\ CRC Press. Boca Raton FL (2010)    
\bibitem{RPM14}     M. B. Green, J. H. Schwarz, E. Witten, {\it Superstring Theory}, Vols. 1 and 2,\\
                    Cambridge University Press, Cambridge (1987)
\bibitem{RPM15}     J. Polchinski, {\it String Theory}, Vols. 1 and 2,\\
                    Cambridge University Press, Cambridge (1998)   
\bibitem{RPM16}     D. Lust, S. Theisen, {\it Lectures in String Theory}, Springer-Verlag, New York (1989) 
\bibitem{RPM17}     K. Becker,  M. Becker, J. H. Schwarz, {\it String Theory and M-Theory},\\ 
                    Cambridge University Press, Cambridge (2007)  
\bibitem{RPM18}     D. Rickles, {\it A Brief History of String Theory From Dual Models to M-Theory},\\ Springer, Germany  (2014)
\bibitem{RPM19}     C. Becchi, A. Rouet, R. Stora,  
                     The Abelian Higgs Kibble Model: Unitarity of the S-Operator, {\it Phys. Lett.} B {\bf 52},  344 (1974)   
\bibitem{RPM20}     C. Becchi, A. Rouet, R. Stora,  Renormalization of the Abelian Higgs-Kibble Model,  {\it Comm. Math. Phys.} {\bf 42},  127  (1975) 
\bibitem{RPM21}     C. Becchi, A. Rouet, R. Stora, Renormalization of Gauge Theories, \\ {\it Ann. Phys.}  (N. Y.) {\bf 98},  287 (1976)  
\bibitem{RPM22}     I. V. Tyutin, Gauge Invariance in Field Theory and Statistical Physics in Operator Formalism, 
                    {\it Lebedev Institute Preprint}, Report Number: {\bf FIAN-39} (1975)\\ (unpublished),
                    {\bf arXiv:0812.0580 [hep-th]} 
\bibitem{nish23}    N. Nakanishi, I. Ojima,  {\it Covariant Operator Formalism of Gauge Theories 
                    and Quantum Gravity},  World Scientific, Singapore (1996) 
\bibitem{nish24}    S. Weinberg, {\it The Quantum Theory of Fields: Modern Applications},  
                    Volume 2, Cambridge University Press, Cambridge (1996)
\bibitem{RPM25}     K. Nishijima, BRS Invariance, Asymptotic Freedom and Color Confinement,  
                    \\ {\it Czechoslovak Journal of Physics} {\bf 46}, 140 (1996) (A Review)
\bibitem{RPM26}     M.  Henneaux,  C. Teitelboim, {\it Quantization of Gauge Systems}, \\Princeton University, New Jersey (1992)
\bibitem{RPM27}     R. Kumar, S. Krishna, A. Shukla, R. P. Malik,  Abelian $p$-Form $(p = 1, 2, 3)$\\ 
                      Gauge Theories as the Field Theoretic Models for the Hodge Theory,\\
                    {\it Int. J. Mod. Phys.} A {\bf 29}, 1450135 (2014) ({\it A Brief Review}) 
\bibitem{RPM28}     S. Gupta, R. P. Malik, A Field-Theoretic Model for Hodge Theory, \\ {\it Eur. Phys. J.} C {\bf 58}, 517 (2008)
\bibitem{RPM29}     R. P. Malik, Abelian 2-Form Gauge Theory: Superfield Approach, \\ {\it Eur. Phys. J.} C {\bf 60}, 457 (2009)
\bibitem{RPM30}     S. Krishna, R. Kumar, R. P. Malik, A Massive Field-Theoretic Model for Hodge Theory,
                    {\it Ann. Phys.} {\bf 414}, 168087 (2020)
\bibitem{RPM31}     E. Witten,  Supersymmetric Quantum Mechanics on the Lattice:
                    I. Loop Formulation,  {\it Nucl. Phys.} B  {\bf 202}, 253 (1982)
\bibitem {RPM32}    A. S. Schwarz, On Quantum Fluctuations of Instantons,\\ 
                    {\it Lett. Math. Phys.}  {\bf 2}, 217 (1978)
\bibitem{RPM33}    R. P. Malik,  New Topological Field Theories in Two Dimensions, \\ {\it J. Phys. A: Math. Gen.} {\bf 34},  4167(2001)
\bibitem{RPM34}    R. P. Malik, Abelian 2-Form Gauge Theory: Special Features, \\ {\it J. Phys. A: Math. Gen.} {\bf 36}, 5095 (2003)
\bibitem{RPM35}    A. K. Rao, R. P. Malik,  Modified Massive Abelian 3-Form Theory: Constraint Analysis,
                   Conserved Charges and BRST Algebra,     {\bf arXiv: 2207.11738  [hep-th]}
\bibitem{RPM36}    B. Chauhan, A. K. Rao, R. P. Malik, Constraints, Symmetry Transformations and Conserved Charges for Massless Abelian 3-Form Theory,\\   
                   {\it Nucl. Phys.} B {\bf 996},  116366 (2023)  
\bibitem{RPM37}    A. K. Rao, R. P. Malik, Nilpotent Symmetries of a Modified Massive Abelian 3-Form Theory: Augmented Superfield Approach,  
                   {\it Nuclear Physics} B {\bf 983}, 115926  (2022)  
\bibitem{RPM38}     A. K. Rao, R. P. Malik, Modified Proca Theory in Arbitrary and Two Dimensions, 
                     {\it Euro. Phys. Lett.} {\bf 135}, 21001 (2021)
\bibitem{RPM39}    B. Chauhan, S. Kumar, A. Tripathi, R. P. Malik,  Modified 2D Proca Theory: Revisited Under BRST and 
                   (Anti-)Chiral Superfield Formalisms, \\ {\it Adv. High Energy Phys.} {\bf 2020}, 3495168 (2020)    
\bibitem{RPM40}    A. K. Rao, A. Tripathi, B. Chauhan, R. P. Malik, Noether Theorem and Nilpotency Property of the (Anti-)BRST
                   Charges in the BRST Formalism: A Brief Review.\\ {\it Universe} {\bf 8},  566 (2022)
\bibitem{RPM41}    H. Ruegg, M. Ruiz-Altab, The Stueckelberge Field. \\ {\it Int. J. Mod. Phys.} 
                   A {\bf 19}, 3265 (2004)
\bibitem{RPM42}    P. Mitra, R. Rajaraman, New Results on Systems with Second-Class Constraints,\\
                   {\it Ann. Phys.} {\bf 203}, 137 (1990)
\bibitem{RPM43}    P. Mitra, R. Rajaraman, Gauge-Invariant Reformulation of Theories with Second-Class      
                   Constraints, {\it  Ann. Phys.} {\bf 203}, 157 (1990)
\bibitem{RPM44}    R. Kumar,   S. Krishna, Augmented Superfield Approach to Gauge-Invariant Massive 2-Form Theory, 
                   {\it Eur. Phys. J.} C {\bf 77}, 387 (2017)
\bibitem {RPM45}     T. Eguchi, P. B. Gilkey, A. Hanson, Gravitation, Gauge Theories and Differential Geometry. 
                    {\it Physics Reports} {\bf 66}, 213 (1980)
\bibitem {RPM46}   S. Mukhi, N.  Mukunda, {\it Introduction to Topology, Differential  Geometry 
                    and Group Theory for Physicists}, {Wiley Eastern Private  Limited, New Delhi} (1990)   
\bibitem{RPM47} 	K. Nishijima, The Casimir Operator in the Representations of BRS Algebra,\\
					{\it Prog. Theor. Phys.} {\bf 80}, 897 (1988)  
\bibitem{RPM48}     J. W. van Holten, The BRST Complex and the Cohomology of Compact \\Lie Algebras,
                    {\it Phys. Rev. Lett.} {\bf 64}, 2863 (1990)  
\bibitem {RPM49}    M. G{\"o}ckeler, T. Sch{\"u}cker,  {\it Differential Geometry, Gauge Theories   
                    and Gravity},\\ {Cambridge University Press, Cambridge} (1987)
\bibitem {RPM50}    A. K. Rao, R. P. Malik, Modified St$\ddot u$ckelberg Formalism:
                    Free Massive Abelian 2-Form Theory in 4D, 
                   {\it Universe} {\bf 9}, 191 (2023)  
\bibitem{RPM55}    P. J. Steinhardt, N. Turok, A Cyclic Model of the Universe, \\{\it Science} {\bf 296}, 1436 (2002) 
\bibitem{RPM56}    Y. F. Cai, A. Marcian, D.-G. Wang, E. Wilson-Ewing, Bouncing Cosmologies with Dark Matter and Dark Energy,
                   {\it Universe} {\bf 3}, 1 (2017)
\bibitem{RPM57}    K. Koyama, Ghost in Self-Accelerating Universe, \\{\it Classical and Quantum Gravity} 
                   {\bf 24}, R231 (2007) 
\bibitem{SK26}     V. M. Zhuravlev, D. A. Kornilov, E. P. Savelova, The Scalar
                   Fields with Negative Kinetic Energy, Dark Matter and Dark
                   Energy,\\ {\it General Relativity and Gravitation} {\bf 36}, 1736 (2004)
\bibitem{SK27}     Y. Aharonov, S. Popescu, D. Rohrlich, L. Vaidman, Measurements,
                   Errors, and Negative Kinetic Energy, {\it Phys. Rev.} A {\bf 48}, 4084 (1993)                     
\bibitem {RPM51}    R. P. Malik, BRST Cohomology and Hodge Decomposition Theorem in Abelian Gauge Theory, 
                   {\it Int. J. Mod. Phys.} A {\bf 15}, 1685 (2000)
\bibitem {RPM52}    E. Harikumar, R. P. Malik, M. Sivakumar, Hodge Decomposition Theorem for Abelian 2-Form           
                    Theory, {\it J. Phys. A: Math. Gen.} {\bf 33}, 7149 (2000) 
\bibitem{RPM53}      R. Vinze, T. R. Govindarajan, A. Misra, P. Ramadevi,  Stuckelberg SUSY QED and Infrared
                    Problem, {\it Mod. Phys. Lett.} A {\bf 35}, 2050303 (2020)  
\bibitem{RPM54}     T. R. Govindarajan, N. Kalyanapuram, Infrared Effects and the Soft Photon Theorem in
                    Massive QED, {\it Mod. Phys. Lett.} A {\bf 34}, 1950009 (2019)
\bibitem{RPM55}     T. R. Govindarajan, N. Kalyanapuram, Stueckelberg Bosons as an Altralight Dark Matter   
                    Candidate, {\it Mod. Phys. Lett.} A {\bf 33}, 1950309 (2019)
\bibitem{RPM58}     R. P. Malik, {\it et.al.},  in preparation           
\bibitem{RPM56}     L. Bonora, R. P. Malik, BRST, Anti-BRST and Gerbes, {\it Phys. Lett.} B {\bf 655},  75 (2007) 
\bibitem{RPM57}     L. Bonora, R. P. Malik, BRST, Anti-BRST and Their Geometry,\\ {\it J. Phys.} A: {\it Math. Theor.} {\bf 43}, 375403 (2010)              






                 
                    

                
\end{thebibliography}
\end{document}